\documentclass[a4paper,british,useAMS,usenatbib]{mn2e}
\pdfoutput=1
\usepackage{lmodern}

\usepackage[T1]{fontenc}
\usepackage[latin9]{inputenc}
\usepackage{listings}
\usepackage{babel}
\usepackage{array}
\usepackage{amstext}
\usepackage{graphicx}
\usepackage[authoryear]{natbib}
\usepackage[unicode=true,
 bookmarks=true,bookmarksnumbered=false,bookmarksopen=false,
 breaklinks=false,pdfborder={0 0 1},backref=false,colorlinks=false]
 {hyperref}

\makeatletter

\pdfpageheight\paperheight
\pdfpagewidth\paperwidth

\newcommand{\noun}[1]{\textsc{#1}}
\providecommand{\tabularnewline}{\\}

\renewcommand\[{\begin{equation}}
\renewcommand\]{\end{equation}}
\usepackage{aas_macros}

\makeatother

\begin{document}
\title[GAMA: SIGMA]{Galaxy And Mass Assembly (GAMA): Structural Investigation of Galaxies via Model Analysis (SIGMA)}

\author[L.~S.~Kelvin et al.]{
%%%%% Authors start here %%%%%
\parbox{\textwidth}{
\raggedright
Lee~S.~Kelvin,$^{1,2}$\thanks{E-mail: lee.kelvin@st-andrews.ac.uk}
Simon~P.~Driver,$^{1,2}$
Aaron~S.~G.~Robotham,$^{1,2}$
David~T.~Hill,$^{1}$
Mehmet~Alpaslan,$^{1,2}$
Ivan~K.~Baldry,$^{3}$
Steven~P.~Bamford,$^{4}$
Joss~Bland-Hawthorn,$^{5}$
Sarah~Brough,$^{6}$
Alister~W.~Graham,$^{7}$
Boris~Häussler,$^{4}$
Andrew~M.~Hopkins,$^{6}$
Jochen~Liske,$^{8}$
Jon~Loveday,$^{9}$
Peder~Norberg,$^{10}$
Steven~Phillipps,$^{11}$
Cristina~C.~Popescu,$^{12}$
Matthew~Prescott,$^{3}$
Edward~N.~Taylor,$^{5,13}$
and Richard~J.~Tuffs$^{14}$
}\vspace{0.4cm}\\
\parbox{\textwidth}{
$^{1}$School of Physics and Astronomy, University of St. Andrews, North Haugh, St. Andrews, Fife, KY16 9SS, UK\\
$^{2}$International Centre for Radio Astronomy Research, 7 Fairway, The University of Western Australia, Crawley, Perth, Western Australia 6009, Australia\\
$^{3}$Astrophysics Research Institute, Liverpool John Moores University, Twelve Quays House, Egerton Wharf, Birkenhead, CH41 1LD, UK\\
$^{4}$The Centre for Astronomy and Particle Theory, The School of Physics and Astronomy, University of Nottingham, University Park, Nottingham, NG7 2RD, UK\\
$^{5}$Sydney Institute for Astronomy, University of Sydney, School of Physics A28, NSW 2006, Australia\\
$^{6}$Australian Astronomical Observatory, PO Box 296, Epping, NSW 1710, Australia\\
$^{7}$Centre for Astrophysics and Supercomputing, Swinburne University of Technology, Hawthorn, Victoria 3122, Australia\\
$^{8}$European Southern Observatory, Karl-Schwarzschild-Str. 2, 85748 Garching, Germany\\
$^{9}$Astronomy Centre, University of Sussex, Falmer, Brighton, BN1 9QH, UK\\
$^{10}$Institute for Computational Cosmology, Department of Physics, Durham University, South Road, Durham, DH1 3LE, UK\\
$^{11}$Astrophysics Group, H.H. Wills Physics Laboratory, University of Bristol, Tyndall Avenue, Bristol, BS8 1TL, UK\\
$^{12}$Jeremiah Horrocks Institute, University of Central Lancashire, Preston, PR1 2HE, UK\\
$^{13}$School of Physics, The University of Melbourne, Parkville, VIC 3010, Australia\\
$^{14}$Max-Planck-Institut fuer Kernphysik, Saupfercheckweg 1, D-69117 Heidelberg, Germany\\
}
%%%%% Authors stop here %%%%%
\vspace{-0.75cm}
}

\date{Accepted 2011 December 7}

\pagerange{\pageref{firstpage}--\pageref{lastpage}} \pubyear{2011}

\maketitle

\label{firstpage}
\begin{abstract}
We present single-Sérsic two-dimensional model fits to $167,600$
galaxies modelled independently in the $ugrizYJHK$ bandpasses using
reprocessed Sloan Digital Sky Survey Data Release Seven (SDSS DR7)
and UKIRT Infrared Deep Sky Survey Large Area Survey (UKIDSS-LAS)
imaging data available from the GAMA database. In order to facilitate
this study we developed SIGMA, an R wrapper around several contemporary
astronomy software packages including Source Extractor, PSF Extractor
and GALFIT 3. SIGMA produces realistic 2D model fits to galaxies,
employing automatic adaptive background subtraction and empirical
PSF measurements on the fly for each galaxy in GAMA. Using these results,
we define a common coverage area across the three GAMA regions containing
$138,269$ galaxies. We provide Sérsic magnitudes truncated at $10$
$r_{e}$ which show good agreement with SDSS Petrosian and GAMA photometry
for low Sérsic index systems ($n<4$), and much improved photometry
for high Sérsic index systems ($n>4$), recovering as much as $\Delta m=0.5$
magnitudes in the $r$ band. We employ a $K$ band Sérsic index/$u-r$
colour relation to delineate the massive ($n>\sim2$) early-type galaxies
(ETGs) from the late-type galaxies (LTGs). The mean Sérsic index of
these ETGs shows a smooth variation with wavelength, increasing by
$30\%$ from $g$ through $K$. LTGs exhibit a more extreme change
in Sérsic index, increasing by $52\%$ across the same range. In addition,
ETGs and LTGs exhibit a $38\%$ and $25\%$ decrease respectively
in half-light radius from $g$ through $K$. These trends are shown
to arise due to the effects of dust attenuation and stellar population/metallicity
gradients within galaxy populations.
\end{abstract}
\begin{keywords}
galaxies: structure -- galaxies: fundamental parameters -- astronomical data bases: miscellaneous -- catalogues
\vspace{-0.5cm}
\end{keywords}

\section{Introduction}

\label{sec:intro}The shapes and sizes of galaxies are not random
but are defined by the orbital motions of their constituent stellar
populations, arranging themselves into elliptical, bulge, disk, and
bar like structures. Exactly why and how these structures come about
is somewhat a mystery which no doubt relates to a complex formation
history involving collapse, merging, infall, secular evolution, and
feedback processes as well as the precise nature of the coupling between
the dark matter, gas, dust and stars and the influence of the larger
halo in which the galaxy might reside (group, cluster, etc.), and
the broader environment (filament, void, nexus, etc.). The combination
of variations in, for example, galaxy structure, formation history,
evolution and relative environment lead to distinct measurable effects
on global galaxy parameters such as colour, concentration and size.
The ultimate goal of structural analysis is to inform this discussion
by robustly isolating and quantifying these parameters and exploring
correlations between these properties and those obtained by other
means, such as dynamical information.

Once the underlying structure of a galaxy is understood the overarching
morphological class may be determined, and from this we can explore
correlations with, for example, environment through the well known
morphology-density relation \citep{Dressler1980}, i.e., the apparent
preference for red, passive galaxies in the dense cores of galaxy
groups and clusters. Several mechanisms have been suggested to explain
this feature, most notably the combined effects of strangulation \citep{Larson1980,Kauffmann1993,Diaferio2001},
ram pressure stripping \citep{Gunn1972}, harassment \citep{Moore1996}
and tidal interactions and merging \citep{Park2008}. Recent studies
by, e.g., \citet{VanderWel2008,Welikala2008,Welikala2009,Bamford2009}
and others confirm this morphology-environment connection; however,
they suggest that the relation between structure and morphology is
less apparent. Indeed, it appears that the mass of a galaxy rather
than the environment in which it resides is more influential in determining
its structure, highlighting the importance of stellar mass estimates.

As an example of the connection between galaxy structure and the physical
processes of galaxy formation \citet*{Dalcanton1997} and independently
\citet*{Mo1998}, both following on from \citet*{Fall1980}, relate
the scale length of the disk to the angular momentum of a galaxy's
dark matter halo. In addition, numerous properties of the bulge component
are now known to relate to the mass of the super-massive black hole
\citep*[e.g.,][]{Haring2004,Novak2006,Graham2007}. Variations in
structural properties as a function of wavelength \citep[e.g.,][]{LaBarbera2010a}
enable the extraction of colour gradients, potentially implying the
direction of disk growth (e.g., inside out, \citealt{Barden2005,Bakos2008,Trujillo2009,Wang2010}),
or arguing for the redistribution of populations from the inner to
outer regions \citep{Roskar2008}, possibly coupled with bar formation
\citep{Debattista2006}.

The physics underpinning galaxy structure is relatively immature,
despite the very long history dating back to Knox-Shaw, Reynolds and
Hubble and essentially consists of spot-check simulations which focus
on a particular phenomena in a non-cosmological context (for recent
developments see \citealt{Roskar2010} and \citealt{Agertz2010}).
For example, numerical models can readily produce bar, pseudo-bulge,
spiral patterns, and spheroidal structures through coupled rotation,
secular evolution, shock-wave propagation, and merging history. Until
recently the very thin nature of the spiral disks has presented a
particular challenge for numerical models, with numerical simulations
in particular forming small thick disks, mainly because of the high
level of merging \citep{Navarro1997}. This is in stark contrast to
a number of independent empirical studies \citep[e.g.,][]{Driver2007,Gadotti2009,Tasca2011}
which estimate that approximately $60\%$ of the stellar mass in the
universe today lies within disk systems, suggesting a more quiescent
merger history (however, see \citealp{Hopkins2010} on the stability
of gas rich disks). In addition, studies by \citet{MenendezDelmestre2007}
suggest that up to $67\%$ of spiral galaxies contain a barred structure,
further complicating simulation efforts. However, numerical simulations
are now starting to produce realistic disk systems \citep{Governato2007,Governato2009,Agertz2009,Agertz2010}
albeit with heavily controlled initial conditions, more quiescent
merger histories and a greater degree of gas infall.

Beyond a distance of $\sim100$ Mpc detailed structural studies have
been relatively rare and mostly confined to the deep yet very narrow
pencil beam surveys from the Hubble Space Telescope (HST). It was
only following the refurbishment of HST that structural analysis once
again became a mainstream study \citep{Driver1995a,Driver1995b,Driver1998}.
HST provides kpc resolution across the full path length of the Universe,
which is now also becoming possible with AO ground-based systems \citep{Huertas-Company2007}.
The conjunction of development in numerical models and this new ability
to resolve the shapes and sizes of galaxies at any distance has led
to a dramatic renewed interest in structural analysis. One interesting
claim is the apparent remarkable growth of galaxy sizes since intermediate
redshifts (e.g., \citealp{Trujillo2005,Barden2005,McIntosh2005,Trujillo2006,Trujillo2007,Weinzirl2011}),
potentially supporting the notion of recent growth in disk systems
following an earlier aggressive merger phase at $z\sim2$ \citep[see][]{Driver1996,Driver2005}.
An alternative suggestion which does not require galaxy growth through
mergers is the transformation of some of these so-called {}``red
nuggets'' into the bulges of disk galaxies via the accretion of a
cold gas disk \citep{Graham2011}.

However, structural analysis is not trivial to implement and interpret
correctly, and is plagued by a number of key issues. In particular:
\begin{enumerate}
\item Wavelength bias. At different wavelengths, light traces varying stellar
populations (\citealp{Block1999}). Typically this is a young stellar
population at shorter wavelengths and an older stellar population
at longer wavelengths. For this reason, it is vital when comparing
structural properties to compare properties measured at the same rest
wavelength.
\item Dust attenuation. Dust is predicted to modify not only the recovered
total flux as a function of wavelength (e.g., \citealp{Tuffs2004,Pierini2004})
but also galaxy sizes, shapes and concentrations (see for example
\citealp{Mollenhoff2006,Graham2008b}). Dust can vary enormously from
system to system with significant environmental dependencies and strong
evolution with redshift. Each individual galaxy ultimately requires
either a dust correction or analysis at rest-NIR wavelengths where
dust will have a smaller impact (see photon escape fraction curve
in \citealt{Driver2008}). It is obvious that any attempt to model
the dust in galaxies raises the larger problem of degeneracies appearing
between additive and subtractive flux components.
\item Local minima during the minimisation process. For a single profile
fit there are often 7 free parameters, with that number rising for
multi-component fits. The surface within this parameter space is known
to be complex, containing multiple local minima representing potentially
non-physical results, e.g., a bulge which contributes significantly
more flux to the outer regions of a galaxy than the disk (\citealt{Graham2001}).
Other than manual checks of the output, various methods may be employed
to reduce the risk of divergence on an incorrect result including;
constraints applied during the minimisation routine and employing
an automated logical filter (e.g., \citealt{Allen2006}).
\item What lies below the limiting isophote. Whilst the surface brightness
profile of some galaxies behaves as expected out to very faint magnitudes
(e.g., NGC 300:\citealp{Bland-Hawthorn2005,Vlajic2009}, NGC 7793:
\citealp{Vlajic2011}), the potential myriad of phenomena present
in the outer wings of many systems may cause deviations away from
a typical light profile. These include truncated and anti-truncated
disks \citep{Erwin2005,Pohlen2006}, UV excesses \citep{Bush2010},
tidal debris, halos \citep{Barker2009,McConnachie2009} and minor
merger fossil records \citep{Martinez-Delgado2010}. In fact, the
outer regions of galaxies may defy any systematic profile fitting
into a restricted number of structures. The accuracy of any estimation
of the background sky and gradients therein will also no doubt affect
analyses of these outer structures.
\item The number of components required. When considering the structure
of very nearby galaxies, the deeper one looks the more one finds.
Some galaxies, even in the dust free $3.8$ micron bands, require
up to six components \citep{Buta2010} before a satisfactory fit can
be obtained. In many cases there is uncertainty as to how many components
are required, how to quantitatively decide this in an automated and
repeatable fashion, and which components are fundamental and which
perhaps secondary. For example, should bar and pseudo-bulge flux be
incorporated into a single disk model or kept distinct.
\item Sky estimation. Understanding the background sky level at the position
of your primary object of interest is crucial in producing meaningful
measurements of that galaxy. Considerations must be made in regards
accuracy and speed of estimating the background.
\item Systematic selection bias. Sample bias will be introduced due to size,
resolution, orientation, profile shape and smoothing scale limitations
\citep{Phillipps1986,Driver1999}. Samples of galaxies are usually
selected based on global criteria, such as magnitude. However, it
becomes non-trivial to transcribe these global limits into appropriate
limits for galaxy sub-components, e.g., a certain type of disk may
only have been detected because it also contains a prominent bulge.
\end{enumerate}
There are several publicly-available galaxy modelling codes in common
usage including GIM2D \citep{Simard2002}, BUDDA \citep{DeSouza2004},
GASPHOT \citep{Pignatelli2006} and GALFIT 3 \citep{Peng2010a}. In
addition, there are a number of software pipelines, wrappers around
contemporary astronomy software, that aim to automate the process
of galaxy modelling including GALAPAGOS (Barden et al., 2011, submitted)
and \noun{PyMorph} \citep{Vikram2010}. These packages all have their
advantages and disadvantages and have been compared in a number of
external studies (e.g. \citealp{Haussler2007,Hoyos2011}) in addition
to their own internal comparisons, and so we refer the reader to these
papers for discussions of the pros and cons between 1D v 2D fitting
and the actual minimisation algorithms employed. For this body of
work GALFIT was chosen for its ease of use and high-quality realistic
model outputs, plus the ability to perform simultaneous modelling
of nearby neighbours to the primary galaxy.

In this series of papers we introduce and utilise SIGMA, an automated
code designed to produce single-Sérsic and multi-component profile
fits for galaxies in the GAMA dataset. Using SIGMA, this paper presents
one of the largest catalogues of multi-wavelength single-Sérsic model
fits; $167,600$ galaxies modelled independently across $9$ bandpasses.
This catalogue is currently in use to aid in measurement of the evolution
in the size-(stellar mass) distribution of galaxies (Baldry et al.,
2011); explore star formation trends as a function of morphology (Bauer
et al., in prep.); to further understand the cosmic SED from 0.1 micron
to 1 mm (Driver et al., 2011); to apply dust corrections to galaxies
observed at multiple inclinations (Grootes et al., 2012); to explore
the dust properties and star-formation histories of local submillimetre
selected galaxies \citet{Rowlands2011}; better constrain stellar
mass measurements by providing total flux corrections \citep{Taylor2011};
comment on the quenching of star formation in the local universe (Taylor
et al., in prep.); explore the relation between galaxy environments
and their star formation rate variations (Wijesinghe et al., in prep.);
provide a new method for automatic morphological classification (Kelvin
et al., in prep.); and further explore the relation between environment
(i.e., halo mass; \citealp{Robotham2011}), morphology and structure
(Kelvin et al., in prep.). Future studies building on SIGMA will incorporate
advanced logical filtering and profile management to produce multi-component
fits for a low redshift sample, allowing full structural decomposition
into bulge-disk-bar, etc. (Kelvin et al., in prep). 

This paper is organised as follows. We first outline the GAMA data
in Section \ref{sec:data}. We describe the SIGMA (Structural Investigation
of Galaxies via Model Analysis) pipeline developed to process this
data and produce robust 2D galaxy models in Section \ref{sec:sigma}
and present SIGMA single-Sérsic results for $167,600$ objects modelled
independently in $ugrizYJHK$ from the GAMA Phase I study in Section
\ref{sec:output}. From this large catalogue we establish a common
coverage sample of $138,269$ galaxies. Finally, we further explore
the wavelength dependence on recovered structural parameters in Section
\ref{sec:wavelength}. A standard cosmology of $H_{0}=70$ km s$^{-1}$
Mpc$^{-1}$, $\Omega_{m}=0.3$, $\Omega_{\Lambda}=0.7$ is assumed
throughout.

\section{Data}

\label{sec:data}
\begin{figure*}
\includegraphics[width=1\textwidth]{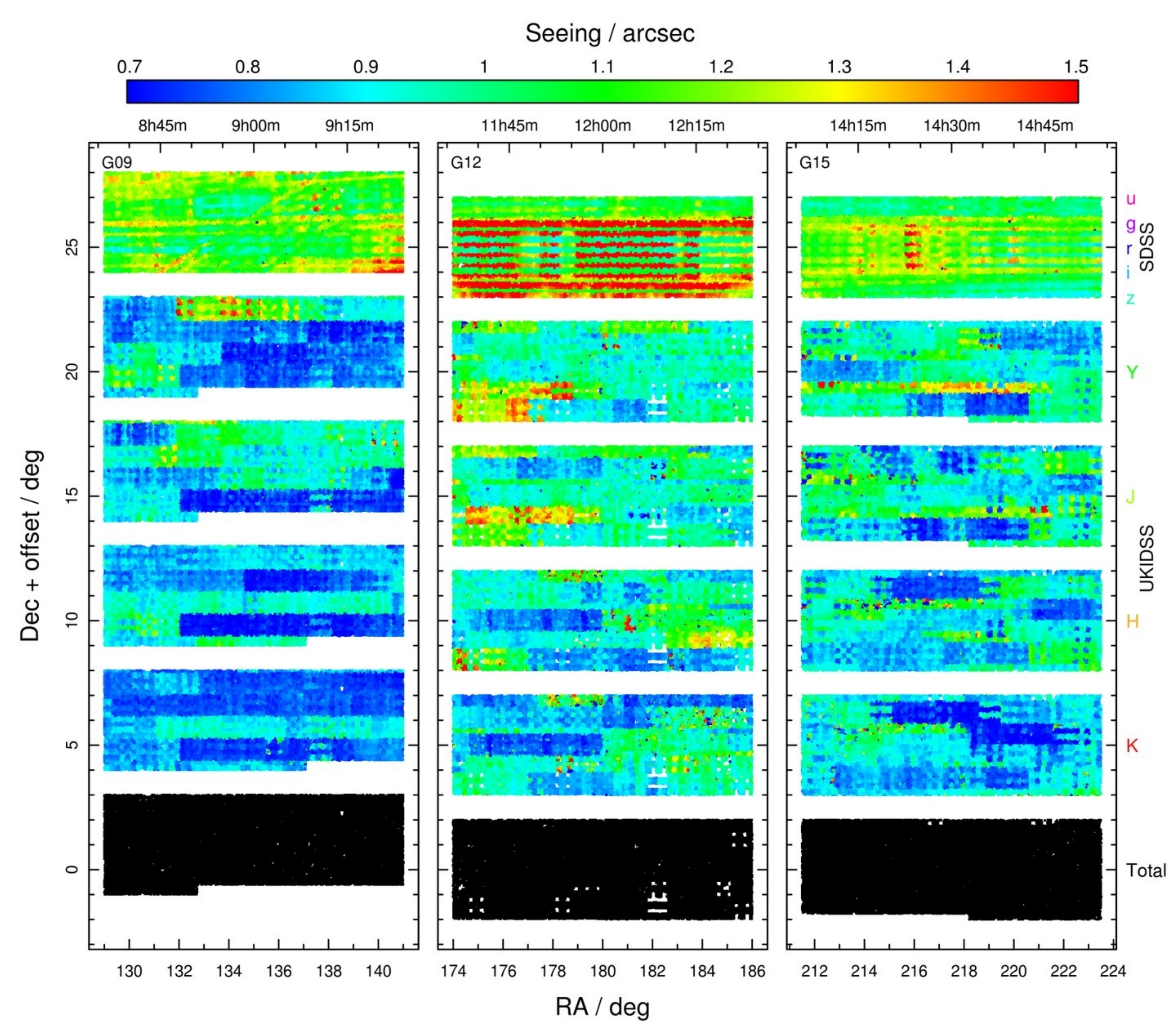}

\caption{\label{fig:coverage}Coverage in each bandpass across the three GAMA
regions is shown, where white space indicates a lack of coverage at
that position. Different bandpasses are offset in declination for
plotting purposes, with the bottom row (black data points) positioned
at the correct GAMA coordinates. The five SDSS bands ($ugriz$) were
taken simultaneously and therefore are represented together, whilst
the four UKIDSS bands ($YJHK$) are shown separately. The data points
are coloured according to the local value of the PSF FWHM at that
position, giving an indication of the variation in seeing (SDSS data
is taken from the $r$ band). See Section \ref{sub:psfpipe} for further
details on the derivation of the PSF's. The bottom row represents
a common coverage area, containing galaxies that have been observed
across all nine bands ($92.8\%$ of the total). A common coverage
sample is derived from this area, and used for all subsequent figures.}
\end{figure*}

The GAMA survey is a combined spectroscopic and multi-wavelength
imaging programme designed to study spatial structure in the nearby
($z<0.25$) Universe on scales of 1kpc to 1Mpc (see \citealt{Driver2009}
for an overview). The survey, after completion of Phase I, consists
of three regions of sky each of $4$ deg (Dec) $\times12$ deg (RA),
close to the equatorial region, at approximately 9$^{h}$ (G09), 12$^{h}$
(G12) and 14.5$^{h}$ (G15) Right Ascension (see Table \ref{tab:regions}
and Figure \ref{fig:coverage}). The three regions were selected to
enable accurate characterisation of the large scale structure over
a range of redshifts and with regard to practical observing considerations
and constraints. They lay within areas of sky scheduled for survey
by both SDSS (\citealt{Abazajian2009}) as part of its Main Survey,
and UKIRT as part of the UKIDSS Large Area Survey (UKIDSS LAS; \citealp{Lawrence2007}).
These data provide moderate depth and resolution imaging data in $ugrizYJHK$
suitable for analysis of nearby galaxies. The accompanying spectroscopic
input catalogue was derived from the SDSS PHOTO parameter \citet{Stoughton2002}
as described in \citet{Baldry2010}. The GAMA spectroscopic programme
\citep{Robotham2010} commenced in 2008 using AAOmega on the Anglo-Australian
Telescope to obtain distance information (redshifts) for all galaxies
brighter than $r<19.8$ mag. The survey is $\sim99$ percent complete
to $r<19.4$ mag in G09 and G15 and $r<19.8$ mag in G12 (see Table
\ref{tab:regions}, column 6), with a median redshift of $z\sim0.2$.
Full details of the GAMA Phase I spectroscopic program, key survey
diagnostics, and the GAMA public and team databases are given in \citet{Driver2011}.

\begin{table*}
\begin{tabular}{|c|c|c|c|c|c|}
\hline 
Region & RA (J2000) & Dec (J2000) & $r_{lim}$ & $N_{obj}$ & $z_{comp}$\tabularnewline
\hline 
G09 & $129^{\circ}.0<\alpha<141^{\circ}.0$ & $-1^{\circ}.0<\delta<3^{\circ}.0$ & $19.4$ ($19.8$) & $30,289$ ($48,548$) & $99.23\%$ ($79.36\%$)\tabularnewline
G12 & $174^{\circ}.0<\alpha<186^{\circ}.0$ & $-2^{\circ}.0<\delta<2^{\circ}.0$ & $19.8$ ($19.4$) & $50,868$ ($32,747$) & $99.12\%$ ($99.39\%$)\tabularnewline
G15 & $211^{\circ}.5<\alpha<223^{\circ}.5$ & $-2^{\circ}.0<\delta<2^{\circ}.0$ & $19.4$ ($19.8$) & $33,205$ ($51,217$) & $98.95\%$ ($79.27\%$)\tabularnewline
\hline 
\end{tabular}

\caption{\label{tab:regions}GAMA region definitions. The GAMA main survey
definitions are based on SDSS extinction corrected r-band Petrosian
magnitude limits, the depth of which varies between $r=19.4$ mag
in G09/G15 and $r=19.8$ mag in G12. Comparison magnitude limits are
shown in brackets for reference. Number counts and redshift completeness
are based on objects which passed star-galaxy separation in the GAMA
TilingCatv11 (see \citet{Baldry2010} for further details).}
\end{table*}
The data used in this paper are obtained from the GAMA Database \citep{Driver2011},
and include reprocessed imaging from the SDSS ($ugriz$) and UKIDSS
LAS ($YJHK$) archive as described in \citet{Hill2011}. The reprocessing
involves the creation of large single image mosaics for each region
in each filter, commonly referred to as SW\noun{arp}ed images (\emph{swpim})
due to the \noun{SWarp} software used in their creation \citep{Bertin2002}.
Associated weight map mosaics (\emph{swpwt}) are also constructed.
The mosaicing process is described in full in \citet{Hill2011}. In
brief, all native reduced frames are downloaded from the respective
archives (SDSS DR7 and ROE/WFAU) and scaled to a single uniform zero
point. For SDSS, the input data are the corrected (\emph{fpC}) Data
Release 7 (DR7) files, with the data having already been bias subtracted
and flat-fielded as part of the SDSS \emph{frames} pipeline (\citealp{Stoughton2002},
Section 4.4). The UKIDSS LAS data has been collected from the UKIDSS
Early Data Release (EDR; \citealp{Dye2006}) and data releases 1 and
2 (DR1; \citealp{Warren2007a}, DR2; \citealp{Warren2007b}). The
UKIDSS project is defined in \citet{Lawrence2007}. UKIDSS uses the
United Kingdom Infrared Telescope Wide Field Camera (WFCAM; \citealp{Casali2007}).
The photometric system is described in \citet{Hewett2006}, and the
calibration is described in \citet{Hodgkin2009}. The pipeline processing
and science archive are described in Irwin et al (in prep.) and \citet{Hambly2008}.

Once these input data have been obtained and calibrated, \noun{SWarp}
is then used to combine them into a single image mosaic at a resolution
of $0.339''$ arcseconds per pixel in the TAN projection system \citep{Calabretta2002}
centred within each GAMA region as appropriate. Note that we are using
version 2 \noun{SWarp} mosaics scaled to a slightly higher resolution
($0.339''$)%
\footnote{This increased resolution has been chosen to match that which is expected
for future VISTA VIKING data releases, allowing easy cross-wavelength
cross-facility comparison of data. Original SDSS and UKIDSS resolutions
of $0.396''$ and $0.4''$ respectively place a limit on how high
one is able to artificially increase the resolution of mosaiced data,
requiring increasing amounts of interpolation with increasing artificial
resolution. Further details may be found in Liske et al. (in prep.).%
} than the version 1 mosaics ($0.4''$) described in \citet{Hill2011}.
Version 2 mosaics are a minimum of $193900\times79700$ pixels each,
with each individual FITS file $\sim60$GB in size. The process used
to create the version 2 mosaics is identical except the regions have
been expanded in preparation for GAMA Phase II operations and at higher
resolution in preparation for matching to VISTA data in due course%
\footnote{These larger, higher-resolution version 2 mosaics will be released
shortly via the GAMA website: http://www.gama-survey.com.%
}.

As part of the \noun{SWarp} mosaicing process the background is removed
on each individual frame prior to merging using a $256\times256$
pixel median filtered mesh which itself is median filtered within
a $3\times3$ mesh. The original SDSS and UKIDSS data are typically
held in chunks of $2048\times1489$ pixels and $2072\times2072$ pixels
respectively, at comparable pixel scales (SDSS: $0.396''$/pixel,
UKIDSS: $0.4''$/pixel). At the native pixel resolution the mesh size
therefore equates to $101.4''\times101.4''$ and $102.4''\times102.4''$
respectively and so structures with half-light radii less than $\sim17''$
should be unaffected by the background smoothing%
\footnote{UKIDSS $J$ band data and selected UKIDSS EDR fields in both $H$
and $K$ bands were microstepped. These data are typically stored
in chunks of $4103\times4103$ pixels at a native resolution of $0.2''/pixel$,
giving a mesh size of $51.2''\times51.2''$.%
}.

In addition to the science image frames are the associated weight
maps. Because of the zero-point normalisation across all data, and
overlapping edge duplication in the SDSS data, the actual weight map
values produced by SW\noun{arp} are an approximation of their correct
value. However, the weight maps remain useful as a record of which
stars can be associated with which pre-mosaiced frame for the purposes
of detailed PSF modelling (described in section \ref{sub:psfpipe}). 

To create our sample of galaxies for modelling, we extracted $167,600$
galaxies from the GAMA Tiling Catalogue version 11 (\emph{TilingCatv11}),
selecting all galaxy-like objects using the GAMA catalogue flag SURVEY\_CLASS$>1$%
\footnote{SURVEY\_CLASS is a flag present in many GAMA catalogues allowing one
to quickly select pre-defined subsets of the GAMA data.%
}. The output from these galaxies is stored in the catalogue \emph{SersicCatv7},
presented in Section \ref{sec:output}. These galaxies, the mosaiced
images and the weight maps constitute our input dataset and are all
available from the GAMA database%
\footnote{The GAMA database can be found at http://www.gama-survey.org/database.%
} as: \emph{TilingCatv11}, x.mosaic.v2.fits and x.weight.v2.fits, where
x=$ugrizYJHK$.

\section{SIGMA: Automated Galaxy Modelling}

\label{sec:sigma}SIGMA (Structural Investigation of Galaxies via
Model Analysis) is an automated front-end wrapper which utilises a
wide-range of image analysis software and a series of logical filters
and handlers to perform bulk structural analysis on an input catalogue
of galaxies. This is primarily achieved through the use of Source
Extractor \citep{Bertin1996}, PSF Extractor (Bertin and Delorme,
priv. comm.) and GALFIT 3 \citep{Peng2010a}, with additional packages
also created and utilised to aid in the fitting process. Key to this
process is the galaxy modelling software GALFIT. GALFIT is able to
create a realistic model of each input galaxy by fitting one or more
analytical functions (e.g. Sérsic, exponential, Ferrer, Moffat, Gaussian)
in multiple combinations.

\begin{figure}
\includegraphics[width=1\columnwidth]{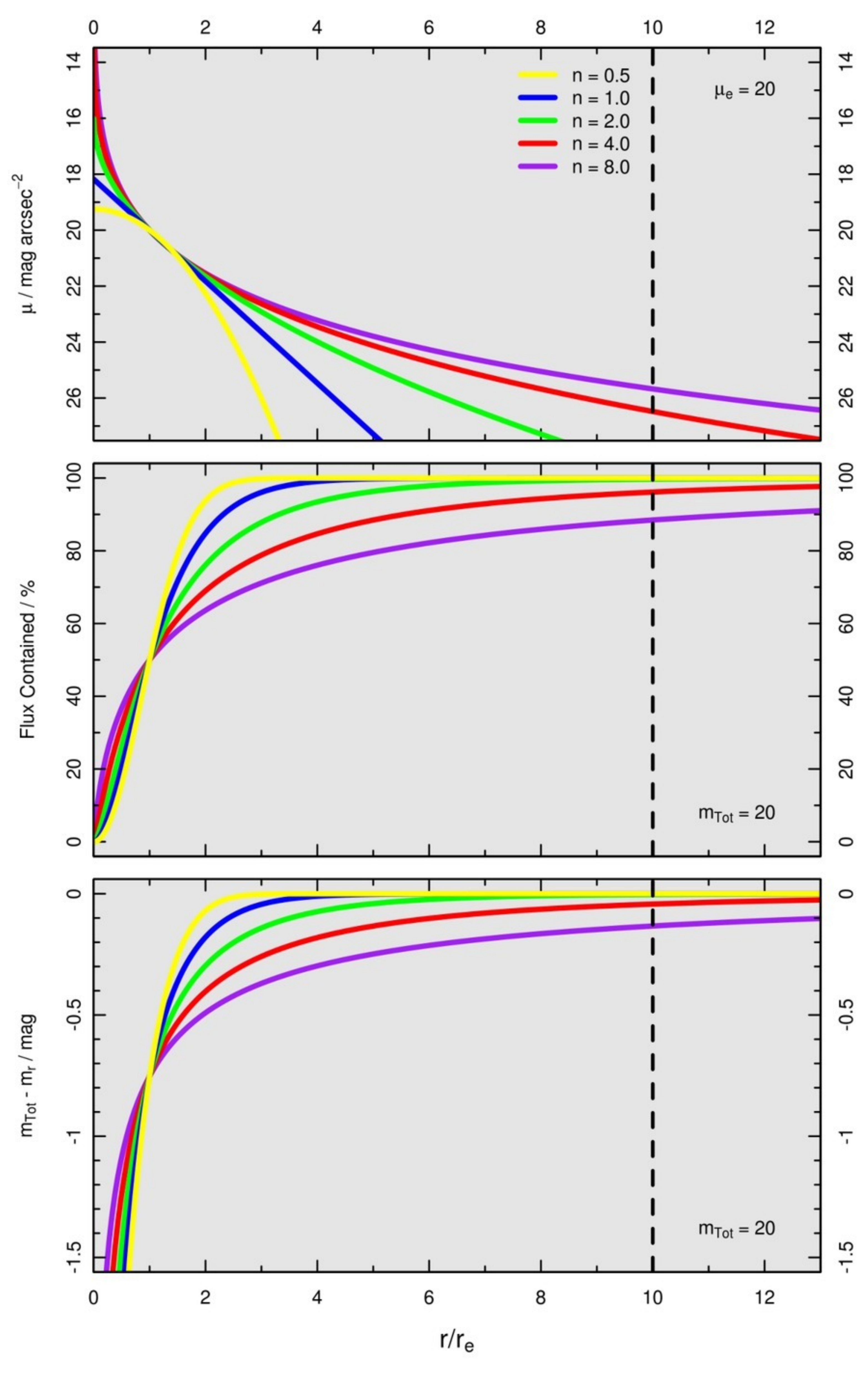}

\caption{\label{fig:sersic}The Sérsic profile (Equation \ref{eq:sersic})
describes how a galaxy light profile varies as a function of radius,
shown here for five distinct values of Sérsic index $n$. (top) Surface
brightness at a given radius. (middle) Flux contained within a given
radius. (bottom) The magnitude offset between the total magnitude
of the galaxy and the Sérsic magnitude at a given radius.}

\end{figure}

Principle in the available GALFIT fitting functions used throughout
this paper is the Sérsic profile \citep{Sersic1963,Sersic1968,Graham2005}
which describes how the galaxy light profile varies as a function
of radius. The Sérsic equation provides the intensity $I$ at a given
radius $r$ as given by:
\begin{equation}
I\left(r\right)=I_{e}\exp\left[-b_{n}\left(\left(\frac{r}{r_{e}}\right)^{1/n}-1\right)\right]\label{eq:sersic}
\end{equation}
where $I_{e}$ is the intensity at the effective radius $r_{e}$,
the radius containing half of the total light, and $n$ is the Sérsic
index which determines the shape of the light profile (see Figure
\ref{fig:sersic}). The value of $b_{n}$ is a function of Sérsic
index and is such that $\Gamma(2n)=2\gamma(2n,b_{n})$%
\footnote{$b_{n}$ can trivially be calculated within R using the relation $b_{n}=\mathrm{qgamma}(0.5,2n)$,
where $\mathrm{qgamma}$ is the quantile function for the Gamma distribution.%
}, where $\Gamma$ and $\gamma$ represent the complete and incomplete
gamma functions respectively\citep{Ciotti1991}. Varying the Sérsic
index parameter $n$ allows one to model a wide range of galaxy profile
shapes, with $n=0.5$ giving a Gaussian profile, $n=1$ an exponential
profile suitable for galactic disks, and $n=4$ a de Vaucouleurs profile
commonly associated with massive spheroidal components such as elliptical
galaxies.

The only inputs required by SIGMA are the imaging data itself and
the locations of the primary galaxies within them which are to be
modelled. All additional parameters and starting values for extra
neighbouring objects in the field of view (secondary objects), including
PSF evaluation, are determined by SIGMA on the fly on a per-galaxy
basis. All scripting and additional programming is written in the
open source and freely available R programming language \citep{RDevelopmentCoreTeam2010}.
Further information on the invocation of SIGMA may be found in Appendix
\ref{app:startup}.

\begin{figure*}
\includegraphics[width=1\textwidth]{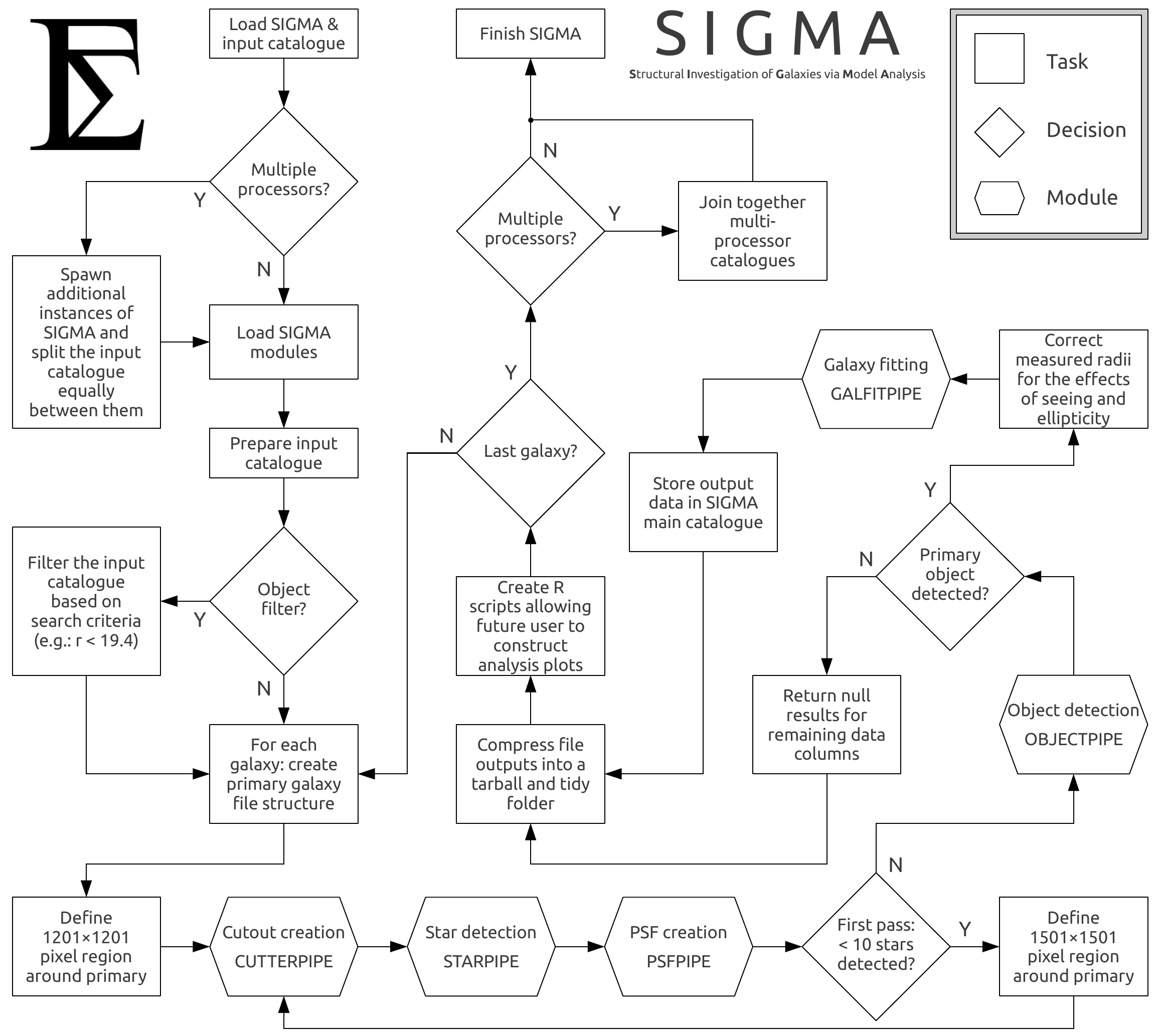}

\caption{\label{fig:flowchart}A flowchart describing SIGMA's operation. Required
inputs are imaging data in the form of GAMA mosaics (including weight
maps), and an input catalogue with a list of galaxy coordinates.}
\end{figure*}

SIGMA operates in a semi-modular fashion, with an overarching master
script calling and linking several key modules within. Each module
is specialised in performing and handling a different task. A summary
of each module and its purpose is shown in Table \ref{tab:modules},
and a schematic of the SIGMA data-flow process is shown in Figure
\ref{fig:flowchart}. The average run-time to profile a single primary
object is 15 seconds per processor%
\footnote{Using current computer hardware at the University of St Andrews. This
consists of a 16 core Intel Xeon E5520 server with 48GB RAM.%
} sustained over several hundred thousand objects.

\begin{table}
\begin{tabular}{|>{\raggedright}p{0.18\columnwidth}|>{\raggedright}p{0.53\columnwidth}|>{\raggedright}p{0.13\columnwidth}|}
\hline 
Module & Description & File Outputs\tabularnewline
\hline 
\hline 
\noun{cutterpipe}

Section \ref{sub:cutterpipe} & Creates a science and weight map cutout from the master GAMA mosaics
using the CFITSIO routine \noun{fitscopy (}Figure \ref{fig:cutouts})
and performs an additional local background sky subtraction using
Source Extractor (Figure \ref{fig:simpweight}). & \emph{cutim}

\emph{cutwt}\tabularnewline
\hline 
\noun{starpipe}

Section \ref{sub:starpipe} & Determines which frames contributed flux to the primary galaxy, and
creates a catalogue of stars that lie within these frames using Source
Extractor. & \emph{psfws}

\emph{psfwt}

\emph{psfct}\tabularnewline
\hline 
\noun{psfpipe}

Section \ref{sub:psfpipe} & Generates an empirical 2D PSF at the primary object position using
PSF Extractor (Figure \ref{fig:psfcutouts}). & \emph{psfss}

\emph{psfim}

\emph{psfsr}\tabularnewline
\hline 
\noun{objectpipe}

Section \ref{sub:objectpipe} & Calculates starting parameters for size, brightness, position angle
and ellipticity for the primary galaxy and any secondary neighbours
(galaxies and stars) using Source Extractor. A dynamic search algorithm
is used to attempt to detect the primary galaxy. Elongated objects
(such as satellite trails) are removed from the secondary catalogue
and instead added to a bad pixel mask. & \emph{objct}

\emph{segim}\tabularnewline
\hline 
\noun{galfitpipe}

Section \ref{sub:galfitpipe} & Fits an analytical function in 2D to the science image using GALFIT.
Both primary and secondary objects are modelled, with any detected
errors/crashes flagged and a fix attempted on a per-galaxy basis (Figures
\ref{fig:detail} and \ref{fig:objim}). & \emph{segfr}

\emph{extct}

\emph{objim}\tabularnewline
\hline 
\end{tabular}

\caption{\label{tab:modules}Summary of the modules that comprise SIGMA, a
brief description of their purpose, and a list of the file outputs
produced by each.}
\end{table}

\subsection{SIGMA Master Script}

\label{sub:startup}When initialising SIGMA a number of options are
specified. In addition to several expected inputs such as band, naming
conventions and the combination and types of GALFIT functions to be
modelled, SIGMA also allows the workload to be split over several
processors. Using a greater number of processors directly decreases
the amount of time required to analyse every primary object in the
input catalogue, and is limited only by available hardware. A full
summary of SIGMA input options is given in Appendix \ref{app:startup}.

To begin, SIGMA's master script loads into memory the entirety of
the GAMA input catalogue and defines a naming convention for each
primary object based on its own unique identifier (SIGMA\_INDEX).
A template master CSV catalogue is created into which all of the output
data will accumulate as SIGMA loops across each primary galaxy. Once
the setup is complete, the master script will loop across every primary
object in turn. If for any reason a primary galaxy causes a software
crash, with attempted fixes as detailed in subsequent sections unsuccessful,
SIGMA will report how far it was able to progress and record a NULL
result before proceeding on to the next primary galaxy in the input
catalogue. We now discuss each module from Table \ref{tab:modules}
in turn.

\subsection{\noun{cutterpipe}: Image Cutout and Preparation}

\label{sub:cutterpipe}The \noun{cutterpipe} module creates and prepares
the fitting image to be fed into GALFIT. Version 2 mosaics of the
three GAMA regions are used as an input to \noun{cutterpipe}, with
a full description of the construction and manipulation of these files
given in \citet{Hill2011} and summarised in Section \ref{sec:data}.

\noun{cutterpipe}'s first task is to create the core cutout of the
science image and its associated weight map. Using the WCS information
stored in the FITS header of the appropriate mosaiced image, \noun{cutterpipe}
converts the input RA/DEC into an x/y pixel coordinate. The upper
and lower limits of a $1201\times1201$ pixel ($\sim400''\times400''$)
region centred on the primary galaxy are determined. Using the NASA
HEASARC package's CFITSIO subroutine library, namely the routine \noun{fitscopy},
cutouts centred on the primary galaxy on both the mosaiced science
image, \emph{swpim}, and mosaiced weight map, \emph{swpwt}, are created.
These cutouts are named \emph{cutim} and \emph{cutwt} respectively.
\noun{fitscopy} was found to be the most efficient routine at dealing
with the large mosaic files in use, able to quickly analyse the input
file and read into memory only the relevant area of interest, thereby
reducing file handling time significantly. 

The process of creating the GAMA mosaics alters a number of keywords
in the FITS header in order to better describe the nature of the mosaiced
data. The mosaic headers are copied over to \emph{cutim} and \emph{cutwt}
during their creation. Several of these keywords are required later
in the fitting process by GALFIT in order to generate a sigma-map
(an image showing the $1\sigma$ confidence interval at every pixel).
\noun{cutterpipe} reverts these to typical pre-mosaic values which
are more appropriate for a smaller single image rather than a larger
mosaic. GAIN, RDNOISE, NCOMBINE and EXPTIME are set to values of $0.5$,
$3$, $1$ and $1$ respectively%
\footnote{These typical values are averages taken from pre-mosaiced data frames.%
}.\noun{ }

An estimate of the local background sky is then made with Source Extractor
(v2.8.6; \citealp{Bertin1996}) using a variable background grid in
a $3\times3$ mesh configuration. Possible grid sizes are $32\times32$,
$64\times64$ and $128\times128$ pixels. The size of the chosen background
grid is dependent upon the size of the primary galaxy: larger galaxies
will lead to a larger background grid being used so as not to contaminate
the sky estimate with galaxy flux. An initial basic estimate of the
total size of the primary is given by:
\[
r_{tot}=2\times r{}_{99}
\]
where $r_{99}$ is the radius of the primary galaxy which contains
99\% of the flux. This is obtained from the Source Extractor parameter
FLUX\_RADIUS setting PHOT\_FLUXFRAC$=0.99$%
\footnote{Elsewhere in this paper, Source Extractor FLUX\_RADIUS will typically
refer to $r_{e}$, a radius containing $50\%$ of the flux of the
primary galaxy. It is worth noting however that a size estimate produced
by Source Extractor in this manner is known to be smaller than the
true galaxy value, scaling as a function of Sérsic index and thus
absolute magnitude. This effect has been accounted for, and does not
adversely affect any of the analysis or results presented in this
paper.%
}. If $r_{tot}<128$,\noun{ cutterpipe} rounds\emph{ }up\emph{ }$r_{tot}$
to the nearest available grid size and performs a background subtraction
on the science image as appropriate. If $r_{tot}\geq128$, no background
subtraction is necessary, as the master GAMA mosaics have a $256\times256$
pixel background grid subtraction already applied. Although the value
of actual subtracted sky varies with position on the cutout image,
the specific value at the position of the primary galaxy, $\rho_{\mathrm{sky}}$
is recorded through the Source Extractor parameter BACKGROUND. The
error on the background sky estimate is then given by: 
\[
\Delta\rho_{sky}=\frac{\sigma_{\mathrm{sky}}}{\sqrt{0.9\times n_{x}\times n_{y}}}
\]
where $\sigma_{\mathrm{sky}}$ is the RMS of background sky counts
across the cutout, and $n_{x}$ and $n_{y}$ are the dimensions of
the cutout in the $x$ and $y$ dimensions respectively. The background
sky typically encompasses $\sim90\%$ of any given cutout, and hence
a factor of $0.9$ is introduced into the above calculation to account
for this. After extensive testing, this variable background mesh method
for sky estimation was found to be the most robust at removing small-scale
sky fluctuations in the data without subtracting real galactic light
from objects (see Section \ref{sub:sky} for sky results discussion). 

\begin{figure*}
\includegraphics[width=1\textwidth]{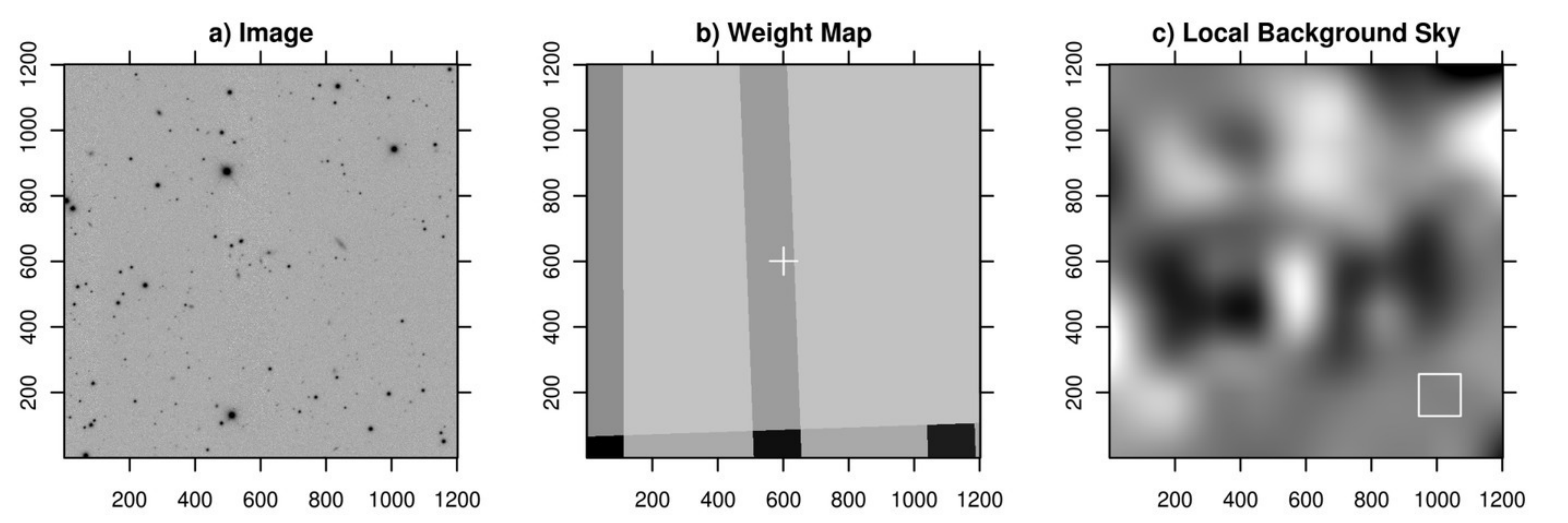}

\caption{\label{fig:cutouts}A series of setup images centred on G00196053.
$a$) A science image cutout of $1201\times1201$ pixels ($\sim400''\times400''$)
in the $r$ band, with the primary galaxy centred at $601\times601$.
The image is scaled logarithmically from $-1$ $\sigma_{\mathrm{sky}}$
to $40$ $\sigma_{\mathrm{sky}}$, where $\sigma_{\mathrm{sky}}$
is the typical RMS of the sky in the $r$ band. $b$) A weight map
cutout of $1201\times1201$ pixels ($\sim400''\times400''$) in the
$r$ band, with the primary galaxy located in an overlap region at
the white cross-hairs. Darker shades indicate larger values, therefore
a greater number of overlapping frames. $c$) The additional local
background sky removed from this cutout. The z-axis range shown is
$\pm0.5$ counts. For this galaxy, a background mesh size of $128\times128$
pixels (white square) was chosen in addition to the coarse $256\times256$
background mesh already applied in creating the GAMA mosaics.}
\end{figure*}

An example image cutout, weight map and background estimation map
are shown in Figure \ref{fig:cutouts}.

\subsection{\noun{starpipe}: Star Detection}

\noun{\label{sub:starpipe}starpipe }uses Source Extractor to create
a catalogue of star-like objects with which to create a point-spread
function (PSF) in the subsequent \noun{psfpipe} module (section \ref{sub:psfpipe}). 

The first step is to determine which of the original pre-mosaic frames
contain the primary galaxy. This step is non-trivial, as a single
cut-out image (\emph{cutim}) may contain data from several pre-mosaiced
frames overlapping at random angles to each other, with only some
of the frames contributing flux (and therefore seeing information)
to the primary galaxy. Calculating frame ownership is crucial in PSF
determination, as using stars from non-contributing frames would skew
the PSF estimate away from its true shape at the position of the primary
galaxy. A method was devised to determine contributing frames using
the information within the GAMA weight maps. Each pre-mosaiced frame
is assigned a numerical value based on the global variance of the
data for that frame. This value, repeated for each pixel, becomes
the weight-map for that individual frame. Weight-map values are essentially
unique to several significant figures, and therefore useful in identifying
that particular frame. During the \noun{SWarp} process, overlapping
imaging data is median combined (setting the \noun{SWarp} argument
COMBINE\_TYPE to MEDIAN) whereas weight-maps are co-added to produce
a global weight-map representing the change in the variance across
the data. When two or more frames overlap, their individual weight
map values are summed. Larger values indicate a greater number of
overlapping frames.

The value of the weight map at the primary position is determined,
with all pixels of that value clearly contributing flux to the primary
galaxy. This defines the initial primary region. However, since this
primary region may be an overlap region itself, parent frames must
also be determined. The weight-map values of all bordering pixels
to the primary region are determined. Higher pixel values indicate
a region which contains data from additional frames that did not contribute
flux to the primary region, and so these pixels are discarded. Lower
values (if any) indicate parent frames of the primary region, and
(if they exist), their pixel positions are added to the primary region.
This process will continue until all pixels are accounted for across
the cutout. As an example of frame determination, contrast Figure
\ref{fig:cutouts}b with the shaded red regions in Figure \ref{fig:detail}.

\begin{figure}
\includegraphics[width=1\columnwidth]{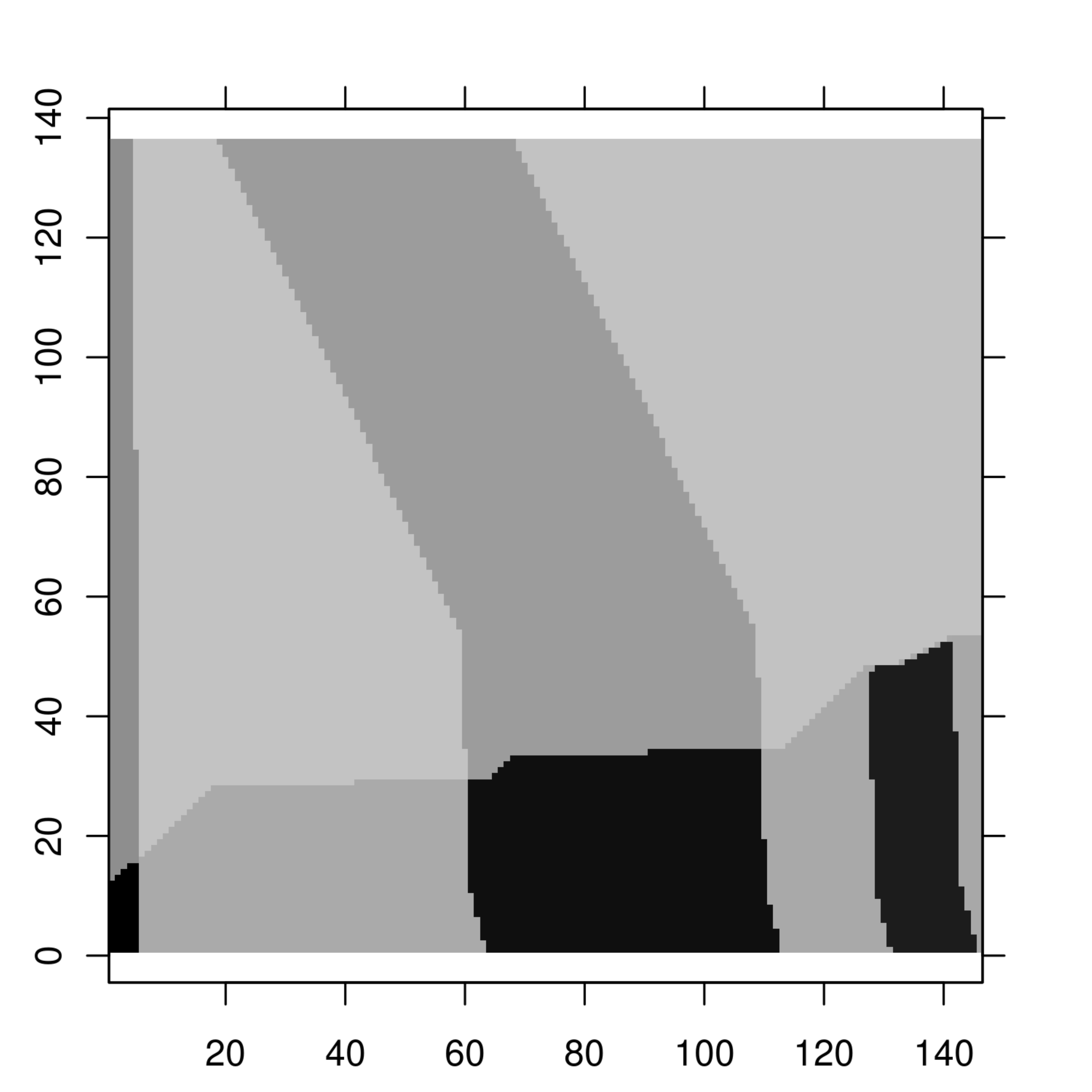}

\caption{\label{fig:simpweight}A simplified version of the weight map shown
in Figure \ref{fig:cutouts}b. A single cutout from a GAMA mosaic
may contain data from several frames observed on different nights
in different seeing conditions. The weight map allows us to determine
which frames contributed flux to the primary galaxy and which did
not, thereby allowing a PSF to be constructed representing the seeing
for that galaxy. Simplification of the weight map removes superfluous
information contained within the original cutout weight map, reducing
the number of pixels to analyse typically by a factor of $\sim150$,
and significantly speeding up \noun{starpipe}.}
\end{figure}

Pixel determination via this technique is time intensive for the full
$1201\times1201$ cutout region, as it requires analysis of $1.4$
million pixels for each galaxy. A more efficient method is to reduce
the number of pixels that require analysis by simplifying the weight
map to its minimal number of pixels which still describe the nature
of the data. Duplicate rows and columns in the weight map are removed,
producing a simplified weight map, \emph{psfws}, typically of order
$\sim100\times100$ pixels. This thereby reduces the number of pixels
needing to be analysed by a factor of $\sim150$, significantly speeding
up the primary region determination. Figure \ref{fig:simpweight}
is one such simplification of the cutout weight map shown in Figure
\ref{fig:cutouts}b, in this case reducing the number of pixels from
$1.4$ million to $\sim20,000$.

Once a primary region is determined, a local star catalogue must be
created. A modified version of \emph{cutwt}, \emph{psfwt}, is created,
setting all non-primary pixels to a weight value of zero. This will
bar Source Extractor detecting any objects in those regions. The central
$25\times25$ pixel region is also set to a weight value of zero to
ensure against the primary galaxy being falsely classified as a star
and used in later PSF analysis. 

A Source Extractor parameter file is created to output NUMBER, X\_IMAGE,
Y\_IMAGE, FLUX\_RADIUS, FLAGS, FLUX\_APER(1), FLUX\_MAX, ELONGATION,
VIGNET(25,25) and BACKGROUND. The numeric values following VIGNET
determine the ultimate size in pixels of the 2D PSF created subsequently
in \noun{psfpipe.} A detection threshold of $2\sigma$ above the background
is specified, along with a minimum object detection area of 10 pixels,
a $25,000$ count saturation level and a fixed sky pedestal value
of zero counts. The image is filtered through a $5\times5$ pixel
Gaussian convolution kernel with $\Gamma=2$ pixels. Source Extractor
defaults are used everywhere else. Using these settings, Source Extractor
is run across the cutout image \emph{cutim} using the weight map image
\emph{psfwt}. An output catalogue, \emph{psfct}, is created in the
\emph{FITS\_LDAC} format, a format which saves the image data as well
as the catalogue.

\subsection{\noun{psfpipe}: PSF Estimation and Creation}

\label{sub:psfpipe}The point spread function (PSF) describes the
blurring effect of both the atmosphere and the telescope optics on
our imaging data. Observed galaxy images have had their flux redistributed
according to this PSF. The galaxy flux most affected by the PSF blurring
is that which emanates from the core regions, where the gradient of
the light profile is at its steepest. It is therefore crucial to have
a good understanding of the PSF when considering fitting smooth analytical
galaxy models to imaging data. Furthermore, most current galaxy modelling
software weights model fitting towards higher signal-to-noise regions
(typically the same core regions), increasing the importance of accurate,
reliable PSF estimation. 

The \noun{psfpipe} module is a wrapper around the PSF extraction software
PSF Extractor (PSFEx v3.3.4; Bertin, priv. comm.%
\footnote{More information on the PSFEx software may be found at http://www.astromatic.net/software/psfex.%
}), and produces a 2D PSF model to be taken into account at the later
galaxy modelling stage. PSFEx extracts precise models of the PSF from
images pre-processed by Source Extractor, allowing for a wide range
of PSF's to be quickly and accurately constructed, including arbitrary
non-parametric features present in the PSF.

In brief, the sample of objects from the \emph{psfct} catalogue created
by \noun{starpipe} is initially used for analysis. PSFEx reduces this
object list to a star sample based on a set of pre-defined criteria.
A signal-to-noise limit of at least 10 is required, and objects with
an eccentricity of $\left(\frac{\left(a-b\right)}{\left(a+b\right)}\right)>0.05$
are removed, where $a$ and $b$ refer to the semi-major and semi-minor
axes respectively%
\footnote{PSFEx refers to this quantity as ellipticity rather than eccentricity,
however its definition is more akin to that of the latter. We adopt
the terminology eccentricity here to avoid confusion with the standard
definition of ellipticity used throughout the remainder of this paper,
namely $e=1-\frac{b}{a}$. An eccentricity of $0.05$ therefore corresponds
to an ellipticity of $e\sim0.095$.%
}. Each star's full-width-half-maximum, $\Gamma$, is estimated, with
only stars in the pixel range $2<\Gamma<10$ accepted. Furthermore,
variability in the star sample is limited to the central 50\% quantile.
After extensive testing on the variation in PSF quality with star
sample size, and communication with the authors of PSFEx, we found
that a star sample size of at least $10$ stars is necessary to ensure
that the resultant PSF is not adversely affected by small-number biases.
Therefore, if fewer than $10$ stars remain in the star sample after
selection criteria have been applied, SIGMA will loop back to \noun{cutterpipe}
and expand the cutout region to $1501\times1501$ pixels ($\sim500''\times500''$).
The mean number of stars used for PSF estimation in the $r$ band
is $24.4$, with $10.2\%$ of cutouts containing fewer than $10$
stars after the cutout region has been expanded.

\begin{figure*}
\includegraphics[width=1\textwidth]{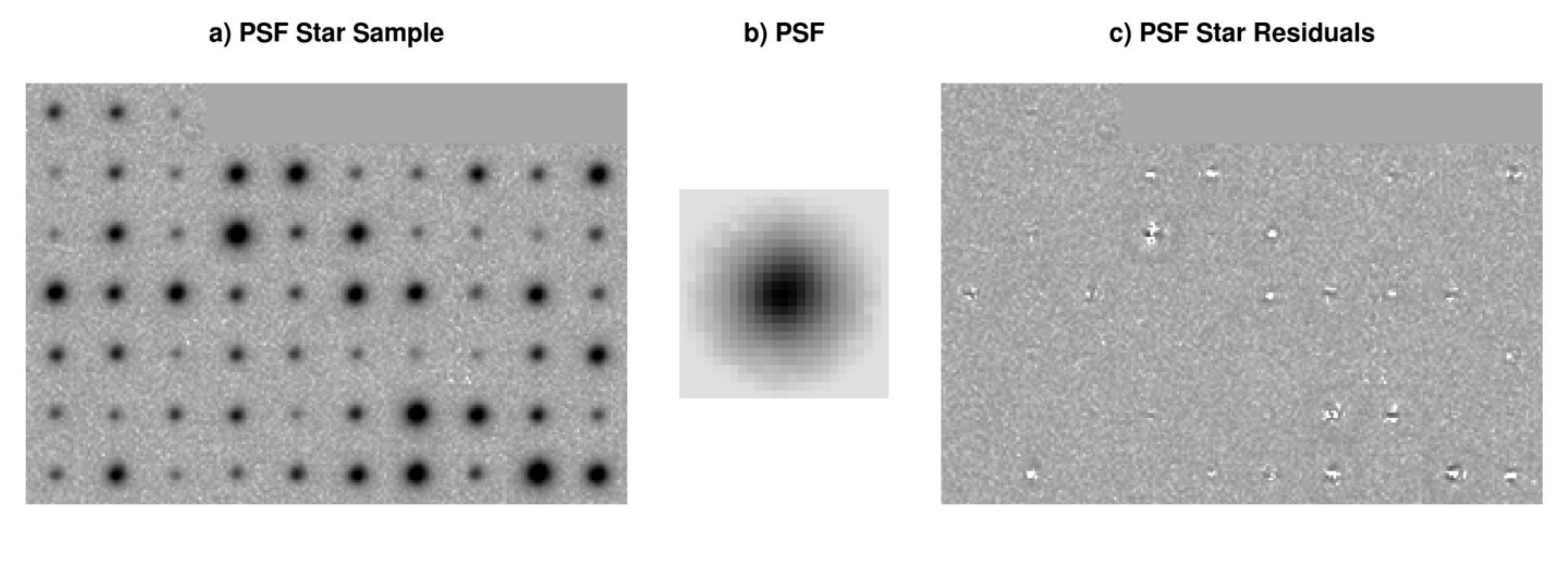}

\caption{\label{fig:psfcutouts}PSFEx generates an empirical PSF (b) from a
host sample of representative stars (a). This figure shows 63 sample
star cutouts of $25\times25$ pixels each chosen from around GAMA
object G00196053, whose real sky positions may be noted in Figure
\ref{fig:detail} (orange circles). Panel (c) represents the residual
of each star sample with a scaled form of the PSF subtracted from
each. (a) and (c) are scaled logarithmically from $-1$ $\sigma_{\mathrm{sky}}$
to $40$ $\sigma_{\mathrm{sky}}$, where $\sigma_{\mathrm{sky}}$
is the typical RMS of the sky in the $r$ band.}
\end{figure*}

Cutout images of each star are pre-stored in the \emph{FITS\_LDAC}
format of \emph{psfct}, the size of the cutout having been specified
at the Source Extraction stage. PSFEx uses the positional information
from Source Extractor to mask nearby neighbours to the final star
sample, and presents this sample in the output \emph{psfss} FITS image
(Figure \ref{fig:psfcutouts}a).

The variation in the shape of the stars in the star sample is then
modelled in both $x$ and $y$ as a function of position in the field
by a 2D $n^{\text{th}}$ order polynomial function. Higher order terms
in the fit (i.e. $x$, $x^{2}$, etc.) describe the variation in the
PSF at positions away from the centre of the frame. Within SIGMA,
the primary galaxy is always centred in the cutout image, and so a
zeroth order polynomial was found to adequately describe the PSF.
The best-fit polynomial is sampled at a 1:1 ratio relative to the
input data, and an output PSF image \emph{psfim} is produced of the
same size as the input cutout stars, $25\times25$ pixels (Figure
\ref{fig:psfcutouts}b). As a consistency check, scaled models of
\emph{psfim} are fit to each of the input stars in \emph{psfss}, and
a residual map \emph{psfsr} produced (Figure \ref{fig:psfcutouts}c).
Note that some of the PSF residuals still show noticeable structure
once the PSF model has been subtracted from the star sample. This
is as expected when subtracting a zeroth order PSF model (only accurate
at the location of the primary galaxy in the centre) from a star sample
taken over a large area on the sky. Those stars with noticeable residuals
therefore are typically either significantly spatially separated in
the field of view from the primary galaxy or approaching saturation
(or both). Both of these factors are accounted for by PSFEx when constructing
the model PSF. The stars chosen as part of the star sample are shown
in orange circles in Figure \ref{fig:detail}, with each circle numbered
according to their position in Figure \ref{fig:psfcutouts}a, starting
at $1$ in the bottom left and increasing horizontally left-to-right
and then bottom-to-top.

\subsection{\noun{objectpipe}: Object Detection}

\label{sub:objectpipe}A second catalogue of objects optimised for
galaxy detection, \emph{objct}, is created in \noun{objectpipe}, to
be later fed into \noun{galfitpipe}. This catalogue provides the basic
starting parameters for the primary galaxy and any secondary galaxies
and stars in the frame. \noun{objectpipe} also creates a segmentation
map of the frame, modified to mask any erroneous regions of flux in
the image which may cause fitting problems (e.g., satellite trails).
A Source Extractor parameter file is created containing X\_IMAGE,
Y\_IMAGE, MAG\_AUTO, FLUX\_RADIUS, KRON\_RADIUS, A\_IMAGE, B\_IMAGE,
THETA\_IMAGE, ELLIPTICITY and CLASS\_STAR. These give position ($x$/$y$),
luminosity, size, position angle and ellipticity for the primary and
all secondaries in the field. Source Extractor settings are similar
to those used in \noun{starpipe}, excepting a lower detection threshold
of $1.8\sigma$ above the background (where $\sigma$ is the RMS as
estimated by Source Extractor), and a lower deblending contrast parameter
of $0.0001$. The Source Extractor Neural-Network Weights V1.3 file
is used in the creation of the CLASS\_STAR parameter, as well as a
standard $5\times5$ pixel Gaussian filter, $\Gamma=2$ pixels, used
during object detection.

\noun{objectpipe} calls Source Extractor, and records the results.
If initially the primary galaxy is unable to be located within a 5
pixel radius of the input coordinates, \noun{objectpipe} will in the
first instance decrease the detection threshold in steps of $0.4\sigma$
down to $1\sigma$ above the background until an object is found,
re-running Source Extractor as appropriate. This usually occurs with
faint objects in the field, or in crowded regions. If the primary
object is still unable to be located, the threshold is reset to $1.8\sigma$,
and a larger search radius of up to 15 pixels from the input coordinates
in 5 pixel steps is tried. This stage accounts for large nearby galaxies
whose centroids are not matched to better than 5 pixels, hence requiring
a larger detection area. If multiple-matches are found, the largest
object will be taken to be the primary galaxy. If at this stage the
primary galaxy is still not found, \noun{objectpipe} will report a
null detection, and move on to the next primary in the input catalogue. 

Output parameters from Source Extractor are modified by \noun{objectpipe}
before being fed into \noun{galfitpipe} with the exception of magnitude
which is used unaltered. Position angle is modified to the GALFIT
standard (by adding $90$ degrees), increasing anti-clockwise from
the positive $x$ axis. Ellipticity $e$ is converted to an axis ratio
using the relation:
\[
e=1-\frac{b}{a}
\]
with semi-minor axis $b$ and semi-major axis $a$. Half-light radius
$r_{e}$ is estimated using the relation:

\begin{equation}
r_{e}=\sqrt{\left(r_{50}^{2}\times\frac{a}{b}\right)-\left(0.32\times\Gamma^{2}\right)}\label{eq:effradfix}
\end{equation}
where $r_{50}$ is the (unmodified) Source Extractor half-light radius
as given by FLUX\_RADIUS (setting PHOT\_FLUXFRAC$=0.5$) and $\Gamma$
is the Full-Width Half-Maximum of the PSF of the primary galaxy. A
minimum bound on $r_{e}$ of $1$ pixel is enforced. This conversion
corrects for the fact that Source Extractor's output half-light radii
are circularised and based on PSF convolved data, whereas GALFIT radii
are along the semi-major axis and intrinsic (non-PSF convolved). The
value of 0.32 was derived from simulated test data, see Appendix A
of \citet{Driver2005} for further details. Figure \ref{fig:radiicomp}
shows a comparison of corrected and uncorrected radii against modelled
GALFIT radii for all GAMA objects in the $r$ band. This suggests
that the revised starting value for $r_{e}$ is appropriate and an
accurate first estimate of the true half-light radius of the primary
galaxy. Due to the downhill minimisation employed by GALFIT, it is
important to provide input parameters as close as possible to the
desired solution in order to avoid local minima.

\begin{figure}
\includegraphics[width=1\columnwidth]{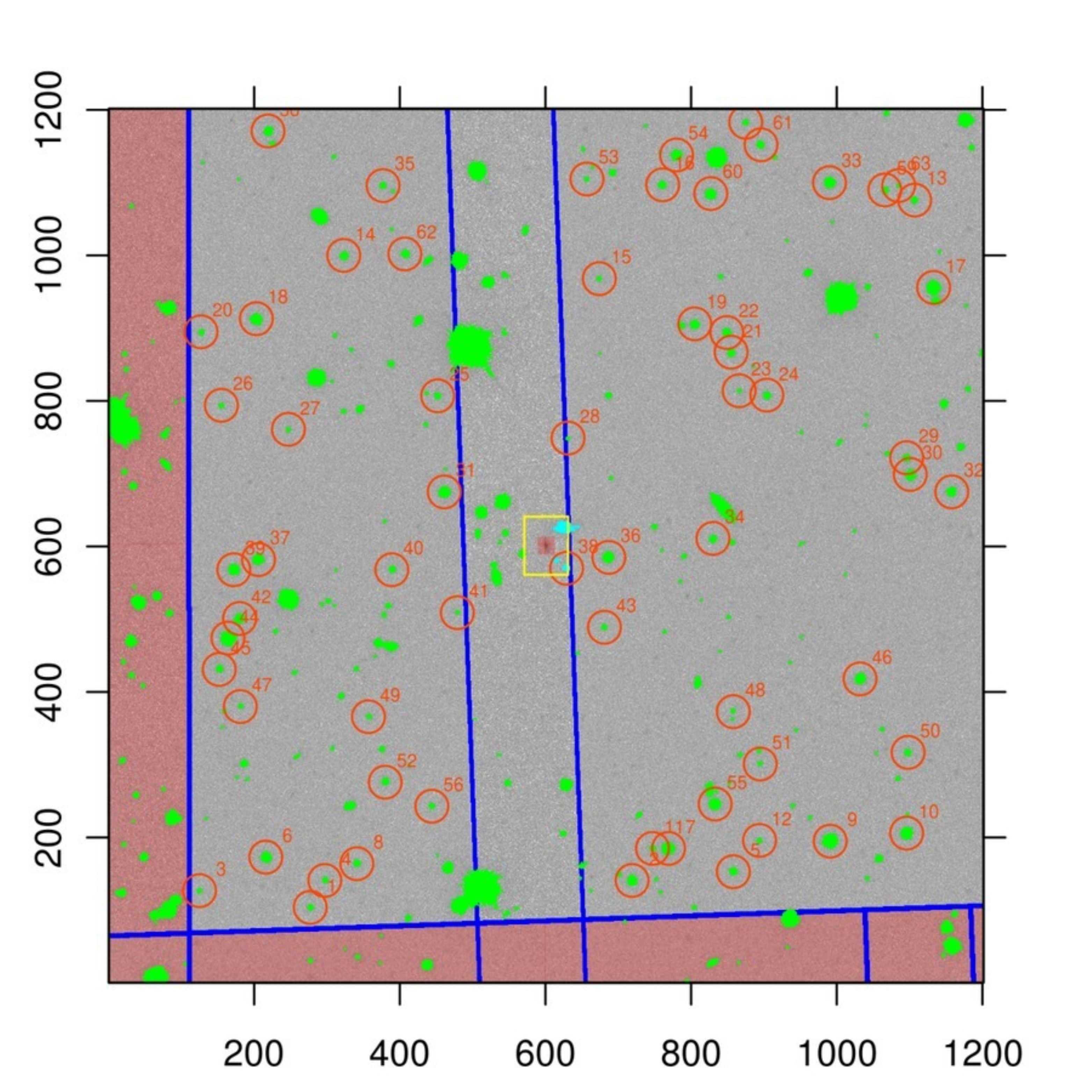}

\caption{\label{fig:detail}Final detail analysis plot for G00196053. Green
shaded areas represent all detected objects, with stars chosen as
part of the PSF star sample circled in orange. The GALFIT fitting
area is outlined in yellow, and the weight map frame edges in blue.
Secondary objects which will be modelled as nearby neighbours in the
fitting process are represented in turquoise. Note how no sample stars
are taken from frames which did not contribute directly to the flux
of the primary galaxy (areas represented with red shading). Consequently,
no stars are taken for PSF analysis from the red shaded areas. The
image is scaled logarithmically from $-1$ $\sigma_{\mathrm{sky}}$
to $40$ $\sigma_{\mathrm{sky}}$, where $\sigma_{\mathrm{sky}}$
is the typical RMS of the sky in the $r$ band.}
\end{figure}

Once physical parameters for the primary galaxy have been determined,
a segmentation map of the frame is created to be used as a potential
mask for secondary features should modelling them fail. Secondary
objects whose ellipticity is greater than 0.95 are excluded from modelling
and will instead be masked, as these are determined to be satellite
trails or bad data, and consequently difficult to model. Similarly,
secondary objects with a stellaricity index of CLASS\_STAR$>0.8$
(see \citealp{Bertin1996}) are modelled by a PSF within \noun{galfitpipe},
with all others being modelled using a single Sérsic function. A relatively
low CLASS\_STAR boundary is chosen as tests have shown that a more
reasonable fit is produced when fitting a PSF rather than a Sérsic
function to ambiguous objects. A graphical representation of detected
galaxies, stars, weight-map areas and secondary neighbour determination
for G00196053 is shown in Figure \ref{fig:detail}.

\subsection{\noun{galfitpipe}: Galaxy Fitting}

\label{sub:galfitpipe}The actual modelling is handled by \noun{galfitpipe},
which is a wrapper around the GALFIT image analysis software (v3.0.2;
\citealp{Peng2010a}) along with several event handlers and logical
filters written in R. GALFIT is a 2D parametric galaxy fitting algorithm
written in the C language. It allows for multiple parametric functions
(such as Sérsic, exponential, Ferrer, Moffat, Gaussian, etc.) to be
modelled simultaneously as either multiple components of a single
object, multiple objects in a single frame, or combinations thereof. 

GALFIT uses a Levenberg-Marquardt algorithm to fit a 2D function to
2D data, in doing so minimising the global $\chi^{2}$ until the gradient
$\Delta\chi^{2}$ has become negligible and convergence is reached.
When a global minimum is thought to be found, GALFIT introduces a
10 iteration cool-down period, sampling the parameter space around
the best-fit parameters in an attempt to overcome the problem of converging
on a local rather than global minimum. 

In this paper we fit each primary galaxy with a single Sérsic function
containing 7 free parameters: object centres $x_{0}$ and $y_{0}$;
total integrated magnitude $m_{tot}$; effective radius along the
semi-major axis $r_{e}$; Sérsic index $n$; ellipticity $e$ and
position angle $\theta$. Secondaries (galaxies and stars) will also
be modelled by either a Sérsic function or a scaled PSF as appropriate.
The PSF contains 3 free parameters; $x_{0}$, $y_{0}$ and $m_{\mathrm{tot}}$.
For additional information on the operation of GALFIT, refer to \citet{Peng2010a}.

All primary inputs to GALFIT are taken from Source Extractor and modified
as described in Section \ref{sub:starpipe}, with the exception of
Sérsic index which starts at $n_{\mathrm{initial}}=2.5$. After extensive
testing on the $r$ band data, it was found that the chosen starting
Sérsic index has little to no effect on the end result, and so choosing
a value in the centre of the expected parameter range was deemed appropriate.
No explicit constraints were put on the range of acceptable Sérsic
indices upon which GALFIT may converge, however, GALFIT has internal
limits of $0.05<n<20$, where the lower limit is a 'soft' limit (indices
scatter around this value) and the upper is a 'hard' limit (indices
may not converge above this value). More conservative limits were
not enforced on Sérsic index so as not to lead the final results and
make presumptions about Sérsic index distributions. More detail on
chosen initial conditions may be found in Appendix \ref{app:initialconditions}.

In order for \noun{galfitpipe} to function correctly, it needs the
cutout science image from \noun{cutterpipe}, \emph{cutim}; the associated
segmentation map and object catalogue from \noun{objectpipe}, \emph{segim}
and \emph{objct} respectively; and a 2D FITS image PSF representing
the PSF at the primary galaxy location from \noun{starpipe} and\noun{
psfpipe}, \emph{psfim}. Note that the weight map (\emph{cutwt}) is
no-longer required at this modelling stage. 

Once the aforementioned files are in place, an initial fitting region
radius on the cutout is defined by:
\[
r_{x}=2r_{\mathrm{Kron}}\left(|\cos\left(\theta\right)|+\left(1-e\right)|\sin\left(\theta\right)|\right)
\]

\[
r_{y}=2r_{\mathrm{Kron}}\left(|\sin\left(\theta\right)|+\left(1-e\right)|\cos\left(\theta\right)|\right)
\]
in order to account for the ellipticity $e$ of the object and its
position angle $\theta$. Objects within the central $2r_{x}\times2r_{y}$
of the fitting region will be convolved with the supplied PSF at the
modelling stage. The segmentation map is modified to unmask all secondary
objects in the fitting region, with the resultant map saved into a
new \emph{segfr} file.

A GALFIT \emph{feedme} file is created containing the starting values
for every object being modelled (primary and secondary) as described
in Section \ref{sub:starpipe} and above. A constraints file is used
to constrain secondary objects. These objects are constrained in order
to reduce the fitting time, and reduce the size of the allowed parameter
space. $x_{0}$ and $y_{0}$ are constrained to $\pm3$ pixels of
their input parameters, ellipticity is constrained to $0<e<0.95$
and half-light radius is constrained to $r_{e\mathrm{,initial}}/4<r_{e\mathrm{,final}}<4r_{e\mathrm{,initial}}$.
A final parameter for \emph{sky} is added to the bottom of the GALFIT
\emph{feedme} file, fixing the value of the sky to zero counts.

GALFIT is then initialised, fixing the sky RMS to that measured in
\noun{objectpipe}.\noun{ T}he time taken to converge on a fit scales
with the size of the fitting region, and the number of secondaries
being fit. Once the GALFIT process has finished, its output (if any)
is read and processed. \noun{galfitpipe} scans the primary galaxy
for a number of problems in this order:
\begin{enumerate}
\item Crash or a segmentation fault
\item Galaxy centre migration of $\sqrt{x^{2}+y^{2}}>r_{e\mathrm{,initial}}$
\item An exceptionally large radius of $log_{10}\left(\frac{r_{e\mathrm{,final}}}{r_{e\mathrm{,initial}}}\right)>3$
\item An exceptionally small radius of $log_{10}\left(\frac{r_{e\mathrm{,final}}}{r_{e\mathrm{,initial}}}\right)<3$
\item A high ellipticity of $e>0.95$
\end{enumerate}
If any of these are detected, a fix will be attempted and GALFIT re-run
as appropriate. Fixes attempted vary depending on the problem encountered.
If a crash or segmentation fault are detected, GALFIT will be re-run
modelling only the primary galaxy, with all secondaries masked. This
usually occurs for large nearby objects with a high number of secondary
neighbours and foreground stars, providing the Levenberg-Marquart
minimisation routine in GALFIT with many local-minima. If the centre
migrates away to fit a secondary feature, GALFIT will be re-run with
the primary centroids fixed to their starting values. Large or small
radii are initially handled by suggesting a lower starting shape parameter
($n=0.5$). This usually assists GALFIT in finding a way out of any
local minima. If this attempt still provides a wildly different size
to the input parameter, the size is fixed to the input and GALFIT
re-run. Finally, a high ellipticity usually indicates the model has
migrated away to fit flocculent secondary features. Re-running GALFIT
with a starting ellipticity of $e=0.1$, i.e., highly circular, in
most cases mitigates this problem. If all fixes have been attempted
and problems persist, \noun{galfitpipe} will record GALFIT's best-guess
model parameters and move on to the next object in its catalogue,
with a flag updated to reflect the fitting history.

\begin{figure*}
\includegraphics[width=1\textwidth]{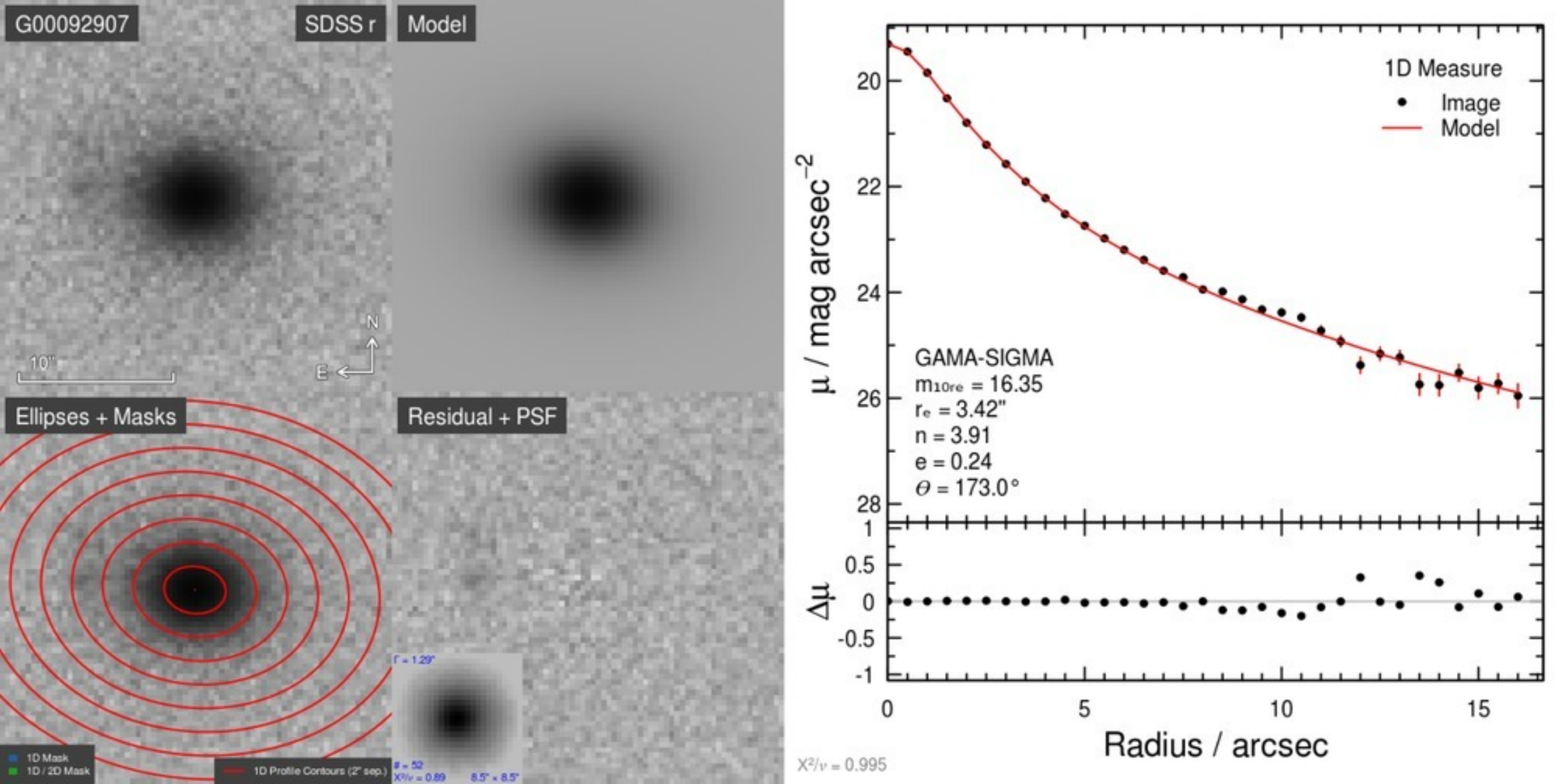}

\caption{\label{fig:objim}An example model output for GAMA galaxy G00092907.
The original SDSS $r$ band image is shown in the top left. The 2D
model of this galaxy and its residual (image - model) are also shown
as indicated. Inset into the residual is a postage stamp of the PSF
constructed for this galaxy. Blue captions within the PSF postage
stamp indicate (anti-clockwise from top-left) the full-width half-maximum
of the PSF; the number of stars used in creating the PSF; the $\frac{\chi^{2}}{\nu}$
for the PSF model fit and the size of the PSF postage stamp ($8.5''\times8.5''$).
The 1D profile for this galaxy, calculated by taking the average counts
along ellipses centred on the primary galaxy and displayed against
the semi-major axis, was created using the IRAF package \noun{ellipse}.
Example ellipses from \noun{ellipse} have been added to the image
in the bottom left to guide the eye, spaced evenly at intervals of
$2''$ along the semi-major axis. Note that . Inset into the 1D profile
are relevant output Sérsic modelling parameters for this galaxy, and
the overall $\frac{\chi^{2}}{\nu}$ for the Sérsic model is shown
in grey at the bottom of the figure. The 1D profile is a 1D measure
of 2D data. Any flux from secondary objects (neighbouring galaxies
and stars) lying outside the 1D mask and overlapping with the primary
galaxy will be counted as belonging to the primary galaxy by \noun{ellipse}.
The 1D profile therefore should chiefly be used as a guide as to the
true light distributions of both the 2D image and 2D model. For plotting,
the image data is divided by some scaling constant ($150$ counts
in the $r$ band) and scaled using the arctan function with cut levels
at $-\frac{\pi}{4}$ and $\frac{\pi}{2}$.}
\end{figure*}

If the fit has been successful, the output multi-HDU FITS file from
GALFIT is saved as \emph{objim}, with a catalogue of final modelled
secondary objects saved as \emph{extct}.\emph{ }An example model output
for GAMA galaxy G00092907 is shown in Figure \ref{fig:objim}. A series
of value added measurements are calculated and added to the structural
measurements already taken for the primary galaxy. These include $\mu_{0}$
(central surface brightness), $\mu_{e}$ (surface brightness at the
half-light radius), $\langle\mu_{e}\rangle$ (average surface brightness
within the half-light radius) and $r_{90}$ (radius along the semi-major
axis that contains $90\%$ of the total%
\footnote{integrated to infinity%
} Sérsic flux), amongst others. These values, along with the output
parameters from GALFIT and previous SIGMA modules are added to a comma-separated
variable (CSV) catalogue, allowing SIGMA to move to the next primary
galaxy in the input catalogue.

\section{SIGMA Output}

\label{sec:output}The SIGMA master catalogue, \emph{SersicCatv7},
provides measurements of Sérsic index, half-light radius, position
angle, ellipticity and magnitude in addition to extra pre-modelling
sky estimation, Source Extraction and PSF Extraction measurements
and post-modelling value added measurements as detailed in Section
\ref{sec:sigma}. Magnitudes contained within this catalogue are defined
according to the AB magnitude system, and have not been corrected
for the effects of foreground Milky Way dust extinction. The catalogue
is an output of the GAMA \emph{SersicPhotometry} data management unit,
and contains $527$ columns of data; $58$ columns per passband, and
$5$ additional common descriptive columns. Here we discuss the results
of the modelling pipeline.

\subsection{Sample Definitions}

\label{sub:sample}Table \ref{tab:sample} summarises the various
sample definitions in use throughout this paper. Our initial input
is the GAMA tiling catalogue, \emph{TilingCatv11}, which contains
$169,850$ sources. Of these sources, $167,600$ are classed as galaxy-like
(as defined in \citealp{Baldry2010}). SIGMA was run across this galaxy-like
subset independently in all nine bands, the output of which defines
the SIGMA master catalogue, \emph{SersicCatv7}, which is available
via the GAMA database. However, we define additional sub-samples in
order to facilitate further analysis of the data throughout the remainder
of this paper. This is to ensure that selection bias does not adversely
affect our conclusions. \emph{SersicCatv07} contains sources fainter
than the deepest nominal GAMA limit of $r_{\mathrm{petro}}=19.8$,
and so a cut was made limiting sources to $r_{\mathrm{petro}}<19.8$.
A common coverage sample was constructed so as not to compare galaxies
between bands whose observations are incomplete or have missing data.
This was defined using the Source Extractor Auto magnitude SEX\_MAG\_X
from \emph{SersicCatv7}, a product of the \noun{objectpipe} module,
where X=$UGRIZYJHK$. A common region is defined as having a detected
Source Extractor magnitude in any of the SDSS bands ($ugriz$) as
well as in each of the UKIDSS bands ($YJHK$). Incompleteness mainly
affects the NIR bands, with noticeable UKIDSS footprint gaps visible
in the final common coverage area shown in Figure \ref{fig:coverage}.
The number of detected sources in individual SDSS bands is typically
very high, $>97\%$, with the exception of the $u$ band. The $u$
band data has a detection percentage via this method of $50.8\%$,
indicating the poorer quality of the data in that band. For this reason,
$u$ band data is excluded from further fits to the data, with relations
instead extrapolating into the $u$ band wavelength for reference.
The common coverage area reduces the sample to $138,269$ galaxies
and is used throughout Section \ref{sec:output} and the beginning
of Section \ref{sec:wavelength} (with the exception of Figure \ref{fig:sersicturnoff}).
A full listing of detected and modelled sources used in each band
may be found in Table \ref{tab:numbers}. In any analysis that makes
use of data from the \emph{StellarMassesv03} catalogue (rest-frame
K-corrected $u-r$ colours or stellar masses), a reduced matched coverage
sample of $116,951$ objects is defined.

\begin{table}
\begin{tabular*}{1\columnwidth}{@{\extracolsep{\fill}}|>{\raggedright}p{0.25\columnwidth}|c|>{\raggedright}p{0.45\columnwidth}|}
\hline 
Name & Number & Definition\tabularnewline
\hline 
\hline 
TilingCatv11 & $169,850$ & Complete GAMA tiling catalogue\tabularnewline
\hline 
SersicCatv07 & $167,600$ & Removes star-like objects\tabularnewline
\hline 
Survey & $150,633$ & Removes $r_{\mathrm{petro}}<19.8$\tabularnewline
\hline 
Common & $138,269$ & Requires SIGMA Source Extractor coverage in ($ugriz$)+$Y$+$J$+$H$+$K$\tabularnewline
\hline 
Matched & $116,951$ & Requires a match in the \emph{StellarMassesv03} catalogue\tabularnewline
\hline 
\end{tabular*}

\caption{\label{tab:sample}Table defining various sample definitions in use
throughout this paper. Cuts are sequential, and include the definitions
from previous rows.}

\end{table}

\begin{table}
\begin{tabular}{|c|c|c|}
\cline{2-3} 
\multicolumn{1}{c|}{} & Detected & Modelled\tabularnewline
\hline 
$u$ & $85,138$ & $81,120$\tabularnewline
\hline 
$g$ & $165,367$ & $165,196$\tabularnewline
\hline 
$r$ & $166,506$ & $166,384$\tabularnewline
\hline 
$i$ & $166,675$ & $166,377$\tabularnewline
\hline 
$z$ & $163,902$ & $160,684$\tabularnewline
\hline 
$Y$ & $156,702$ & $156,280$\tabularnewline
\hline 
$J$ & $152,316$ & $151,612$\tabularnewline
\hline 
$H$ & $159,464$ & $158,797$\tabularnewline
\hline 
$K$ & $157,537$ & $156,662$\tabularnewline
\hline 
\end{tabular}

\caption{\label{tab:numbers}The number of detected and modelled galaxies in
\emph{SersicCatv07} for each band. \emph{SersicCatv07} contains $167,600$
galaxies in total.}
\end{table}

\subsection{Analysis}

\subsubsection{Additional Sky Subtraction}

\label{sub:sky}As part of the cutout creation process, SIGMA uses
a variable background mesh to estimate and subtract the background
sky for each galaxy in each band before any other image analysis takes
place. Sky correction distributions are mostly Gaussian in shape,
with a small bias to recovering positive sky values most likely owing
to background source contamination at the sky estimation stage. The
additional correction on top of that already applied at the GAMA mosaicing
stage is usually small. In the $r$ band for example, the sky correction
distribution has a $3\sigma$-clipped mean of $0.56$ ADU's ($\sim0.01\%$
of the sky pedestal) and a standard deviation of $\sigma=2.06$ ADU's.
Longer wavelengths produce larger corrections as expected. The accuracy
with which we were able to estimate the sky using this method was
found to produce good quality sky estimates in an efficient and relatively
fast manner.

An additional spike feature at zero counts relates to objects whose
determined preferred background mesh size was larger than that already
used in the creation of the GAMA mosaics. If this has occurred, SIGMA
performs no sky subtraction, and returns zero counts. This feature
affects $0.39\%$ of galaxies in the $r$ band, and so whilst a larger
mosaicing background mesh may be preferred for future surveys, it
is not believed to be a major issue affecting these data.

\subsubsection{Astrometry}

\begin{figure*}
\includegraphics[width=1\textwidth]{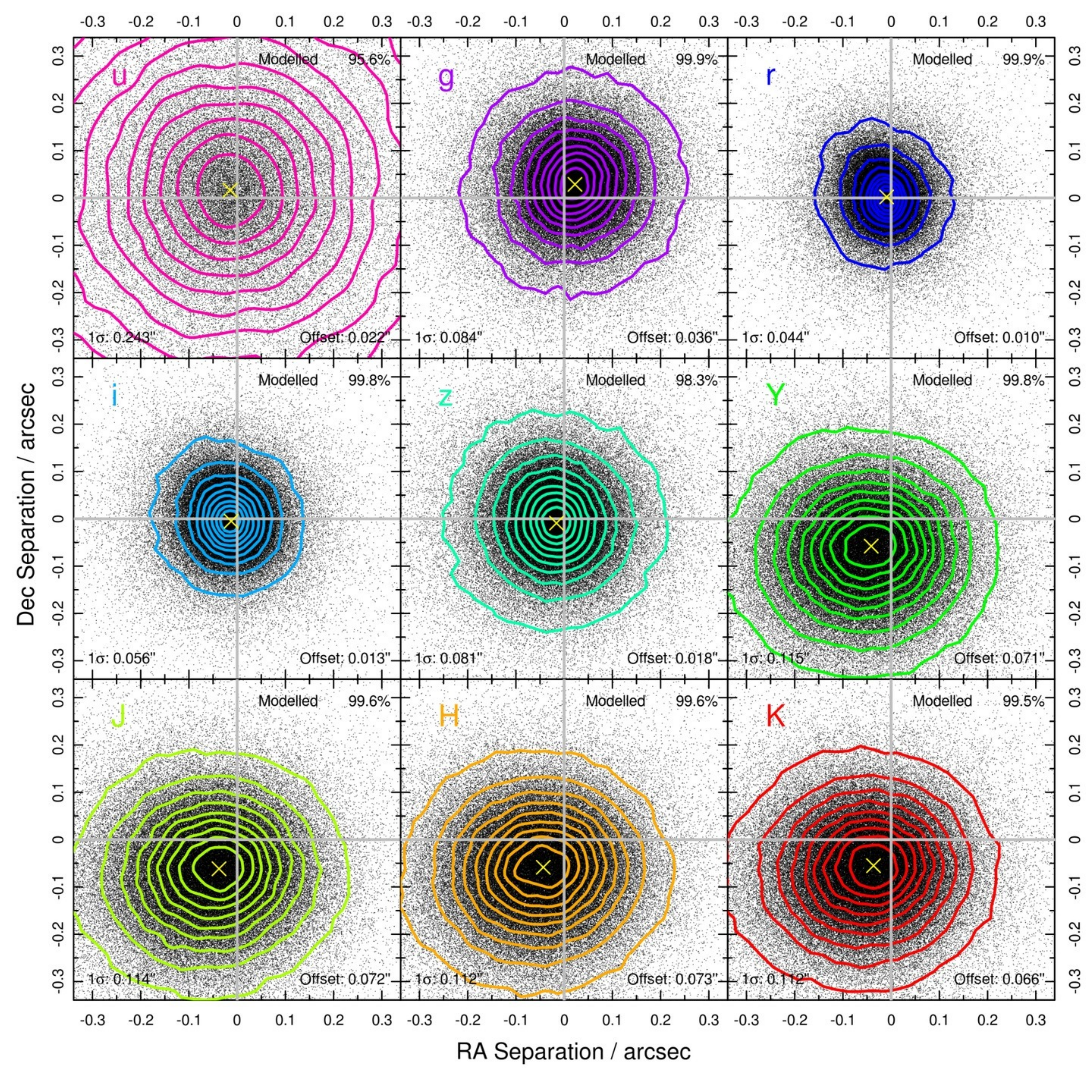}

\caption{\label{fig:astrometry}Astrometric offsets in RA and Dec between the
input SDSS positions and the modelled SIGMA positions for all nine
bands. Contours range from the $10^{th}$ to the $90^{th}$ percentile
in steps of $10\%$ with the peak density of each distribution represented
by a yellow cross. Each sub-plot is exactly $2\times2$ pixels in
dimension. The global systematic offsets in the NIR UKIDSS data ($YJHK$),
typically $\sim0.07''$ ($\sim0.2$ pixels), are caused by minor variations
in the WCS definitions between SDSS and UKIDSS. SIGMA accounts for
this during the modelling phase.}
\end{figure*}

\label{sub:astrometry}Initial checks were made on the output astrometric
accuracy of the SIGMA models. Figure \ref{fig:astrometry} shows the
computed astrometric offsets between the input SDSS positions and
their GALFIT modelled SIGMA positions for all bands, with each sub-plot
representing $2\times2$ pixels. Generally speaking, the astrometry
is in good agreement with SDSS, with the $r$ band offset $c_{r}=0.010''$
($0.029$ pixels) and a $1$-sigma spread of $1\sigma_{r}=0.044''$
($0.130$ pixels). The one exception to this is the $u$ band data,
showing a much larger spread in the recovered centroids owing to the
poor quality and depth of the data in this band. There are however
two interesting features worthy of note in this figure. First, the
apparent asymmetry in the SDSS astrometry, particularly noticeable
in the higher quality $r$ and $i$ bands. Second, the global systematic
offsets in the NIR UKIDSS bands ($YJHK$) of approximately $0.07''$. 

The asymmetry present in the SDSS astrometric data is found to be
associated with an individual SDSS strip%
\footnote{In SDSS imaging data, a single run covers a strip. Two strips constitute
a stripe, with the second strip offset from the first in order to
cover a continuous area.%
} that crosses the G09 field at a large angle of incidence with respect
to the equatorial plane. Galaxies whose input imaging data lie in
this strip appear to have centroids scattered around $\Delta\mathrm{RA}\sim-0.05''$,
$\Delta\mathrm{Dec}\sim0.10''$ rather than the origin. This feature
is less prominent in the lower quality SDSS bands as it becomes lost
in the random scatter, and consequently the effect is most noticeable
in the $r$ and $i$ SDSS bands. Since this asymmetry affects only
one strip of an SDSS stripe, the error must have been introduced at
the splicing stage within the SDSS pipeline. These offsets remain
small however, and are not believed to seriously affect this study
as they are accounted for during the SIGMA modelling pipeline.

Global systematic offsets in the NIR bands represent minor differences
in the WCS calibration between the SDSS and UKIDSS data. Any discrepancy
between the imaging data would be carried through to the larger GAMA
mosaics. This feature also varies according to GAMA region, with measured
offsets of approximately $0.05''$, $0.11''$ and $0.09''$ in G09,
G12 and G15 respectively. As with the previous feature, whilst consistent,
offsets remain small sub-pixel variations ($\sim0.2$ pixels) and
therefore are not believed to be a major factor affecting cross-band
matching between sources within GAMA. These features do not arise
at the GALFIT modelling stage, as similar plots comparing input SDSS
positions against pre-modelling Source Extractor centroids from SIGMA
show similar results albeit with larger spreads. On the contrary,
SIGMA should do a better job of recovering true centroids due to GALFIT's
model extrapolation method in estimating centroids. This makes SIGMA
robust against astrometric errors such as this by recentring every
galaxy at the modelling stage, emphasising the strength of full modelling
against basic source extraction.

\subsubsection{Seeing}

\label{sub:seeing}
\begin{figure}
\includegraphics[width=1\columnwidth]{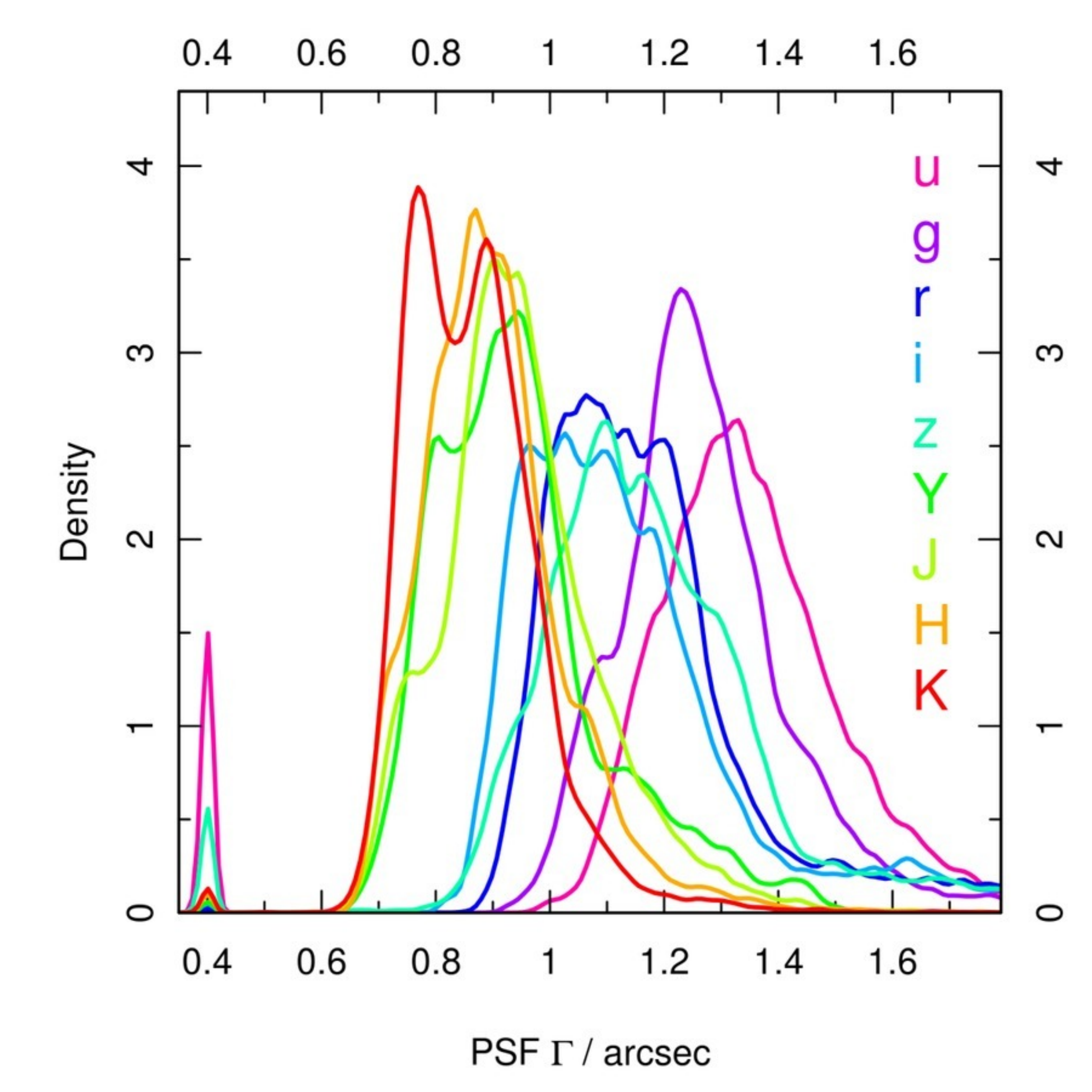}

\caption{\label{fig:seeing}Recovered full-width half-maximum PSF values from
the SIGMA common coverage sample. }

\end{figure}

An independent measure of the seeing and the form
of the PSF for each galaxy in each band is a necessary requirement
when considering galaxy modelling. Through PSFEx in \noun{psfpipe},
SIGMA is able to provide robust measurements of the PSF for each galaxy
as described in Section \ref{sub:psfpipe} prior to the GALFIT modelling
stage. Figure \ref{fig:seeing} shows the recovered PSF full-width
half-maxima $\Gamma$ for every galaxy within the SIGMA common sample
for all 9 bands. Each density curve has a main peak in the range $0.7''<\Gamma<1.4''$,
and an additional peak at $\Gamma=0.4''$, which shall be discussed
later. 

We note that on average the NIR data is of better seeing than the
optical, with the former in the range $0.6''<\Gamma<1.3''$ and peaking
at around $\Gamma=0.9''$, and the latter in the range $0.8''<\Gamma<1.7''$
with variable peaks. These ranges are in good agreement with UKIDSS
($K$ band) and SDSS ($r$ band) seeing targets of $\Gamma_{\mathrm{UKIDSS,K}}<1.2''$
and $\Gamma_{\mathrm{SDSS,r}}<1.5''$ respectively. Some of the NIR
data displays a secondary peak, particularly in the $K$ band, possibly
due to the use of microstepping in the taking of some of the UKIDSS
data. The worst quality seeing data is in the $u$ band, exhibiting
the largest width distribution, and the highest seeing data on average.
This distribution of its mean across the GAMA regions is represented
in Figure \ref{fig:coverage}, with the data points coloured according
to the measured seeing at that location. This figure shows significant
striping in the SDSS $r$ band due to the drift scan mode of collection,
and a measure of consistency coupled with lower average values across
each of the UKIDSS bands. This could cause significant problems for
image analysis routines, with average seeing doubling on the scale
of a few pixels. Modelling the PSF and using that model at the galaxy
modelling stage, as in SIGMA, goes some way towards mitigating this
issue.

An additional peak at $\Gamma=0.4''$ represents those frames where
no stars were detected in order to compute the PSF in that region,
and so a generic value of $\Gamma=0.4''$ is returned. Note that for
the majority of bands this problem is minimal, becoming most noticeable
in the lower quality $u$ and $z$ band data.

\subsubsection{Surface Brightness Limits}

\label{sub:sblims}
\begin{figure}
\includegraphics[width=1\columnwidth]{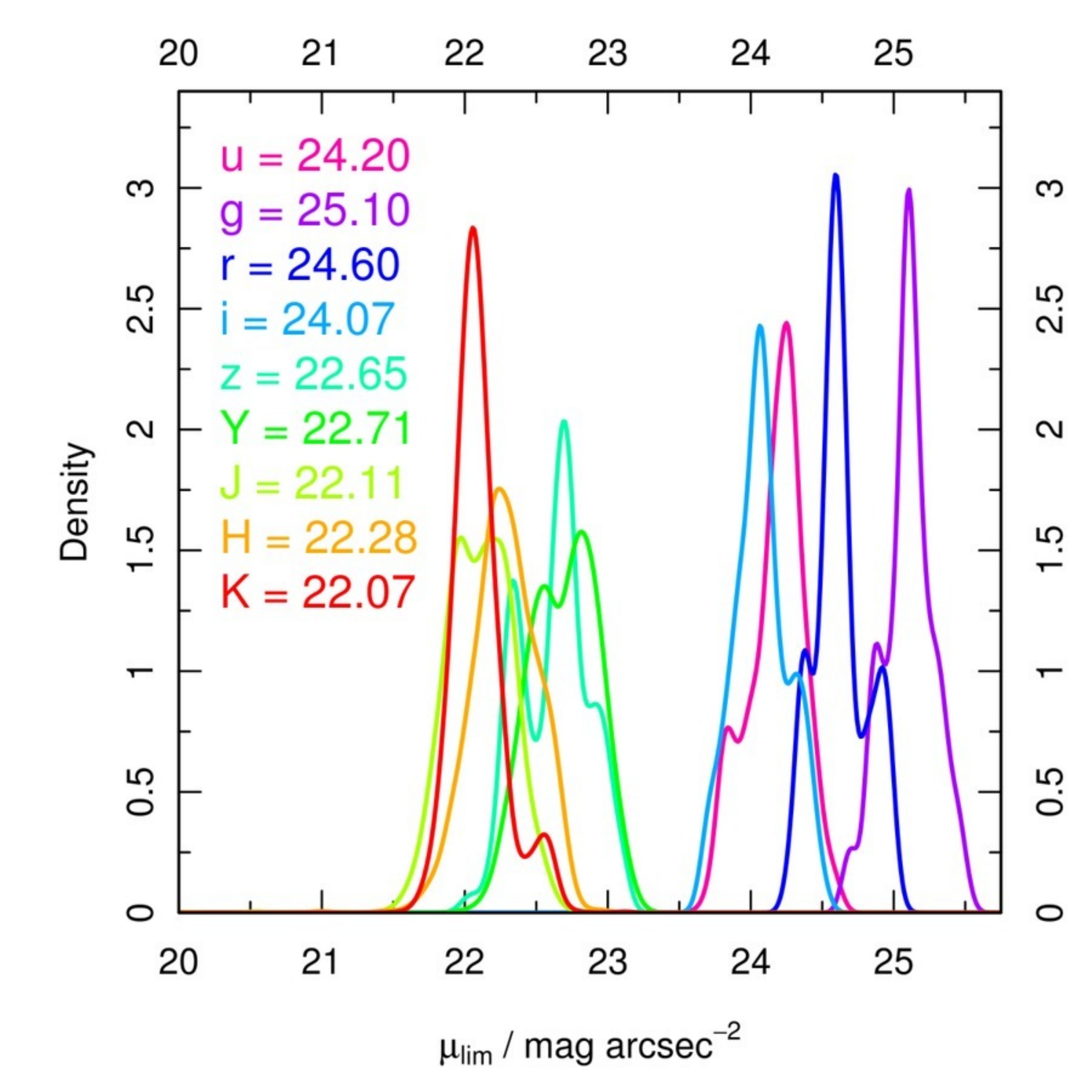}

\caption{\label{fig:sblims}Apparent surface brightness limits for all galaxies
within the SIGMA common coverage sample, with median surface brightness
values for each band inset.}
\end{figure}

\begin{figure*}
\includegraphics[width=1\textwidth]{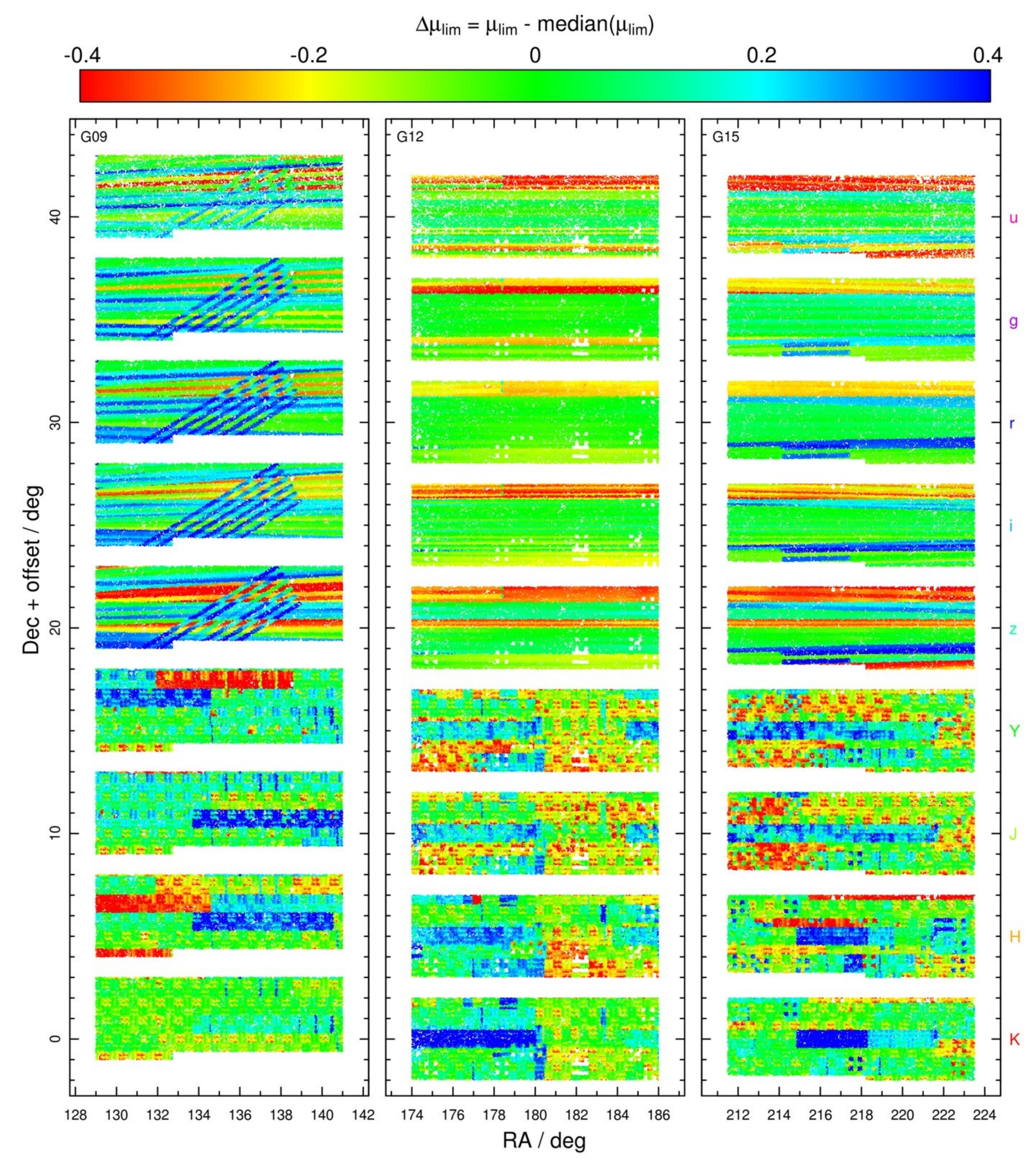}

\caption{\label{fig:sblimsradec}Apparent surface brightness limits for all
galaxies within the SIGMA common coverage sample as a function of
their position on the sky in right-ascension and declination. The
three GAMA regions are displayed, as indicated, with each band labelled
along the right side of the figure. Bands are offset in declination
in order to differentiate them from one another. The $K$ band is
situated at the correct GAMA coordinates. Surface brightnesses are
shown as offsets from the median surface brightness for that band,
the values of which are found in Figure \ref{fig:sblims}. Blue data
points represent the deepest limits and red the shallowest.}

\end{figure*}

Consideration of the surface brightness limit beyond
which data becomes unreliable at the $1\sigma$ level is also important.
An estimate of the surface brightness limit at any given position
may be given by
\[
\mu_{\mathrm{lim}}=\mathrm{ZP}-2.5\log I_{\mathrm{RMS}}
\]
where $\mathrm{ZP}$ is the zero-point of the imaging data, and $I_{\mathrm{RMS}}$
is the root-mean-square of the background sky per square arcsecond.
Note however that this provides a worst-case scenario value to the
surface brightness limit, with the real limit likely to be deeper
on a per-galaxy basis dependent upon the number of pixels ($n$) used
in constructing the 2D model at large radii from the core region,
and scaling as $\sqrt{n}$. Figure \ref{fig:sblims} shows the global
distributions in $\mu_{\mathrm{lim}}$ for the SIGMA common coverage
sample across each bandpass, with the median surface brightness limits
in each band given inset into the figure. We note that the shorter
wavelengths typically exhibit deeper limits, as expected, with a transition
occurring to shallower limits beyond the $i$-$z$ interface. Figure
\ref{fig:sblimsradec} shows the spatial variation of $\mu_{\mathrm{lim}}$
across the GAMA fields. The deepest $\mu_{\mathrm{lim}}$ data is
represented by blue data points, the shallowest by red. The centroid
weighting mechanism employed by GALFIT should minimise the impact
of a spatially varying $\mu_{\mathrm{lim}}$, and therefore should
not heavily affect the output results from SIGMA.

\subsection{Results}

\label{sec:results}

\subsubsection{Case Study Examples}

\label{sub:casestudy}
\begin{figure*}
\includegraphics[height=0.9\textheight]{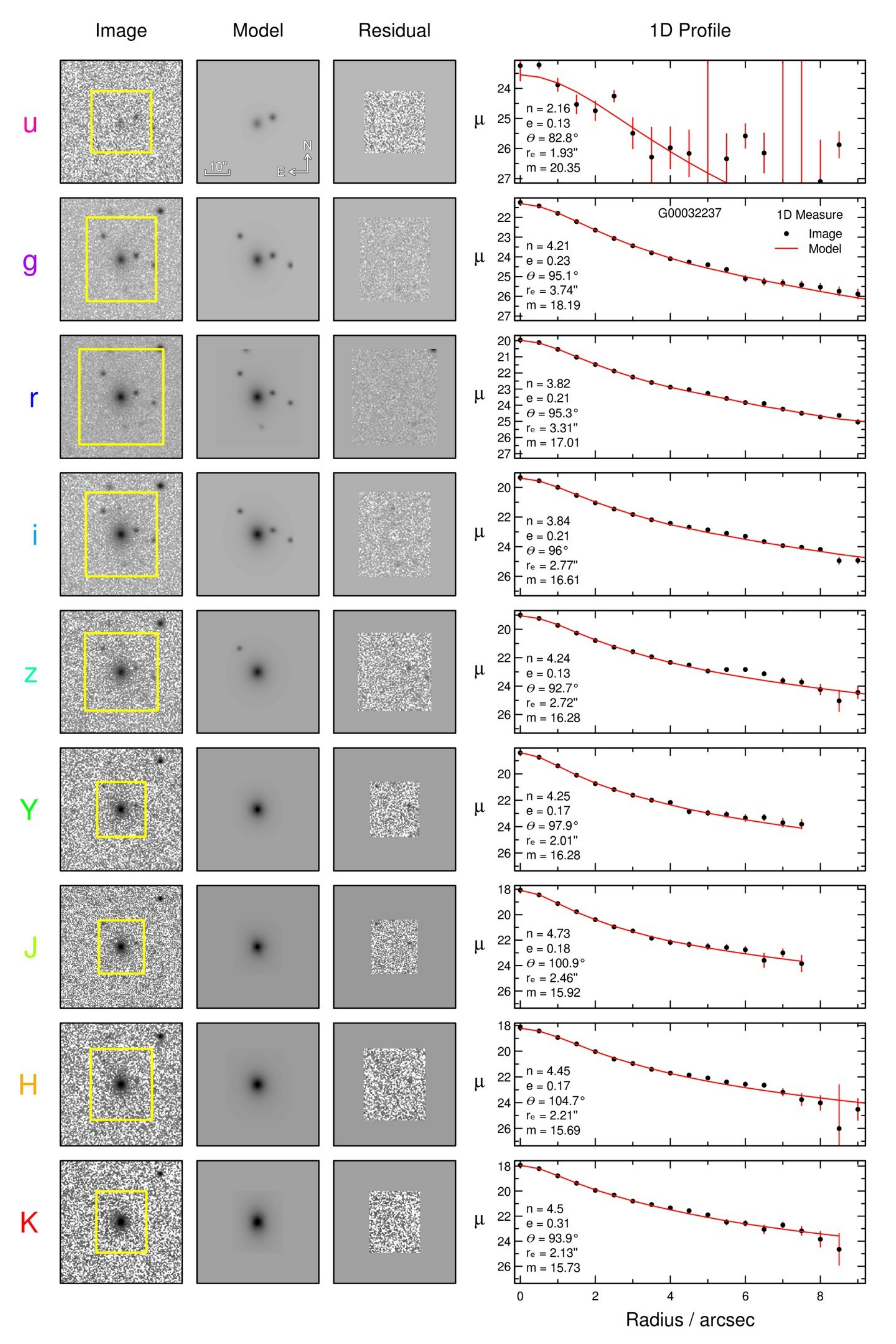}

\caption{\label{fig:caseband}Model fits for G00032237 across all nine bands.
Each column represents (from left to right) the original input image,
the model fit to the input image, the residual image (input - model)
and the 1D surface brightness profile along the semi-major axis (averaged
along the annulus). The fitting region (the region within which 2D
modelling takes place) is represented by a yellow box. Recovered 2D
intrinsic (i.e., prior to PSF convolution) Sérsic parameters are listed
inset into the 1D profile plot. The images are scaled logarithmically
from $-1$ $\sigma_{\mathrm{sky}}$ to $n\sigma_{\mathrm{sky}}$,
where $\sigma_{\mathrm{sky}}$ is the typical RMS of the sky in that
band, and $n$ is some scaling constant (generally, $n\sim40$).}
\end{figure*}

\begin{figure*}
\includegraphics[height=0.9\textheight]{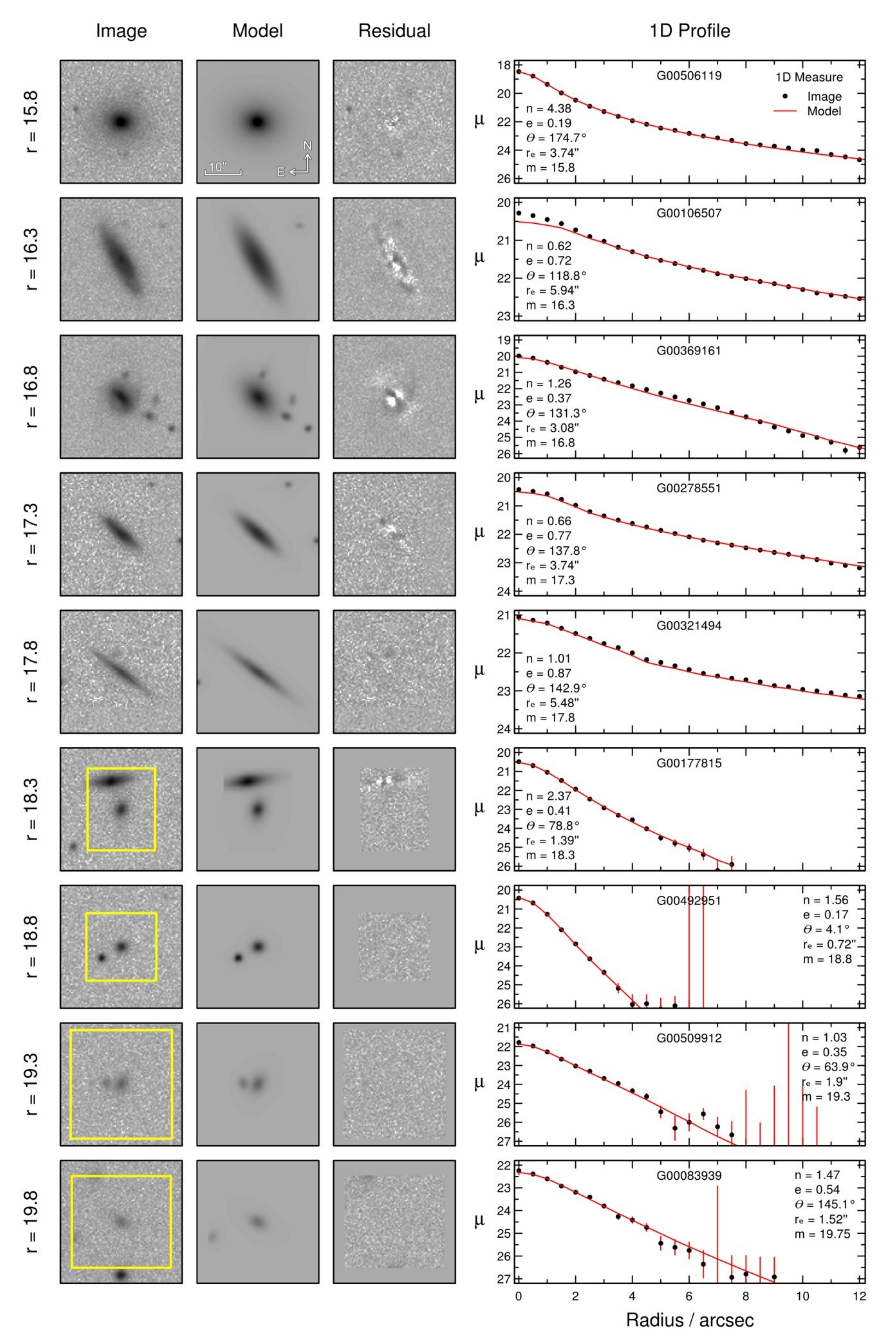}

\caption{\label{fig:casemag}Model fits for nine galaxies in the $r$ band
separated in magnitude space by approximately $\Delta m_{r}=0.5$
in the range $15.8<m_{r}<19.8$. Each column represents (from left
to right) the original input image, the model fit to the input image,
the residual image (input - model) and the 1D surface brightness profile
along the semi-major axis (averaged along the annulus). If the fitting
region (the region within which 2D modelling takes place) lies within
the image thumbnail above, it is represented by a yellow box. If no
yellow box is visible, the fitting region is larger than the thumbnail.
Recovered 2D intrinsic (i.e., prior to PSF convolution) Sérsic parameters
are listed inset into the 1D profile plot. The images are scaled logarithmically
from $-1$ $\sigma_{\mathrm{sky}}$ to $40$ $\sigma_{\mathrm{sky}}$,
where $\sigma_{\mathrm{sky}}$ is the typical RMS of the sky in the
$r$ band.}

\end{figure*}

We present example model fits for an individual
galaxy across all nine bands and various galaxies separated in magnitude
space in Figures \ref{fig:caseband} and \ref{fig:casemag} respectively.
These figures each represent the input 2D science image, model and
residual along with a 1D profile radiating outwards from the core
region of the galaxies along the semi-major axis. The displayed input
image is a postage-stamp sub-region of the background corrected cutout
from the original GAMA mosaiced image. The yellow box in each of the
input image postage stamps represents the size of the fitting region
as determined in \noun{galfitpipe}, the dimensions of which specify
the size of the output model FITS image. Recovered Sérsic parameters
are listed inset into the 1D profile plot.

Figure \ref{fig:caseband} shows the output SIGMA results for the
elliptical galaxy G00032237 across all nine bands. Each image is modelled
independently in each band, leading to a variable fitting region size
dependent upon local conditions including object density and the physical
size of the primary galaxy. The residuals in each band show the high
quality of the fit for this particular galaxy, bar some minor core
disturbance in the higher quality $r$ and $i$ bands. These bands
cover wavelengths that are more sensitive to dust attenuation, in
this case potentially highlighting small quantities of dust in the
centre of the galaxy possibly related to a recent minor-merger or
some form of morphological disturbance. Dust has the effect of perturbing
the light profile slightly away from that of a purely single-Sérsic
object. Interestingly however, no evidence for dust lanes are evident
in the lowest wavelength $u$ band residual. This should not be surprising
considering the lower quality data of the $u$ band, hence these small
perturbations would be lost in the noise of the image. Barring the
$u$ band data, and despite dust attenuation, the recovered Sérsic
indices remain relatively stable, ranging from $n=4.21$ to $n=4.50$
in $g$ to $K$. Sérsic index peaks in the $J$ band ($n_{\mathrm{max}}=4.73$)
and reaches a minimum in the $r$ band ($n_{\mathrm{min}}=3.82$,
excluding $u$ band data). Modelled ellipticity $e$ and position
angle $\theta$ are also in good agreement, with recovered Sérsic
magnitude evolving as expected across this wavelength range. Interestingly,
the recovered half-light radii show a size-wavelength dependence,
ranging from $r_{e}=3.74''$ to $r_{e}=2.13''$ in $g$ to $K$. 

Secondary objects whose object centres lie outside the fitting region
but whose flux leaks into it are masked so as not to effect the model
fit. One such object can be seen in the upper-right corner of the
fitting region in the $r$ band postage stamp in Figure \ref{fig:caseband}.
The \noun{galfitpipe} module creates a bad-pixel mask using the segmentation
map provided by \noun{objectpipe}. Should GALFIT reach an error whilst
trying to converge on a model, a potential additional fix is to mask
all secondary objects in the field of view, and re-run GALFIT. This
dynamic masking, whilst not the first choice for producing a model,
typically allows high-quality and consistent model data to be extracted
from difficult regions where otherwise none would exist.

Figure \ref{fig:casemag} shows nine example galaxies in the $r$
band from the SIGMA common coverage sample, separated in magnitude
space in approximately equal SDSS $r$ band magnitude steps of $\Delta m_{r}=0.5$
from the faintest GAMA limit of $m_{r}=19.8$. These galaxies span
a wide range of morphologies and environments, exhibiting the large
variance in the input data processed by SIGMA. In each case, the residual
images show the quality of the fits are relatively high, more so for
obvious single-component objects such as the huge elliptical galaxy
G00506119 ($m_{r}=15.8$) than multi-component objects such as the
barred-spiral galaxy G00369161 ($m_{r}=16.8$). In the case of the
latter, despite a global single-Sérsic fit to a multi-component object,
the resultant model does a good job at describing global parameters
such as magnitude. Despite a relatively disturbed fit to the secondary
neighbour of G00177815 ($m_{r}=18.3$), the quality of the primary
galaxy model remains high. This outlines the exclusive use of secondary
objects in accounting for additional flux in the wings of primary
galaxies. Note that whilst the quality of model fits reflected through
the residuals appears to become better at fainter magnitudes, this
effect is more likely an example of non-resolved components of a galaxy. 

\subsubsection{Global Results}

\label{sub:global}
\begin{figure*}
\includegraphics[width=1\textwidth]{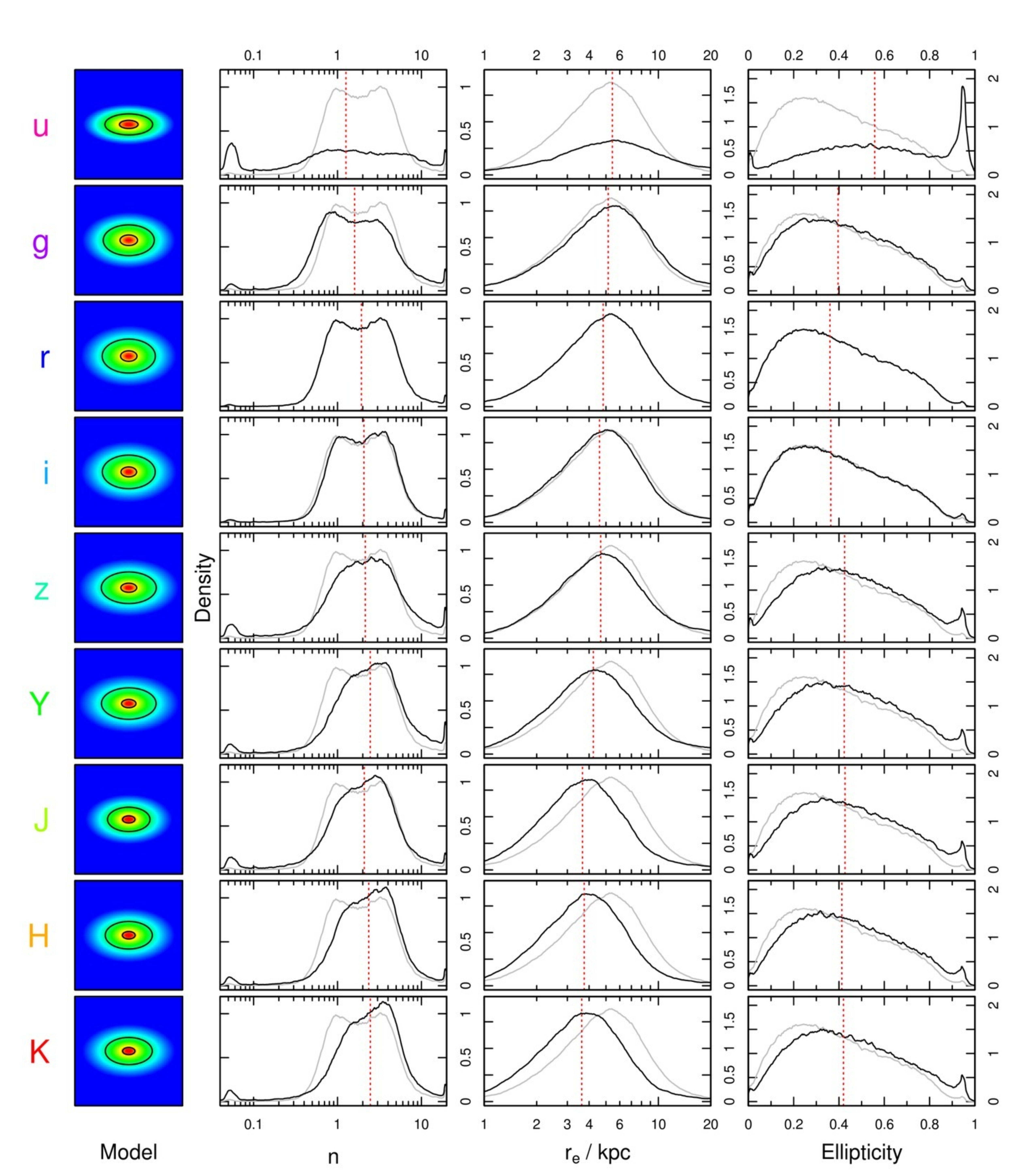}

\caption{\label{fig:fullresults}Global results from the SIGMA common coverage
sample for all nine bands. Each column represents (from left to right)
the average model galaxy based on median values for Sérsic index,
half-light radius and ellipticity and the distributions for all recovered
Sérsic indices, half-light radii (converted into kpc) and ellipticities.
The y-axis for each distribution shows the probability density function
convolved with a rectangular top-hat kernel with standard deviations
of $0.05$, $0.05$ and $0.02$ for index, size and ellipticity respectively.
Median values for each distribution are represented by a red dashed
line and are used in creating the average model galaxy in the left-hand
column. The $r$ band distributions are shown in grey for reference.}
\end{figure*}

Complete distributions for the SIGMA common coverage
sample of $138,269$ galaxies are shown in Figure \ref{fig:fullresults}.
Here we plot 1D density distributions for recovered model Sérsic indices,
half-light radii and ellipticities in each band. Alongside these distributions
are displayed the average model galaxies based on median values from
the aforementioned parameters. 

Recovered Sérsic parameters peak primarily in the range $0.2<n<10$,
with additional peaks at $n\sim0.05$ and $n=20$ arising due to failed
fits, discussed in more detail below. The primary range appears bimodal
in nature, consisting of two approximately Gaussian-like distributions
whose means are $n\sim1$ and $n\sim3.5$. These two peaks, for the
most massive/brightest systems in GAMA, correspond to the two main
galaxy morphologies as originally identified by Hubble, namely, late-type
disk-dominated galaxies and early-type spheroid-dominated galaxies
for $n=1$ and $n=3.5$ respectively. Interestingly, the second of
these two peaks does not appear at $n=4$, which is typically expected
for a classic de Vaucouleurs profile. The relative strength of these
two peaks shifts with increasing wavelength, with the stronger disk-dominated
peak at $n=1$ giving way to the spheroid-dominated $n=3.5$ peak
at wavelengths longer than the $i$ band. This is believed to be an
indicator of the shifting in observed stellar population with wavelength,
however, see Section \ref{sub:wavelength} for further discussion.
In addition, the centroid of the $n=1$ peak at wavelengths longer
than the $i$ band appears to move towards higher Sérsic index values,
merging into an elongated shoulder of the relatively stable $n=3.5$
peak. In the $K$ band, the first peak appears to have a mean centred
on $n\sim1.5$. This should not be surprising, as optical bands are
more likely to probe the young stellar populations in the disks of
galaxies whereas the longer wavelengths pick out the older stellar
populations within the core regions of a spiral galaxy or in elliptical
galaxies. Dust may also be an issue at shorter wavelengths, blocking
light from the core regions of galaxies and therefore biasing recovered
Sérsic indices towards lower values (for example, see \citealp{Pastrav2012}).
It is important to note that these data are derived from the same
$r$ band selected sample of galaxies observed in different wavelengths,
and so these relative shifts in peak positions represent real variances
in observed stellar populations, highlighting a wavelength dependence
on structural measurements.

The additional peaks at $n\sim0.05$ and $n=20$ represent failed
fits. For these galaxies, the fitting procedure drifted into an unrealistic
parameter space during the downhill minimisation routine employed
by GALFIT. Despite attempts to force a better fit from the data within
\noun{galfitpipe}, the fits to the images of these objects remain
corrupted, and are not appropriate for further analysis. Bad fits
occur for many reasons. Typical reasons are over-dense regions introducing
too many free-parameters into the minimisation routine, or bad sky
subtraction for that region. The corrupt peak values of $n\sim0.05$
and $n=20$ arise due to constraints placed by the fitting code GALFIT,
and unchanged for the purposes of this study. The upper peak is a
hard limit, with galaxies unable to obtain a Sérsic index beyond this
value. The lower peak is a result of a consistency check within the
GALFIT code that attempts to force a fit at $n>0.05$, hence leading
to a small distribution around this value. These errors that caused
these additional peaks are also the cause of those found in the ellipticity
distribution, discussed further below. The density of objects within
these failed regions scales with wavelength, with the higher-quality
bands exhibiting fewer cases than poorer quality bands such as the
$u$ band. A conservative estimate ($n<0.07$ or $n>19$) places $1.1\%$
($1,456$) of $r$ band galaxies within these extremely non-physical
regions, rising to $9.1\%$ ($12,630$) in the worst affected $u$
band.

Distributions of recovered effective radii (along the semi-major axis),
converted to kiloparsecs, are also shown. Density profiles at all
wavelengths appear relatively smooth, approximating a skewed Gaussian
distribution. Note that these distributions exhibit no additional
peaks as observed in the Sérsic index and ellipticity plots. When
regarding the median values of these distributions, represented in
Figure \ref{fig:fullresults} by red dashed lines, we note that the
median effective radius of a galaxy ranges from $5.5$ kpc in the
$u$ band to $3.5$ kpc in the $K$ band. This marked decrease in
physical size with observed wavelength is again as expected if one
expects the longer wavelengths to probe core stellar populations in
the bulges of galaxies, whilst shorter wavelengths probe recently
formed populations in the disks of galaxies (i.e., inside-out growth,
see \citet{Trujillo2005}). The transition wavelength appears to be
the $Y$ band, with a median size of $\sim4.5$ kpc in the $z$ band,
and $\sim3.5$ kpc in the $J$ band, highlighting once more the wavelength
dependence on structural measurements and the importance of the right
choice of wavelength when comparing galaxy samples. Indeed, there
appears to be little size variation at wavelengths longer than the
$Y$ band. Clearly, care must be taken when comparing the sizes of
galaxies observed at different rest wavelengths.

Ellipticity ($1-\frac{b}{a}$) measurements remain relatively consistent
across all bands, peaking in the range $0.25<e<0.35$, and displaying
additional peaks at $e=0$ and $e=0.95$. Bands $g$-$K$, excepting
the $r$ and $i$ bands, show very similar distributions, with a consistent
median value of $e\sim0.4$ and a modal value of $e\sim0.35$. The
higher quality $r$ and $i$ bands appear to have, on average, more
circular recovered ellipticities, with median and modal values of
$e\sim0.35$ and $e\sim0.25$ respectively, however the shift is minimal.
The lower quality $u$ band data leads the ellipticity measurements
in that band to be biased towards higher values, caused by the fitting
routine being more susceptible to background noise fluctuations and
random noise in the frame. Ellipticity measurements presented here
are global ellipticities, and there will be variability in ellipticity
with increasing radius from the core on a per-galaxy basis caused
by additional factors, for example, the effect of seeing or the presence
of a bar. The additional peaks directly correspond to those already
discussed previously, and represent failed fits. The values of $e=0$
and $e=0.95$ correspond to internal GALFIT boundaries.

\subsubsection{Sérsic Magnitudes}

\label{sub:sersmags}It is not known exactly how the light profile
of a galaxy behaves at large radii away from the core regions. The
exact nature of any profile will undoubtedly be influenced by many
factors including, but not limited to, recent merger history, star
formation rate, gas and dust content and local environment. Magnitudes
within isophotal radii, not surprisingly, systematically underestimate
the total galaxy light, in particular, relative to the Sérsic magnitude
(e.g., \citealp{Caon1990,Caon1993}). \citet{Graham2005} show for
example that Kron magnitudes may underestimate the total galaxy flux
by as much as $\sim55\%$ dependent upon choice of the multiple of
Kron radii chosen to integrate out to and the profile shape of the
galaxy. The comparative value for Petrosian magnitudes is considerably
worse, underestimating flux by as much as $\sim95\%$ in the extreme
case of a high-Sérsic index object integrating out to thrice the Petrosian
radius. In addition to these considerations, below $\mu_{B}=27$ mag
arcsec$^{-2}$ environmental effects begin to play a large role in
profile shape determination. 

In contrast to traditional aperture methods, studies have repeatedly
shown the strength of Sérsic profiling for the majority of elliptical
galaxies (e.g., \citet{Caon1993,Graham2003,Trujillo2004,Ferrarese2006}).
\citet{Tal2011} support this viewpoint, showing the light profiles
of massive ellipticals are well described by a single Sérsic component
out to $\sim8$ $r_{e}$, with evidence for additional flux beyond
these radii possibly related to unresolved intra-group light. With
regards to disk systems, \citet{Bland-Hawthorn2005} use one of the
deepest imaging studies of spiral galaxy NGC 300 to show that an exponential
profile ($n=1$) is a good descriptor of its light profile out to
$\sim17$ $r_{e}$. From a sample of 90 face-on late-type galaxies,
\citet{Pohlen2006} confirm the accuracy of Sérsic profiling down
to $\mu=27$ mag arcsec$^{-2}$, and suggest up to $10\%$ of their
sample show evidence for a deviation from a standard $n=1$ Sérsic
fit (Type I), instead showing a broken exponential profile. These
breaks appear in the form of either a \emph{downbending} (Type II;
steeper flux drop-off) or \emph{upbending} (Type III; shallower flux
drop-off) with increasing radii. Importantly, this study also suggests
this observed feature is independent of local environment. 

It is clear that opinion is divided amongst the community as to how
a galaxy behaves below the typical limiting isophote, particularly
so in the case of a disk galaxy. Each of these studies does however
suggest a more complex structure at large radii than a Sérsic profile
extrapolated out to infinity would imply. In order to account for
the lack of reliable profile information at large radii, Sérsic magnitudes
require some form of profile truncation so as not to extrapolate flux
into regimes of which we know little. Two schools of thought exist
in terms of appropriate truncation methods, extrapolating flux down
to a fixed surface brightness limit or integrating under the profile
out to a fixed multiple of the half-light radii. A constant surface
brightness limit is more closely related to galaxy gas content, and
so has physical meaning. However, this method introduces a redshift
dependence on truncated flux, causing different fractions of light
to be missed at different redshifts. Truncating at a given multiple
of the effective radius assumes that the effective radius is well
understood prior to truncation, which owing to the inter-dependency
between output Sérsic parameters, is not always evident. It does not
display any redshift dependence however, and is trivial to subsequently
recorrect if desired. Corrections are typically minor for most galaxies,
becoming most acute in high-index systems (see Figure \ref{fig:sersic}).

\begin{figure}
\includegraphics[width=1\columnwidth]{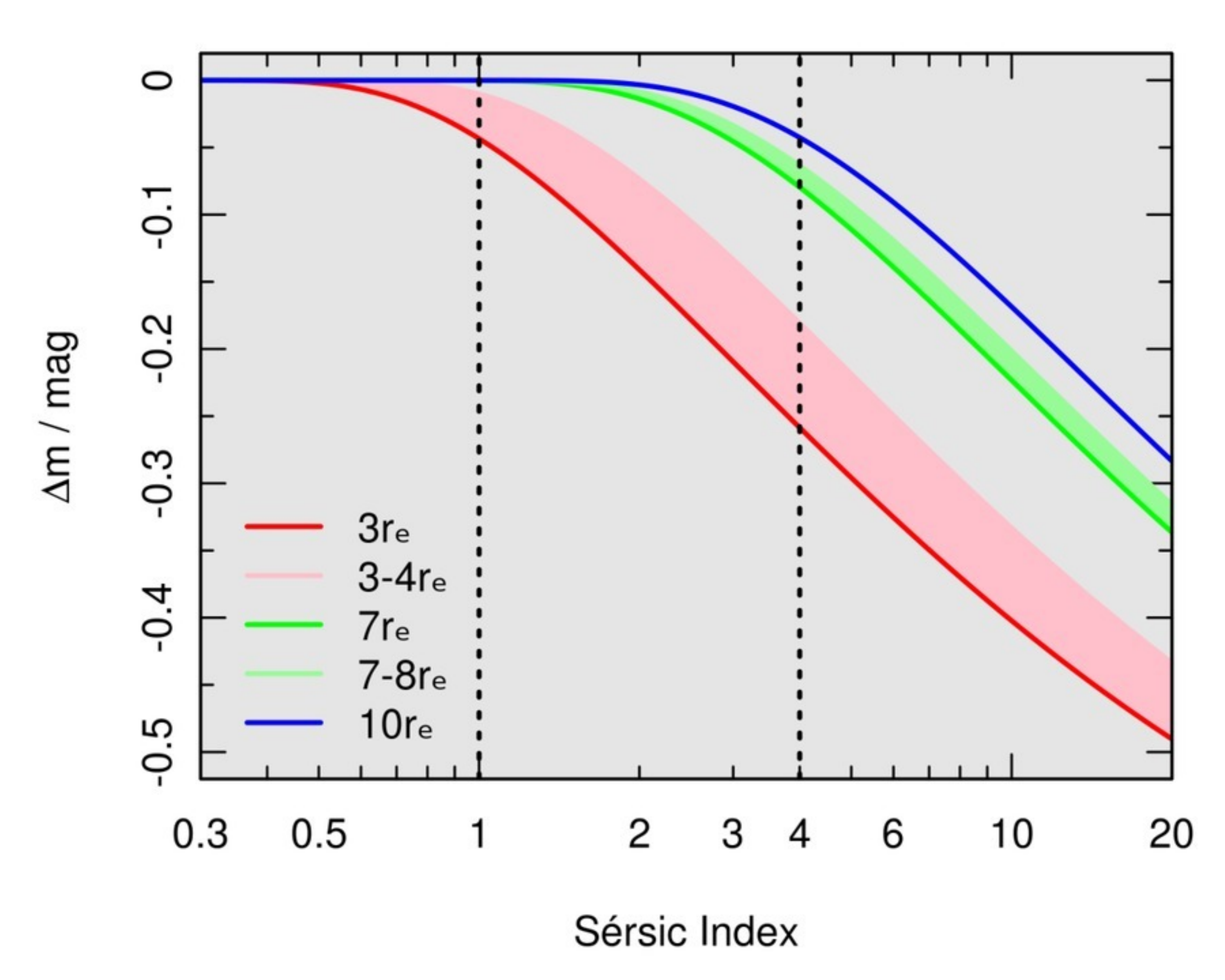}

\caption{\label{fig:sersic10re}Magnitude offset between the Sérsic profile
integrated out to infinity and that truncated at a given multiple
of the effective radius. SDSS model magnitudes (forcing either exponential
($n=1$) or de Vaucouleurs ($n=4$) profile fits, dotted vertical
lines) employ smooth tapering truncation radii, represented by the
shaded red and green areas. SIGMA Sérsic magnitudes within GAMA adopt
a sharp $10r_{e}$ truncation radius, blue line.}
\end{figure}

A sufficiently large truncation radius must be adopted to provide
a close estimate of total flux without extrapolating too deep into
the region of uncertainty. SDSS model magnitudes employ a smooth truncation
at $3r_{e}$ down to zero flux at $4r_{e}$ for exponential ($n=1$)
profiles and $7r_{e}$ down to zero flux at $8r_{e}$ for de Vaucouleurs
($n=4$) profiles. We adopt a sharp truncation radius of $10r_{e}$
for all Sérsic indices, which corresponds to an isophotal detection
limit of $\mu_{r}\sim30$ mag arcsec$^{-2}$, the limit to which galaxy
profiles have been studied. Figure \ref{fig:sersic10re} shows the
magnitude offset between the Sérsic profile integrated to infinity
and that truncated at $3$, $7$ and $10$ $r_{e}$ as red, green
and blue lines respectively, with shaded areas representing the SDSS
tapered limits. A $10$ $r_{e}$ truncation gives a negligible magnitude
offset for $n=1$, and an offset of $\Delta m\sim-0.04$ for $n=4$,
with larger corrections for higher Sérsic indices. Figure \ref{fig:sersic},
middle panel, shows the flux contained within $10$ $r_{e}$ (dashed
vertical line) for various values of $n$. Given a $10$ $r_{e}$
truncation, $\sim100\%$ of the pre-truncation flux is retained for
$n=1$, reducing to $96.1\%$ for $n=4$. Sérsic magnitudes truncated
at $10$ $r_{e}$ for each galaxy processed by SIGMA are adopted as
the standard Sérsic magnitude system throughout the remainder of this
paper, however, full (i.e., integrated out to infinity) Sérsic magnitudes
are provided alongside truncated Sérsic magnitudes in \emph{SersicCatv07}
for reference.

\begin{figure}
\includegraphics[width=1\columnwidth]{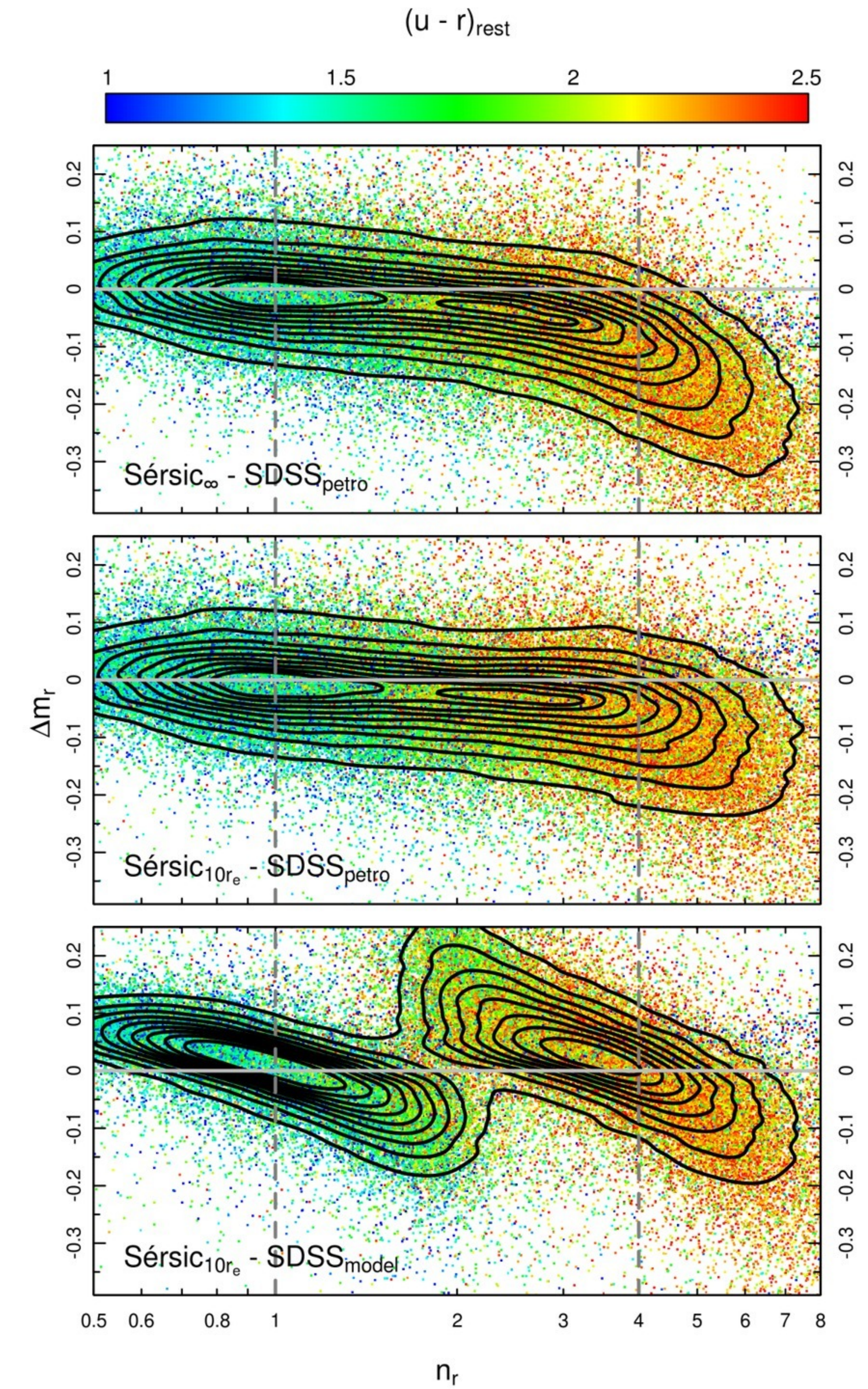}

\caption{\label{fig:sersicturnoff}A series of plots displaying offsets between
various magnitude systems as a function of $r$-band Sérsic index,
with the data points coloured according to their $u$-$r$ rest colour,
as shown. Contours range from the $10^{th}$ percentile to the $90^{th}$
percentile in $10\%$ steps. (top) The Sérsic profile integrated out
to infinity minus the SDSS Petrosian magnitude and; (middle) the Sérsic
profile truncated at $10r_{e}$ minus the SDSS Petrosian magnitude.
These figures show how Sérsic profiling is able to recover flux in
the wings of galaxies that would otherwise be missed by traditional
aperture based methods, such as Petrosian apertures. (bottom) The
Sérsic profile truncated at $10r_{e}$ minus the SDSS model magnitude.
SDSS force fit either an exponential ($n=1$) or de Vaucouleurs ($n=4$)
profile fit to attain their model magnitudes. This figure shows how
model magnitudes provide an inaccurate measure of flux for a galaxy
whose Sérsic index differs significantly from either $n=1$ or $n=4$.
Vertical dashed grey lines at exponential ($n=1$) and de Vaucouleurs
($n=4$) Sérsic indices are added for reference.}

\end{figure}

Figure \ref{fig:sersicturnoff} shows the offsets between various
magnitude systems discussed previously as a function of Sérsic index.
When comparing Sérsic magnitudes integrated to infinity against SDSS
Petrosian magnitudes we see the two systems are in good agreement
until $n_{r}\sim2$, beyond which the magnitude offset relation begins
to turn-off from the $\Delta m=0$ relation. This trend argues that
Sérsic magnitudes are recovering an additional $\sim0.4$ magnitudes
for an $n_{r}=8$ object which would otherwise have been missed by
traditional photometric methods. However, for the reasons previously
discussed, this value should be taken as a rough estimate of the true
amount of flux missed for an object of a given Sérsic index. Truncating
the Sérsic index at $10$ $r_{e}$ reduces the scale of this turn-off,
as expected, keeping the two magnitude systems in agreement out to
$n_{r}\sim3$, however still providing some measure of turn-off beyond
this point. We would expect SDSS Petrosian magnitudes, or indeed any
aperture-based photometry, to underestimate total flux for objects
with large wings, and so this result is not surprising and a good
indication that truncated Sérsic magnitudes are performing as expected.
The final panel in Figure \ref{fig:sersicturnoff} compares truncated
Sérsic magnitudes against SDSS model magnitudes. The SDSS force fit
either an exponential or de Vaucouleurs profile fit (marked on the
figure by dashed grey lines) depending upon which profile an individual
galaxy most approximates. We see clearly the inadequacy of model magnitudes
when a more comprehensive Sérsic magnitude is available, with the
population of galaxies split into two distinct sub-populations based
upon their SDSS forced fit. For a galaxy at $n=2$ for example, the
model magnitude for the galaxy may be offset from its correct magnitude
by as much as $\Delta m=\pm0.3$ magnitudes, with larger offsets observed
for high index galaxies. If one constructs a line of best fit for
each of these two artificial sub-populations the lines pass through
$\Delta m=0$ and $n=1$ or $4$ as appropriate, confirming that SDSS
and SIGMA agree for exponential and de Vaucouleurs type galaxies.
As highlighted previously, the peak Sérsic index for the second sub-population
does not lie at $n=4$ but rather at $n\sim3.5$.

\section{Variations in Structural Parameters with Wavelength}

\label{sec:wavelength}

\subsection{Magnitude Comparisons}

\label{sub:magcomp}The observed nature and form of a galaxy varies
dependent upon the wavelength at which the observation is taken. These
variations reflect physical mechanisms occurring within the galaxy,
including but not limited to; dust attenuation and; intrinsic gradients
in stellar population, age and/or metallicity (e.g., \citealp{Block1999}).
Below, we show the variance in recovered Sérsic parameters with wavelength,
and discuss how this behaviour is characterised.

\begin{figure}
\includegraphics[width=1\columnwidth]{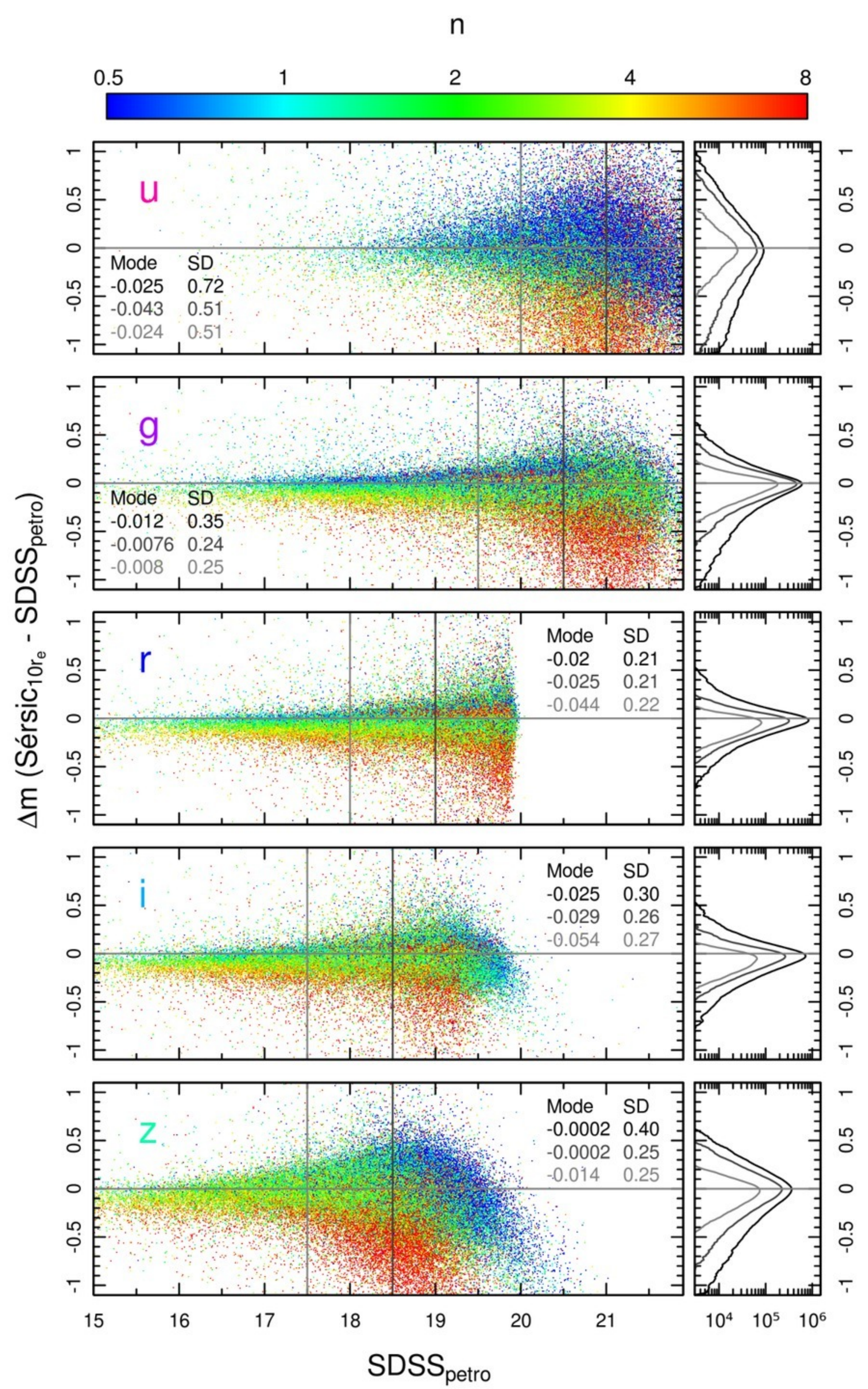}

\caption{\label{fig:trumpetssdss}Comparison between Sérsic magnitudes truncated
at $10r_{e}$ and SDSS Petrosian magnitudes for the five SDSS bandpasses
as a function of SDSS Petrosian magnitude, with the data points coloured
according to their Sérsic index in that band (left column). Vertical
lines define subsets at magnitudes brighter than those values, with
corresponding statistics for mode and standard-deviation inset into
the figure. Density plots (right column) show the relative density
of objects in $\Delta m$-space for each of the aforementioned subsets.}

\end{figure}

Figure \ref{fig:trumpetssdss} compares SDSS $ugriz$ Petrosian photometry
against truncated Sérsic magnitudes as recommended in Section \ref{sub:sersmags}
as a function of SDSS magnitude. Each row represents a different band,
with the mode and standard-deviation for varying magnitude subsets
inset into the left-hand column sub-plots. Across each band we see
a good global agreement between SDSS and recovered Sérsic photometry
at all magnitudes, with the variance between the two photometric methods
increasing towards fainter magnitudes as expected. The global \emph{total}
spread is larger in the lower quality $u$ band than in the higher
quality $r$ band, ranging from $\sigma_{u}=0.72$ magnitudes in the
former and $\sigma_{r}=0.21$ magnitudes in the latter. This trend
should not be surprising, as lower quality data presents a unique
challenge in recovering `correct' structural parameters, with larger
errors expected between different photometric systems for fainter
galaxies. In all cases, the peak modal values are typically less than
$\Delta m=0.03$ magnitudes, re-enforcing the notion of good photometric
agreement between these two different methods. This is also in good
agreement with the offsets previously laid out in \citet{Hill2011}.
The data points are coloured according to the recovered Sérsic index,
and highlights the importance of Sérsic modelling in recovering accurate
structural parameters. At all wavelengths, the largest offset between
SDSS Petrosian and Sérsic magnitudes is observed in those well-resolved
galaxies whose Sérsic indices are large ($n>4$; red data points).
These high-index systems are typified by being highly centrally concentrated
with large extended wings, with the flux in the wings of these galaxies
most likely to be missed by traditional photometric methods such as
Petrosian aperture photometry. By extrapolating the fitted light profile,
Sérsic photometry is able to recover this missing flux and provide
a more accurate measure of `total' magnitude. Unresolved compact high-index
galaxies agree well with the Petrosian aperture method. As an example
of flux recovery, Sérsic photometry for a relatively bright ($r=18$)
high-index%
\footnote{We note here that by high-index we're referring to $n\sim8$ objects,
however, \citet{Caon1993} show that for a deep dataset and employing
good sky-subtraction methods it is possible to find galaxies whose
central concentrations are of the order $n\sim15$.%
} galaxy observed in the $r$ band may recover as much as $\Delta r=0.5$
magnitudes. This missing flux recovery may be seen more clearly in
the turn-off shown in Figure \ref{fig:sersicturnoff} (middle panel),
with the scale of the Petrosian correction a direct function of Sérsic
index. In contrast, intermediate and low Sérsic index galaxies ($n<4$;
blue and green data points) agree much more closely with SDSS Petrosian
photometry. It is interesting to note that high-index galaxies are
also those whose Sérsic magnitudes have been truncated by the largest
amount, and so one must take into consideration the arguments laid
out in Section \ref{sub:sersmags} when analysing these systems.

\begin{figure}
\includegraphics[width=1\columnwidth]{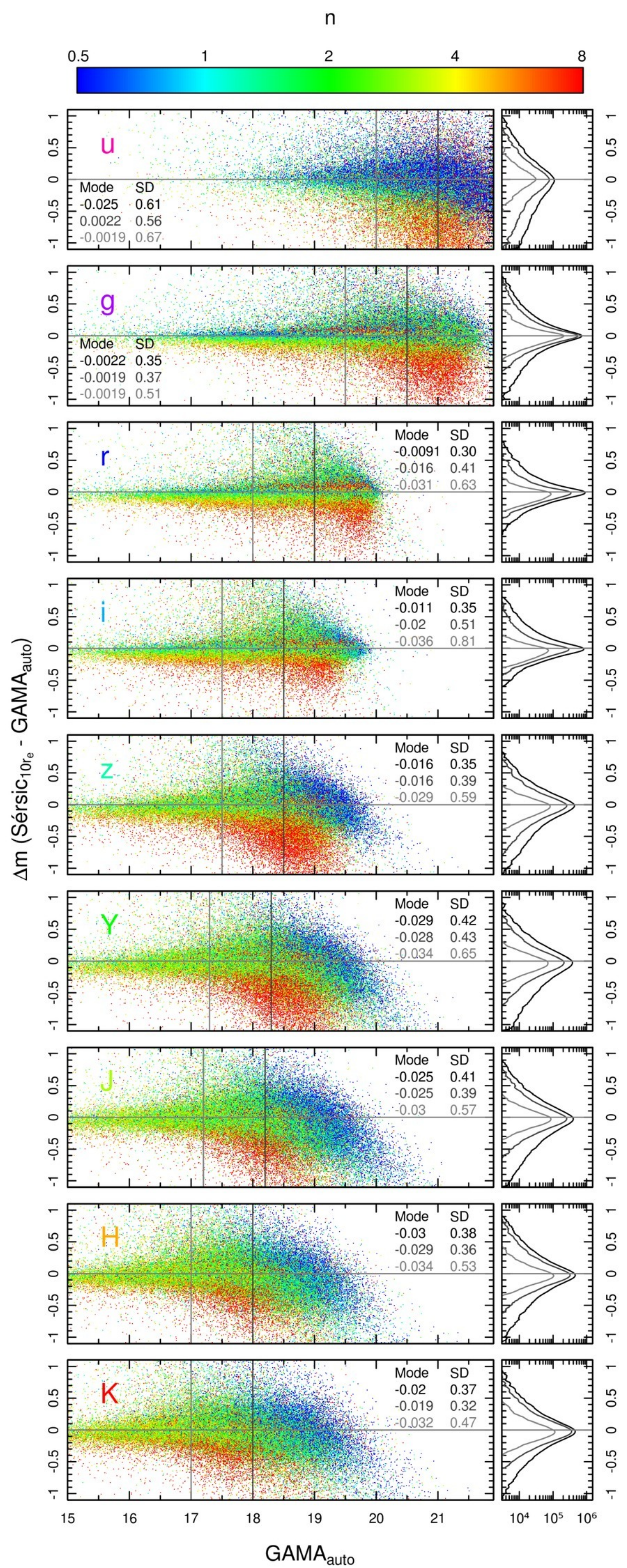}

\caption{\label{fig:trumpets}Comparison between Sérsic magnitudes truncated
at $10r_{e}$ and GAMA AUTO magnitudes for all wavelengths as a function
of GAMA AUTO magnitude, with the data points coloured according to
their Sérsic index in that band (left column). Vertical lines define
subsets at magnitudes brighter than those values, with corresponding
statistics for mode and standard-deviation inset into the figure.
Density plots (right column) show the relative density of objects
in $\Delta m$-space for each of the aforementioned subsets.}
\end{figure}

Figure \ref{fig:trumpets} shows a comparison between GAMA AUTO and
Sérsic magnitudes as a function of GAMA AUTO magnitude. It should
be noted that the GAMA AUTO photometry presented here is version $2$
data and different to the version $1$ data presented in \citet{Hill2011}.
Briefly, version $2$ photometry employs an updated source detection
pipeline over a larger area, with a small fraction of version $1$
input frames discarded due to erroneous data in that region (e.g.,
badly focussed frames). The process used in deriving these renewed
data is similar in approach to that previously employed, a full description
of which may be found in Liske et al. (2011; in prep.). As in Figure
\ref{fig:trumpetssdss}, there is good agreement between the two photometric
systems, with the larger magnitude offsets observed in the resolved
high-index systems. 

\subsection{Two Distinct Populations}

\label{sub:bimodal}
\begin{figure}
\includegraphics[width=1\columnwidth]{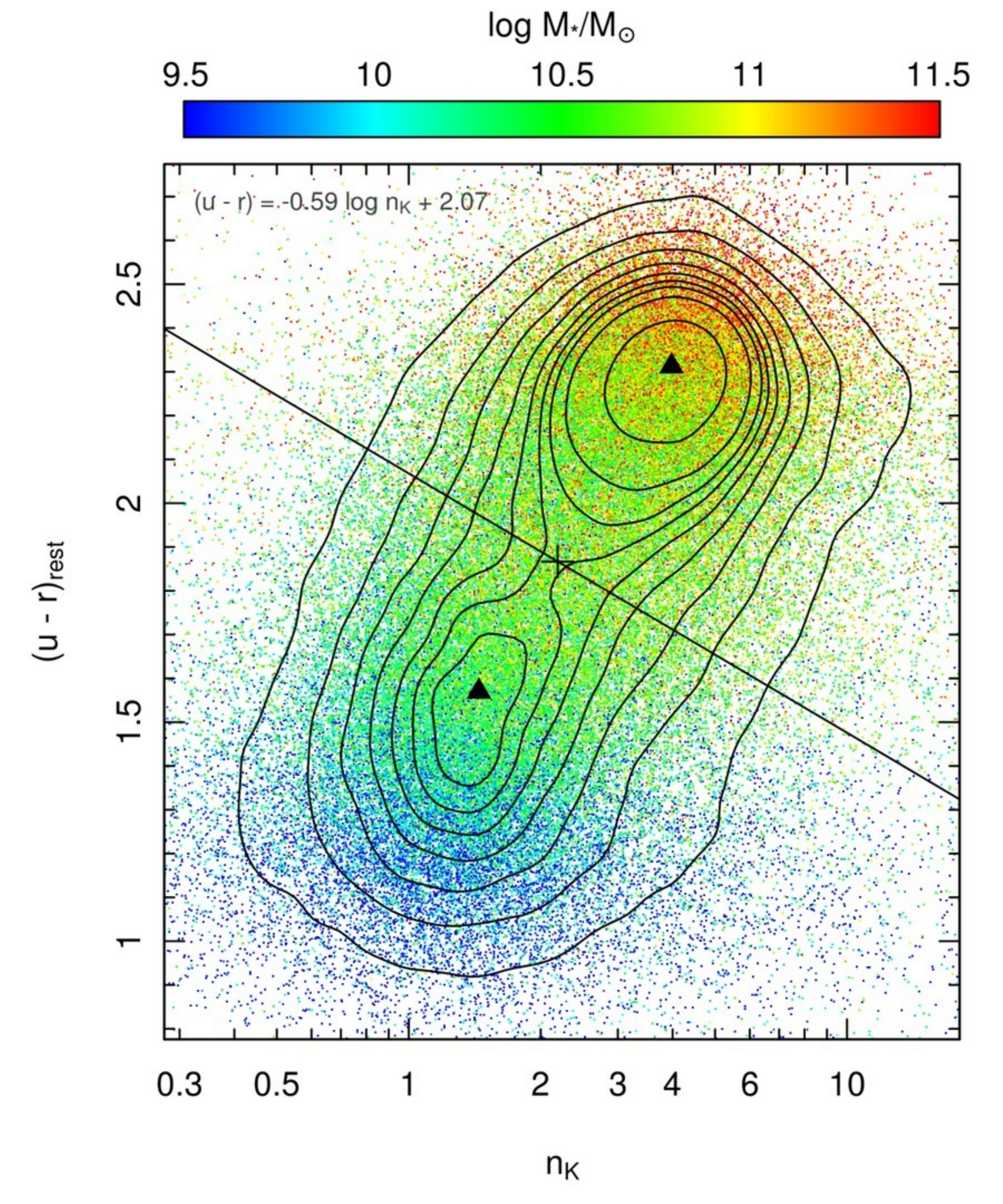}

\caption{\label{fig:colourindex}$K$ band Sérsic index versus $u-r$ rest
frame colour, with the data points coloured according to their galaxy
stellar mass estimates, as shown. Contours range from the $10^{th}$
to the $90^{th}$ percentile in steps of $10\%$. The two highest
peaks in object density, corresponding to two distinct galaxy populations
(late-type and early-type for low-index and high-index respectively),
are represented by filled black triangles. The diagonal line lies
perpendicular to the line connecting these two peaks, and bisects
it at the point of lowest object density along the connecting line,
marked on the figure with a plus sign. This dividing line defines
two sub-samples, which for the most massive galaxies, relate to disk-dominated
systems below the line (LTGs) and spheroid-dominated systems above
(ETGs), the equation of which is inset into the Figure.}

\end{figure}

Figures \ref{fig:trumpetssdss} and \ref{fig:trumpets}
show that Sérsic index plays an important role when considering magnitudes,
with higher index galaxies typically recovering more missing flux
than their lower index counterparts when compared to traditional photometric
methods. In addition to this, there appears to be a wavelength dependence
on the flux difference between high and low index galaxies, which
has implications for variations in other structural parameters with
wavelength. In order to further analyse this wavelength relationship
with structural parameters, we define two galaxy sub-populations based
on Sérsic index and $u-r$ rest frame colour (AUTO aperture defined,
as found in the GAMA catalogue \emph{StellarMassesv03} as described
in \citealp{Taylor2011}). Figure \ref{fig:colourindex} shows this
relation for the $K$-band Sérsic index, with the data points coloured
according to stellar mass. The bulk of the galaxies appear to lie
in two distinct populations, the nature of which have most recently
been explored in \citet{Baldry2006,Driver2006,Allen2006,Cameron2009a,Cameron2009b,Mendez2011},
amongst others, and are typically well described by two overlapping
Gaussians. Blue low-index systems correspond to late-type disk-dominated
galaxies and red high-index systems to early-type spheroid-dominated
galaxies. This is well supported by galaxy stellar mass, with the
least massive galaxies appearing disk-dominated, and the most massive
appearing spheroid-dominated, as expected. The faintest types of galaxy,
namely dwarf systems (dE, dS0), are not represented in our common
coverage sample, and so these two peaks do not relate to those morphological
classes. We used the positions of peak object density for each sub-population
to define a dividing line between them, specifically, the line which
lies perpendicular to one connecting the two peaks in object density,
bisecting it at the point of lowest object density along the connecting
line. The equation of the dividing line is given by:
\begin{equation}
(u-r)_{rest}=-0.59\log n_{K}+2.07\label{eq:indexcolour}
\end{equation}

In order to avoid any potential misclassifications due to the effects
of dust attenuation, our longest wavelength $K$-band data was chosen
as a measure of central concentration. Sérsic indices recovered at
shorter wavelengths return a steeper dividing line, with the gradient
only becoming stable at wavelengths longer than the $z$ band. This
effect is characteristic of the effects of dust, and shall be explored
in more depth in Section \ref{sub:wavelength}. Interestingly, \citet{Mendez2011}
show that the choice of bands used to quantify colour is less important,
and so a standard $u-r$ colour definition is employed in Equation
\ref{eq:indexcolour} for comparison with much of the current literature.
Objects bluer than this dividing line will be referred to as disk-dominated
late-type galaxies (LTGs), whereas objects redder than this line will
be referred to as spheroid-dominated early-type galaxies (ETGs) throughout
the remainder of this paper. It is well known that two 2D Gaussians
are able to aptly describe these two populations. It follows therefore
that a harsh cut of this nature will no doubt introduce a small fraction
of cross-contaminants for galaxies occupying a parameter space in
close proximity to this dividing boundary, namely, those galaxies
that lie in the wings of the opposing Gaussian function. The amount
of contamination will be small however, with the overall trends entirely
sufficient for analysing global trends within each sub-population.
Improvements to the nature of automatic morphological classification
based on global structural measurements exhibited in this paper will
be the focus of future studies presented in Kelvin et al. (2011; in
prep.).

\subsection{Variations with Wavelength}

\label{sub:wavelength}The key galaxy measurements produced by SIGMA,
in addition to improved object centring accuracy are: position angle;
ellipticity; Sérsic magnitude; Sérsic index; and half-light radius.
Understanding how each of these parameters varies with wavelength
is crucial to remove biases when comparing measurements made in different
bandpasses. Wavelength bias may also represent real physical bias
caused by dust attenuation and stellar population gradients.

Recovered position angle should show little variance with wavelength,
instead varying mainly as a function of input data quality. In line
with the cosmological principle, recovered position angle should merely
be a random quantity assuming no detector bias, although for small
area samples it may be coupled with filamentary structure. On a per-galaxy
basis, one might expect minor variations with wavelength to occur
in the presence of stellar population gradients in transient local
features such as star-forming regions and bars, with different bands
being more sensitive to different stellar populations that trace distinct
structural components. However, for the SIGMA common sample of $138,269$
galaxies one does indeed find no noticeable trend with wavelength
for recovered position angle. 

Recovered ellipticity remains relatively stable at all wavelengths,
instead varying primarily as a function of the quality of the input
data, as shown in Figure \ref{fig:fullresults}. The highest quality
$r$ and $i$ bands typically return the most circular galaxy models,
whereas the lowest quality $u$ and $z$ bands return more elongated
profiles across the same galaxy sample. This is as expected as one
reduces the signal-to-noise of the data, with the fitting routine
becoming increasingly sensitive to nearby background noise, however,
further studies and deeper data are required in order to comment further
on this effect.

Finally, recovered Sérsic magnitude is expected to vary as a function
of wavelength as per each galaxy's individual SED, the theory of which
is well understood and will not be discussed further. This leaves
the apparent central concentration (Sérsic index; $n$) and size (half-light
radius; $r_{e}$) of each galaxy as a function of wavelength to be
discussed.

\subsubsection{Sérsic index with wavelength}

\label{sub:wavelengthindex}
\begin{figure*}
\includegraphics[width=1\textwidth]{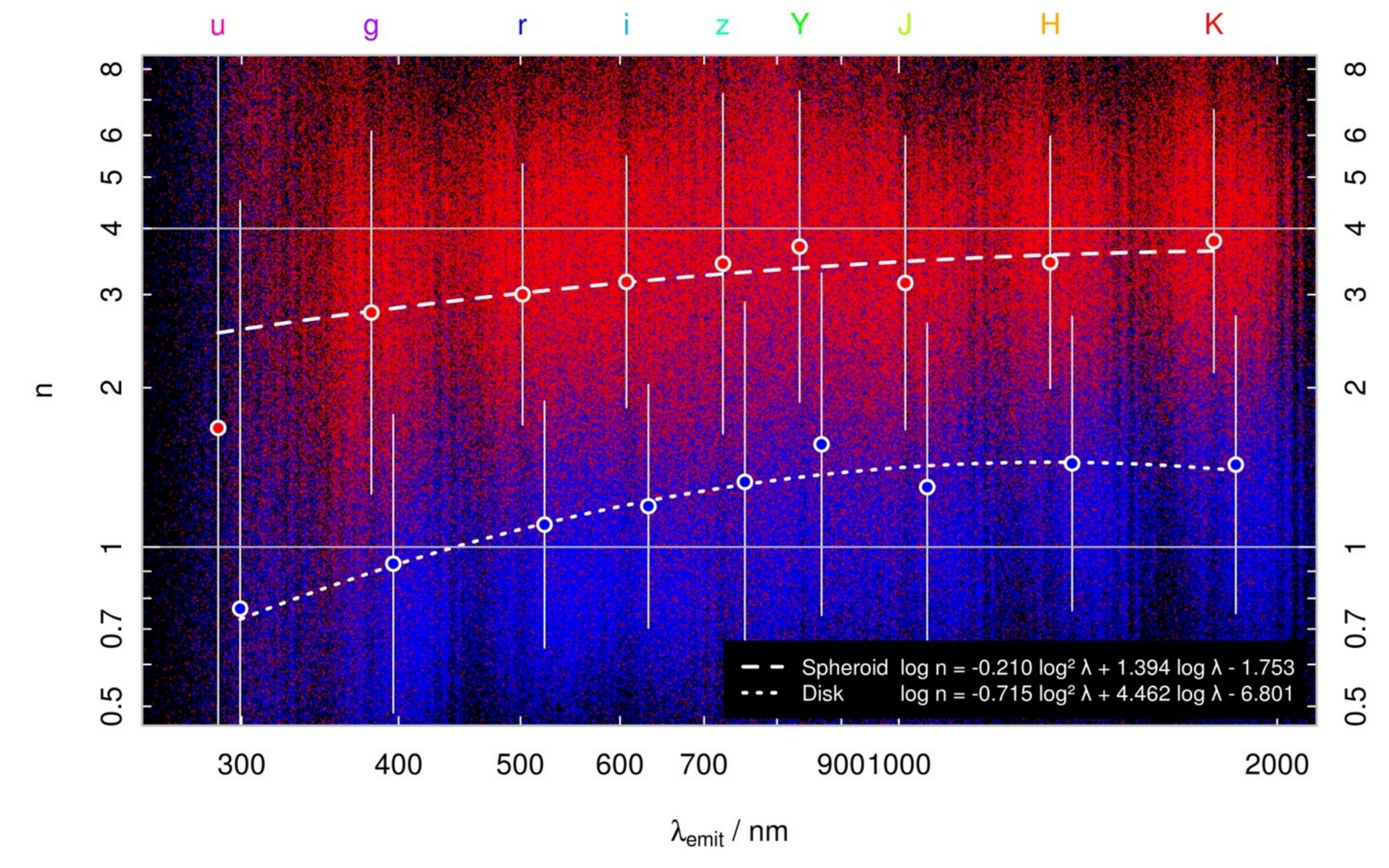}

\caption{\label{fig:indexband}Recovered Sérsic index shown as a function of
rest-frame wavelength in log-log space, coloured according to the
population definitions described in Section \ref{sub:bimodal}. Blue
data-points correspond to disk-dominated galaxies whereas red-data
points correspond to spheroid-dominated galaxies. Large red and blue
circles show the $3\sigma$-clipped mean Sérsic indices for each respective
population in each band, positioned at the median-redshift rest-frame
wavelength for that population. Polynomial fits to these mean Sérsic
indices are shown for both populations, the equations of which are
inset into the figure. Owing to its lower quality imaging data, we
exclude the $u$-band data in the calculation of these lines. Vertical
lines show the $1\sigma$ spread in the data. For reference, grey
horizontal lines at $n=1$ and $n=4$, equating to exponential and
de Vaucouleurs profiles respectively, are added.}

\end{figure*}

Figure \ref{fig:indexband} shows the
recovered Sérsic indices for each galaxy in the SIGMA matched coverage
sample at their rest-frame wavelength, coloured according to their
population classification as described in Section \ref{sub:bimodal}.
Considering the population definitions are based on $K$ band Sérsic
indices it is reassuring to note that the spheroidal population primarily
retain their high Sérsic index values across all wavelengths, and
similarly for the disk population. This indicates a significant level
of consistency in recovered parameters with wavelength, i.e.; a galaxy
that appears disk-like in the $g$ band is likely to appear disk-like
in the $H$ band, for example. $3\sigma$ clipped mean Sérsic indices
are shown for each population in each band, represented by large filled
circles coloured as appropriate. Polynomial fits to these mean data
points, excluding $u$ band values due to their lower quality imaging
data, reveal general trends in Sérsic index with wavelength. The best
fit Sérsic index for the disk-dominated population is given by:
\begin{equation}
\log n_{\mathrm{disk}}=-0.715\log^{2}\lambda_{\mathrm{rest}}+4.462\log\lambda_{\mathrm{rest}}-6.801\label{eq:diskindex}
\end{equation}
and similarly for the spheroid-dominated population:

\begin{equation}
\log n_{\mathrm{sph}}=-0.210\log^{2}\lambda_{\mathrm{rest}}+1.394\log\lambda_{\mathrm{rest}}-1.753\label{eq:sphindex}
\end{equation}
where $\lambda_{\mathrm{rest}}$ is the rest-frame wavelength of the
observation of the galaxy. It is important to remind the reader to
be mindful of our sample selection when considering these relations.
Note that we have adopted log-quadratic relations for Equations \ref{eq:diskindex}
and \ref{eq:sphindex}. Whilst the spheroid-dominated population may
not appear to require a quadratic fit, the disk-dominated population
most-likely does. For this reason, the functional form of both equations
has been kept the same. The linear relation for the disk-dominated
population is given by: 
\[
\log n_{\mathrm{disk}}=0.267\log\lambda_{\mathrm{rest}}-0.676
\]
and the spheroid-dominated population is given by:
\[
\log n_{\mathrm{sph}}=0.170\log\lambda_{\mathrm{rest}}+0.024
\]
These linear relations are provided for reference only and are not
used in any subsequent calculations, with the log-quadratic forms
instead being the preferred descriptors of the two populations.

We find that the spheroid population Sérsic indices remain relatively
stable at all wavelengths, exhibiting slightly lower Sérsic indices
at shorter wavelengths and becoming essentially stable at wavelengths
longer than the $z$/$Y$ interface. Mean ETG Sérsic indices range
from $n_{g}=2.79$ to $n_{K}=3.63$ from $g$ through to $K$, an
increase of $0.11$ dex, equivalent to $30\%$. This increase is consistent
with the $23\%$ increase reported in \citet{LaBarbera2010a} over
a similar wavelength range. However, whilst the fractional increase
is comparable, the absolute values are not; \citeauthor{LaBarbera2010a}
find on average Sérsic indices $n\sim2-3$ larger than those reported
here. Whilst it is unclear what causes this difference, a potential
difference in sample definitions may be important. That study defines
ETGs based on a number of SDSS parameters including fracDeV$_{\mathrm{r}}$;
a parameter that describes how well the global light profile of the
galaxy is fit by a de Vaucouleurs profile. A cut of this nature is
somewhat analogous to making a Sérsic index cut alone which, as can
be seen in Figure \ref{fig:sersicturnoff}, and again in Figure \ref{fig:colourindex},
would introduce an element of contamination from the LTG population.
If a relatively large number of the ETG sample in \citeauthor{LaBarbera2010a}
are in fact bulge-dominated systems with a weak underlying disk then
one might expect the Sérsic indices of their bulges to differ somewhat
from a traditional de Vaucouleurs profile. A Sérsic index of $n\sim6$
is the value found in \citet{Simard2011} for bulge+disk systems with
a well-defined bulge, in good agreement with the offset found here.

The apparent stability found in Sérsic index with wavelength is as
expected for relatively dust-free single-component early-type systems,
and is interesting to re-confirm empirically. Since the spheroid-dominated
population is likely to include a small fraction of misclassified
galaxies, as previously discussed, due to the nature of the harsh
cut presented in Section \ref{sub:bimodal}, a small gradient with
wavelength should not be surprising. Recent work by Rowlands et al.
(2011; submitted) suggest that as many as $5.5\%$ of early-type galaxies
contain significant fractions of previously unaccounted-for dust,
introducing an additional secondary deviation in recovered Sérsic
indices with wavelength. Dust in a galaxy is typically centrally concentrated,
and so has the effect of masking stellar light emanating from the
core regions of a galaxy. Since galaxy fitting algorithms such as
GALFIT apply larger weighting to higher signal-to-noise regions, minor
deviations at small radii have the potential to drastically affect
the recovered structural parameters, including the Sérsic index. Therefore,
the addition of dust to the core region of a galaxy would subdue the
\emph{cuspiness} of the galaxy and bias the model towards a lower
Sérsic index. 

The disk population exhibits a larger change in Sérsic index variation
with wavelength than that observed for the spheroid population. The
recovered mean disk Sérsic indices range from $n_{g}=0.92$ to $n_{K}=1.40$
from $g$ through to $K$, an increase of $0.18$ dex, equivalent
to $52\%$. As with the spheroid population, disk Sérsic indices also
become increasingly stable at wavelengths longer than the $z$/$Y$
interface. Since we typically expect disk-dominated systems to be
dustier than their early-type counterparts, owing to the prevalence
of ongoing star-formation in many of these galaxies, then a significant
variation in Sérsic index with wavelength should be expected as a
consequence of the arguments previously laid out. Since the disk Sérsic
index appears stable beyond the $z$/$Y$ interface, we can conclude
that the effect of dust in these regimes is minimal, and therefore
if `intrinsic' disk Sérsic indices are required, one should look to
the longest wavelength data available, typically longwards of the
$z$ band. In addition to the effects of dust attenuation, we may
also consider stellar population gradients. Since this sample is not
a pure-disk sample, and instead contains a host of disk-dominated
systems, a fraction of galaxies in the disk-dominated population will
no doubt contain additional structures such as a bulge and/or a bar.
Bulges tend to contain older, redder stars of a higher metallicity
than the younger, bluer stars found in the disks of galaxies. Shorter
wavelengths are more sensitive to the blue population found in the
disk whereas longer wavelengths become increasingly sensitive to the
red population. Therefore, any real colour gradients that exist in
a galaxy, which are indicative of metallicity and age gradients in
the underlying stellar population distribution, would also lead to
a change in the measured Sérsic index, dependent upon the wavelength
at which that galaxy was observed. A short wavelength is therefore
more likely to probe the disk stellar population than a longer wavelength.
It is unclear whether the effect of dust attenuation or stellar population
gradients are the dominant factor in determining the variation in
Sérsic index with wavelength, with a combination of both likely to
contribute globally.

We note that the disk-dominated and spheroid-dominated populations,
once stabilised, tend towards $n_{\mathrm{disk}}\rightarrow1.4$ and
$n_{\mathrm{sph}}\rightarrow3.6$ respectively. These values differ
from the Sérsic indices typically used to describe late and early-type
systems (excluding dwarf galaxies, for which there is a Magnitude-Sérsic
index relation), namely $n_{\mathrm{late}}=1$ and $n_{\mathrm{early}}=4$
respectively (represented in Figure \ref{fig:indexband} by horizontal
grey lines). This may indicate morphological contamination between
populations as previously discussed, with some galaxies exhibiting
bulge-to-disk ratios away from values of either zero or unity.

\subsubsection{Half-light radius with wavelength}

\label{sub:wavelengthradius}
\begin{figure*}
\includegraphics[width=1\textwidth]{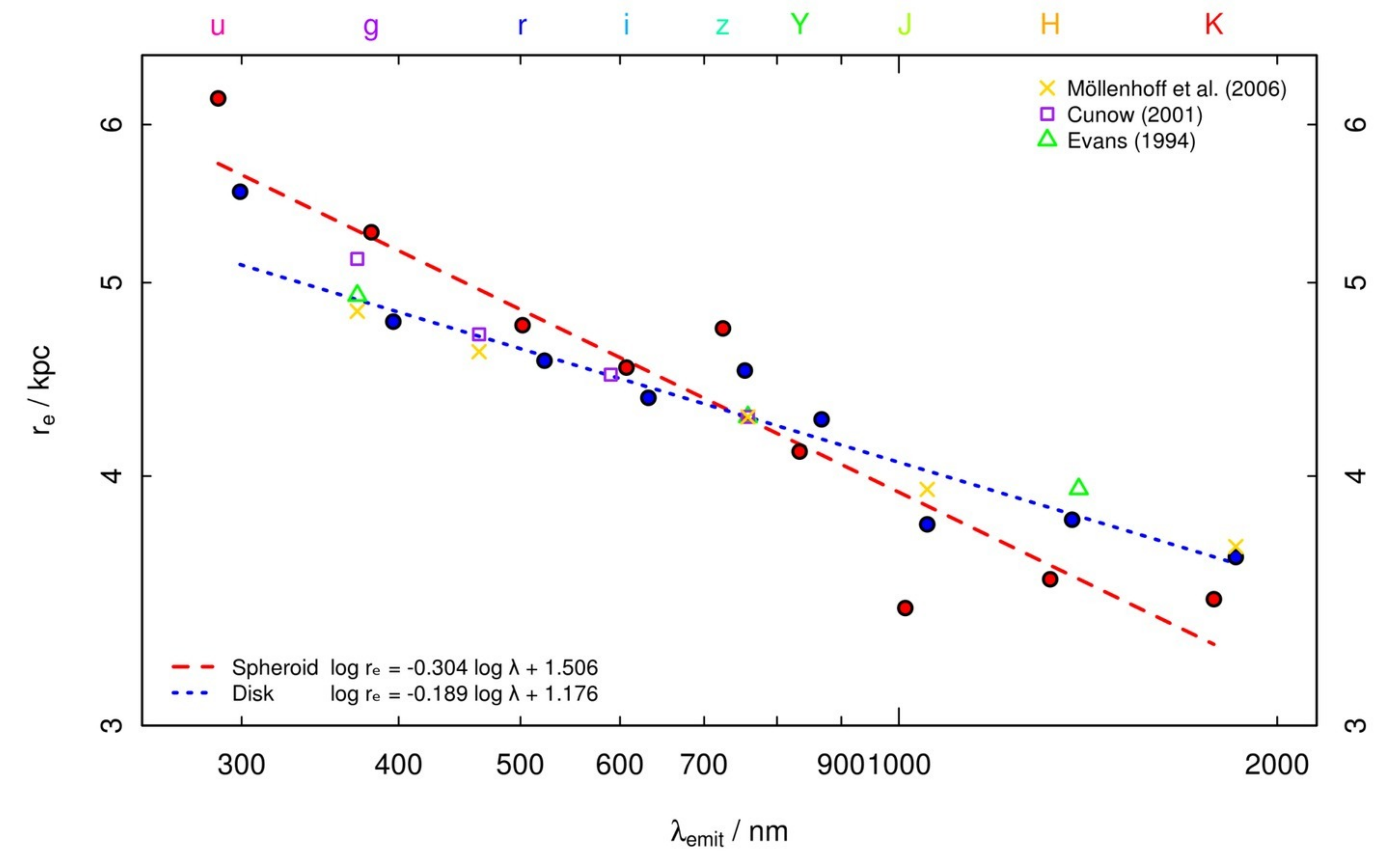}

\caption{\label{fig:reband}Recovered half-light radii in kpc as a function
of rest-frame wavelength. Red and blue circles show the $3\sigma$-clipped
mean half-light radii for spheroid-dominated and disk-dominated galaxies
respectively in each band, positioned at the median-redshift rest-frame
wavelength for that population. Linear fits to these mean half-light
radii are shown for both populations, the equations of which are inset
into the figure. Owing to its lower quality imaging data, we exclude
the $u$-band data in the calculation of these lines. Overlaid are
data from several authors who predict an increase in the measured
half-light radii in late-type systems due to the effects of dust.
Further details are available in the text.}

\end{figure*}

Figure \ref{fig:reband} displays the
recovered half-light radii as a function of their rest-frame wavelength.
$3\sigma$ clipped means are represented by solid red and blue circles
for the disk-dominated and spheroid-dominated systems respectively,
with linear fits to these data shown. The best fit linear relation
describing the half-light radii in physical units (kpc) for the disk-dominated
population is given by:
\begin{equation}
\log r_{\mathrm{e,disk}}=-0.189\log\lambda_{\mathrm{rest}}+1.176\label{eq:diskre}
\end{equation}
and for spheroid-dominated systems:

\begin{equation}
\log r_{\mathrm{e,sph}}=-0.304\log\lambda_{\mathrm{rest}}+1.506\label{eq:sphre}
\end{equation}
where $\lambda_{\mathrm{rest}}$ is the observed rest-frame wavelength
for the galaxy. Again, it is important to remind the reader to be
mindful of our sample selection when considering these relations.

Using these relations, we observe significant variation in the recovered
sizes of galaxies as a function of wavelength. The disk population
mean half-light radii range from $r_{e,g}=4.84$ kpc to $r_{e,K}=3.62$
kpc from $g$ through to $K$, a decrease in size of $0.13$ dex,
equivalent to a drop of $25\%$. The spheroid population exhibits
a larger spread from $r_{e,g}=5.27$ kpc to $r_{e,K}=3.29$ kpc over
the same wavelength range, a decrease in size of $0.20$ dex, equivalent
to a drop of $38\%$. This variation in the spheroid population size
is in good agreement with studies by \citet{LaBarbera2010a,Ko2005},
reporting decreases of $29\%$ and $39\%$%
\footnote{A linear extrapolation of the trends reported in \citet{Ko2005} were
applied in order to convert V-K offsets into the g-K wavelength range
used here.%
} respectively over a similar wavelength range. Explanations for the
variation in recovered size with wavelength include dust attenuation
at shorter wavelengths or metallicity gradients within the galaxy.
The effects of dust on a galaxy light profile have previously been
discussed. Obscuring the central region of a galaxy would shift the
balance of total flux towards larger radii, artificially increasing
the half-light radius. This effect is well understood for late-type
systems, and indeed has been predicted in the literature notably by
\citet{Evans1994,Cunow2001}, and more recently by \citet{Mollenhoff2006,Graham2008b}.
Data from these studies are added into Figure \ref{fig:reband} for
reference (as indicated), normalised to the disk half-light radius
predicted at $\lambda_{\mathrm{obs}}=900$ nm. Where possible, we
adopt or infer a dust face-on optical depth of $\tau_{\mathrm{B}}^{f}=3.8$
(central face-on $B$-band opacity)  as recommended in \citet{Driver2007}.
This value is close to that found in detailed modelling of nearby
galaxies by \citet{Popescu2000,Misiriotis2001,Popescu2011}; Hermelo
et al. (in prep.). We also adopt an average inclination galaxy of
$\cos i=0.5$. \citeauthor{Evans1994} analyses face-on galaxies alone,
whereas \citeauthor{Cunow2001} only produces detailed models for
$\tau_{\mathrm{B}}^{f}=3.0$ systems, and so care should be taken
when comparing these data. In all cases we see an excellent agreement
between the observed size-wavelength variation and that predicted
by these dust model simulations, notably so when compared with the
work of \citeauthor{Mollenhoff2006}, as they employ dust models that
can account for both dust attenuation and emission. As a caveat, we
note that the population shown in Figures \ref{fig:indexband} and
\ref{fig:reband} does not constitute a volume-limited sample, and
so conclusions must remain tentative until further studies can confirm
this relation. However, our initial interpretation is that dust models
more than adequately account for the apparent size-wavelength relation
in late-type disk-dominated galaxies.

It is interesting to note that whilst the spheroid population shows
less variation in central concentration, i.e., Sérsic index, than
the disk population, it exhibits a larger size variation with wavelength.
Dust is not expected to be a dominant factor in the attenuation of
light within these systems (although see the earlier discussion regards
early-type dust fractions). However, higher optical depth values than
that recommended in \citet{Driver2007} would have the effect of skewing
the gradient of the dust attenuated size-wavelength relation to match
the observed distribution. In addition to the possibility of age/metallicity
gradients within spheroids, an alternative explanation for the apparent
size variation with wavelength in early-type systems relies upon the
interdependency between recovered Sérsic index and half-light radius.
A change in the Sérsic index arising due to, e.g. small core dust
components, additional unresolved or disturbed structure in the core
arising due to recent environmental interactions, the influence of
an AGN or uncertainty in the PSF may lead to an equivalent corrective
change in half-light radius. \citet{MacArthur2003} show that uncertainty
on the PSF full-width-half-maximim by as much as $\Gamma=1.5''\pm0.5$,
a range encompassing most of the optical data as shown in Figure \ref{fig:seeing},
would yield an equivalent measured size variation of $25\%$ for the
worst affected compact systems. One concern might be that small numerical
uncertainty in Sérsic index would yield artificial changes in recovered
size. Using the Sérsic index ranges for both disk and spheroid populations
shown in Figure \ref{fig:indexband}, and fixing the effective surface
brightness and total magnitude in the total luminosity variant of
Equation 2 in \citet{Graham2007}, we would expect to see equivalent
changes in Sérsic half-light radii of $\Delta r_{e,\mathrm{disk}}=17\%$
and $\Delta r_{e,\mathrm{sph}}=8\%$ for the disk and spheroid populations
respectively, i.e., far less than that seen here.

\subsubsection{Co-variation of Sérsic index and half-light radius}

\label{sub:wavelengthcovar}We have shown how the Sérsic index and
half-light radius for the spheroid and disk populations vary with
wavelength, however, one must not consider these variations in isolation.
All of the output model parameters have a combined effect on the final
light profile of a galaxy. Several of these parameters including:
Sérsic index; half-light radius; total magnitude; and the background
sky display certain levels of inter-dependence (e.g., \citealp{Caon1993,Graham1996}).
For example, an over-estimation of the sky level would lead to an
under-estimation in the total magnitude of that galaxy, and consequently
Sérsic index and half-light radius also. In the case of sky however,
the signal-to-noise weighting employed by GALFIT somewhat ensures
against sky offsets of a few counts from the true value noticeably
adversely affecting the fit. Since we would not expect the systematic
error in the background sky to show significant trends with wavelength,
and the variation in total magnitude with wavelength is well understood,
we exclude them from our investigation into the co-variation of structural
parameters with wavelength. We now consider the combined effect of
varying the Sérsic index and half-light radius in unison, and how
this impacts on the overall light profile of a test galaxy from $u\rightarrow K$.

\begin{figure}
\includegraphics[width=1\columnwidth]{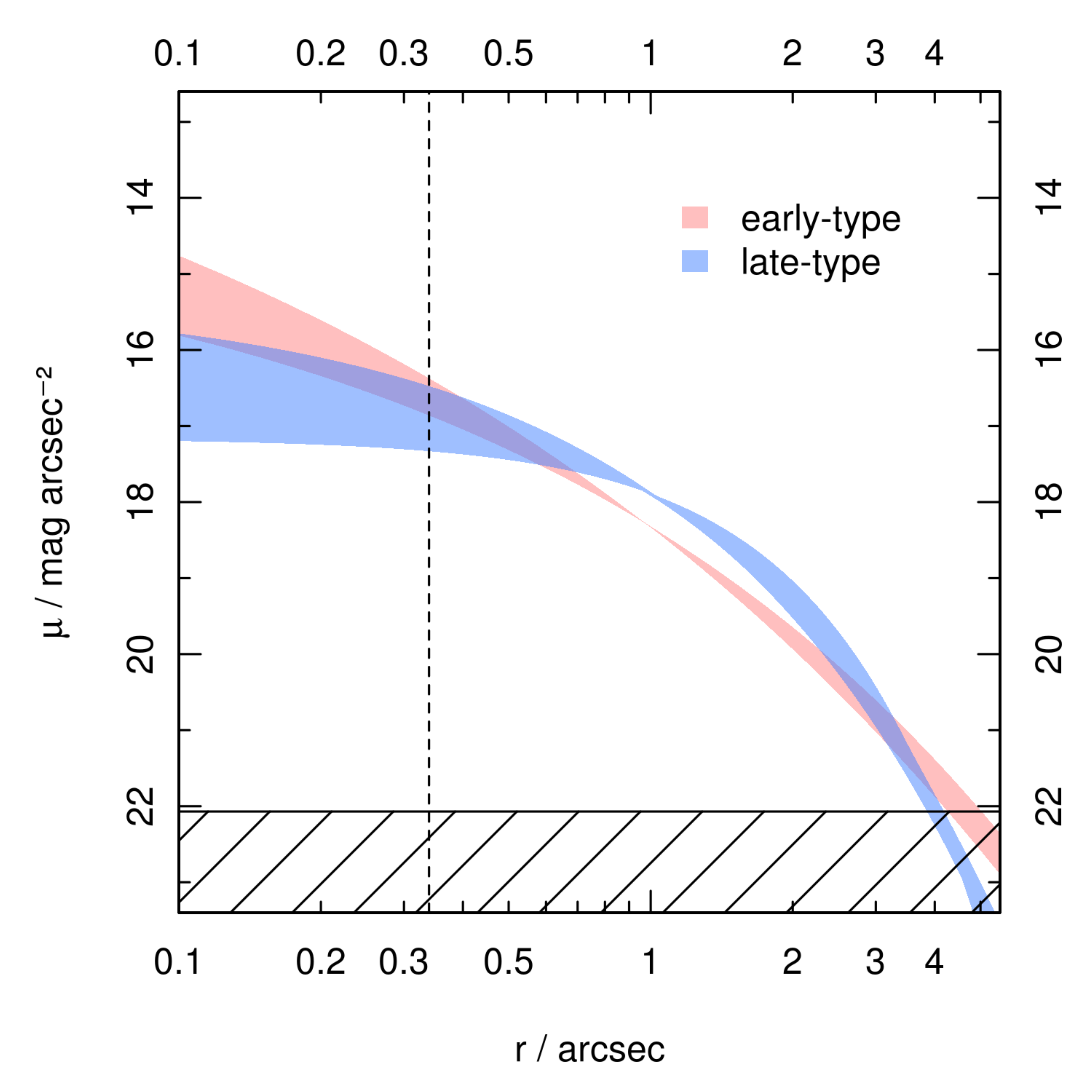}

\caption{\label{fig:trends}Surface brightness variation from $u$ through
to $K$ for the early-type spheroid-dominated and late-type disk-dominated
populations. We generate Sérsic indices and half-light radii in wavelength
steps from $u\rightarrow K$ for each population using the trends
as described in equations \ref{eq:diskindex}, \ref{eq:sphindex},
\ref{eq:diskre} and \ref{eq:sphre}. Using these values, surface
brightness profiles (without PSF convolution) may be constructed for
each wavelength bin. The shaded regions shown here represent the maximal
area swept out by these light profiles along the transition from $u$
through to $K$, and represent how much of an effect the reported
changes in Sérsic index and half-light radius have on the overall
light profiles. The hatched region indicates the brightest limit at
which light profiles may be trusted ($\mu_{\mathrm{lim,}K}=22.07$
mag/arcsec$^{-2}$), and the vertical dashed line represents $1$
pixel in distance from the centre. Profiles are produced assuming
a constant total magnitude of $m_{\mathrm{tot}}=15$ for both the
spheroid and disk populations.}

\end{figure}

Using Equations \ref{eq:diskindex}, \ref{eq:sphindex}, \ref{eq:diskre}
and \ref{eq:sphre} we generate estimates of Sérsic indices and half-light
radii at equal steps in log-wavelength space for both the spheroid
and disk populations. Using the Sérsic relation, and assuming a constant
total magnitude for both spheroids and disks of $m_{\mathrm{tot}}=15$,
we create a series of surface brightness light profiles from $u$
through to $K$. Figure \ref{fig:trends} shows the change in the
recovered surface brightness light profiles over the $u\rightarrow K$
wavelength range, with the shaded areas representing the maximal area
swept out by these light profiles as they vary in wavelength. This
gives us an indication of how changes in recovered structural parameters
affect the underlying surface brightness profile. The hatched region
represents the worst-case limit at which these light profiles may
be accepted as containing significant signal above the background
sky level, as given in Section \ref{sub:sblims}. The vertical dashed
line represents a $1$ pixel distance from the centre.

Despite the relatively large size variation observed in the spheroid
population (a decrease of $38\%$ in $g\rightarrow K$), when considered
in conjunction with the Sérsic index variation (an increase of $30\%$
in $g\rightarrow K$) the combined effect amounts to a relatively
modest impact on the majority of the recovered light profile. It appears
that as the spheroidal size decreases the Sérsic index increases at
a comparative rate. The most noticeable surface brightness variation
is found in the central core region, fluctuating by $0.49$ magnitudes
at the $1$ pixel boundary. Since a significant fraction of total
flux lies in the core regions of high-index systems, it should not
be surprising that a small variation in Sérsic index would produce
a relatively large variation in half-light radius. Despite this effect,
the majority of surface brightness profile out to large radii remains
relatively stable with wavelength, vastly reducing the need for more
complex mechanisms as previously discussed. 

The variation in size for the disk-dominated population (a decrease
of $25\%$ in $g\rightarrow K$) coupled with a relatively large increase
in Sérsic index (an increase of $52\%$ in $g\rightarrow K$) yields
a similar effect on the surface brightness profile variation as previously
described for the spheroid population. Surface brightness fluctuates
by $\sim0.86$ magnitudes at the $1$ pixel boundary, an increase
of $75\%$ on the variation in the spheroid population. Here it appears
that the impact of dust attenuation has a particularly distinct effect
on the light profile in disk-dominated galaxies, agreeing well with
the theoretical predictions for size variation with dust presented
in Section \ref{sub:wavelengthradius}.

Whilst no single mechanism can be shown to be entirely responsible
for the relations between Sérsic index, half-light radius and wavelength
observed across the two populations, it is clear that the large apparent
size fluctuations in the spheroid population appear to be initially
misleading. Only when considering Sérsic index in conjunction with
half-light radius does the true nature of these effects come to the
fore. The spheroid population, despite exhibiting large changes in
half-light radius with wavelength, maintains a relatively stable surface-brightness
profile from $u$ through to $K$. The variation in the disk population
with wavelength appears well described by current dust models, however,
it is most likely a combination of dust attenuation, stellar population/metallicity
gradients, unresolved secondary features in the core region affecting
profile fits, and uncertainty on additional parameters such as the
PSF that affect the underlying physics in these systems. Future studies
aim to further inform this discussion for a limited sub-sample to
be presented in Kelvin et al. (2011; in prep.).

\section{Conclusion}

\label{sec:conclusion}We have produced high-fidelity automated two-dimensional
single-Sérsic model fits to $167,600$ galaxies selected from the
GAMA input catalogue. These have been modelled independently across
$ugrizYJHK$ using reprocessed SDSS and UKIDSS-LAS imaging data. These
data have subsequently been delivered to the GAMA database in the
form of the catalogue \emph{SersicCatv07}. In order to facilitate
the construction of this dataset, SIGMA, an extensive multi-processor
enabled galaxy modelling pipeline, was developed. SIGMA is a wrapper
and handler of several contemporary astronomy software packages, employing
adaptive background subtraction routines and empirical PSF generation
on a per-galaxy per-band basis to tailor input data into the galaxy
modelling software GALFIT 3. Output results from GALFIT are analysed
for pre-determined modelling errors such as positional migration,
extreme model shape and/or size parameters and adverse nearby neighbour
flux. Nearby object masking is employed as a last resort, with secondary
neighbours being preferentially modelled simultaneously with the primary
galaxy in the first instance. 

Using this dataset, we have defined a common coverage area across
the three GAMA regions that encompasses $138,269$ galaxies, $82.5\%$
of the full sample. This common area contains only those galaxies
which have been observed in all nine bands, providing a useful basis
upon which to further explore wavelength trends. We define a Sérsic
magnitude system that truncates Sérsic magnitudes at $10$ $r_{e}$.
This ensures that flux is not extrapolated below the typical limiting
isophote into regions where data quality and quantity is not sufficient
to constrain the form of the galaxy light profile. Truncated Sérsic
magnitudes appear to be a good descriptor of global galaxy colours
and total galaxy flux. For well-resolved disk-like galaxies ($n<2$),
traditional aperture-based methods are in good agreement with truncated
Sérsic magnitudes. For high centrally-concentrated systems however
($n>4$), it appears that traditional aperture-based, such as Petrosian
magnitudes, may miss as much as $\Delta m_{r}=0.5$ magnitudes from
the total flux budget which is only recovered through Sérsic modelling. 

When considering the dataset in n--colour space we find galaxies
appear to exist in two distinct groups. For the most massive systems,
we associate these two groups with the spheroid-dominated early-type
galaxy (ETG) and disk-dominated late-type galaxy (LTG) populations.
Owing to the nature of our input sample selection, these definitions
do not extend down to the fainter dwarf population, and so subsequent
trends will not represent those systems. We use the longest wavelength
$K$ band Sérsic index measurements in conjunction with rest-frame
$u-r$ colour to define these two populations. Using these definitions,
we are able to further probe the variations in recovered structural
parameters with wavelength for each population. 

We find that the Sérsic indices of ETGs remain reasonably stable at
all wavelengths, increasing by $0.11$ dex ($+30\%$) from $g$ to
$K$ and becoming very stable beyond the $z$/$Y$ interface. In contrast
to this, we find that LTGs exhibit larger variations in Sérsic index
with wavelength, increasing by $0.18$ dex ($+52\%$) across the same
wavelength range. Recovered sizes for both the spheroid and disk systems
show a significant variation with wavelength, showing a reduction
in half-light radii of $0.20$ dex ($-38\%$) in ETGs and $0.13$
dex ($-25\%$) in LTGs from $g$ to $K$. Size variation of this scale
for disk systems has been well predicted by dust models, highlighting
the important role dust attenuation plays when considering structural
variations across a broad wavelength range.

We note that spheroidal systems exhibit a larger size variation with
wavelength than that found in disk systems. Possible physical explanations
for this behaviour include low levels of unresolved dust or the effects
of AGN feedback in the core of the galaxy, both of which would affect
Sérsic profiling. Significant amounts of dust, such as an increased
dust attenuation optical depth parameter $\tau_{\mathrm{B}}^{f}$,
may allow current dust models to accurately describe the variation
in half-light radii we find. It is unlikely however that a significant
fraction of our spheroid-dominated population contain sufficient amounts
of dust for this to be the case. Large stellar population/metallicity
gradients present within individual structures of the galaxy would
cause galaxies to look markedly different in different wavelengths,
contributing to any concentration-wavelength/size-wavelength variation.
In addition to these factors, uncertainties on the measured PSF and
background sky must be considered.

However, when considering variations in half-light radius and Sérsic
index together with wavelength we find that the large fluctuations
in spheroidal parameters amount to a relatively modest impact on the
recovered light profile. A comparatively larger effect is noted for
the disk systems, particularly in the core region, supporting the
presence and effect of dust attenuation in addition to stellar population/metallicity
gradients. At a distance of $1$ pixel from the central region, spheroid
systems display a variation in surface brightness of $0.49$ magnitudes
from $u$ through to $K$. In disk systems, the comparative figure
is $0.86$ magnitudes, an increase of $75\%$. This highlights the
importance of not considering recovered parameters in isolation, as
the interplay between them has the possibility of masking underlying
trends.

The effects of dust attenuation appear to be the dominant factor constraining
the variations in structural parameters with wavelength, notably so
for the disk-dominated population. In contrast with this, apparent
large structural variations in the spheroid-dominated population appear
to have a relatively minor effect on the underlying surface-brightness
profile than might have been expected. Future studies in Kelvin et
al. (2011; in prep.), focussing on a limited sub-sample of this dataset,
will provide a deeper understanding of these structural variations
with wavelength, enabling us to comment further on the key mechanisms
involved in varying structural parameters with wavelength for a host
of different morphologies.

\section*{Acknowledgements}

We thank Emmanuel Bertin and Philippe Delorme for providing an early
release version of the PSF Extractor software, and their many useful
conversations on its correct use. We also thank the referee for their
suggestions and comments which helped improve the paper. LSK thanks
the Science and Technology Facilities Council and The University of
Western Australia for their financial support during the writing of
this article. JL acknowledges support from the Science and Technology
Facilities Council (grant numbers ST/F002858/1 and ST/I000976/1).
GAMA is a joint European-Australasian project based around a spectroscopic
campaign using the Anglo-Australian Telescope. The GAMA input catalogue
is based on data taken from the Sloan Digital Sky Survey and the UKIRT
Infrared Deep Sky Survey. Complementary imaging of the GAMA regions
is being obtained by a number of independent survey programs including
GALEX MIS, VST KIDS, VISTA VIKING, WISE, Herschel-ATLAS, GMRT and
ASKAP providing UV to radio coverage. GAMA is funded by the STFC (UK),
the ARC (Australia), the AAO, and the participating institutions.
The GAMA website is http://www.gama-survey.org/ .

\bibliographystyle{mn2e}
\bibliography{biblib}

\appendix

\section{SIGMA Input Options}

\label{app:startup}On starting SIGMA, a number of input options can
be specified. Some of these are essential in its use, whereas others
are designed for testing purposes only. The available input options
can be found in the help document, reproduced below.

\begin{lstlisting}[caption={SIGMA help lists the available input options that may be specified
when starting SIGMA.},basicstyle={\tiny\ttfamily},tabsize=4]
$ sigma -h

----- SIGMA Version 0.9-0 -- 23 Jul 2010

DESCRIPTION
    SIGMA (Structural Investigation of Galaxies via Model Analysis) is a 2
    dimensional fitting code taking inputs from the GAMA SWarped regions and
    producing models using the GALFIT software.   

OPTIONS

    -a A        - append A to output log files
    -b A        - A-band (default: r)
    -c A        - input catalogue [img/csv] (needs at least RA & DEC)
    -d          - show program defaults
    -e #        - error generation method (1=GALFIT, 2=BOOTSTRAP)
    -h          - help (this screen)
    -i          - interactive mode
    -m          - make a plot of output .fits files (png format)
    -n A        - output catalogue name
    -o          - no headers in output catalogue, only data
    -p #        - number of sub-processes to spawn
    -r #        - number of bootstrap runs to generate errors in GALFIT
    -s # #      - subsample, from lower to upper quantile
    -t #,#      - principle allowed multi-component types (eg: 1,2,5,10)
    -v          - version number
    -x #        - GAMA ID
    -y #        - SIGMA ID
    -z #        - SDSS OBJID

CONTACT     
    Lee Kelvin
    University of St Andrews
    lsk9@st-andrews.ac.uk 
\end{lstlisting}

In this study, we initialised SIGMA using the following command:
\begin{lstlisting}
sigma -b x -n sigmacat_x.csv -p 16 -t 1
\end{lstlisting}
where $x$ represents the band for modelling ($ugrizYJHK$). This
command initialises SIGMA on 16 processors, restricting the fits to
single-component (single-Sérsic) only. Running SIGMA across each band
individually, we produced 9 individual catalogues for later matching
in TOPCAT \citep{Taylor2005}.

\section{Initial Conditions}

\label{app:initialconditions}The galaxy modelling phase has been
discussed extensively in Section \ref{sec:sigma}. Most of the input
parameters fed into GALFIT come directly or trivially from the pipeline
\noun{objectpipe} (Section \ref{sub:objectpipe}), a module wrapper
around Source Extractor. The two main exceptions to this are half-light
radius and Sérsic index.

\begin{figure*}
\includegraphics[width=1\textwidth]{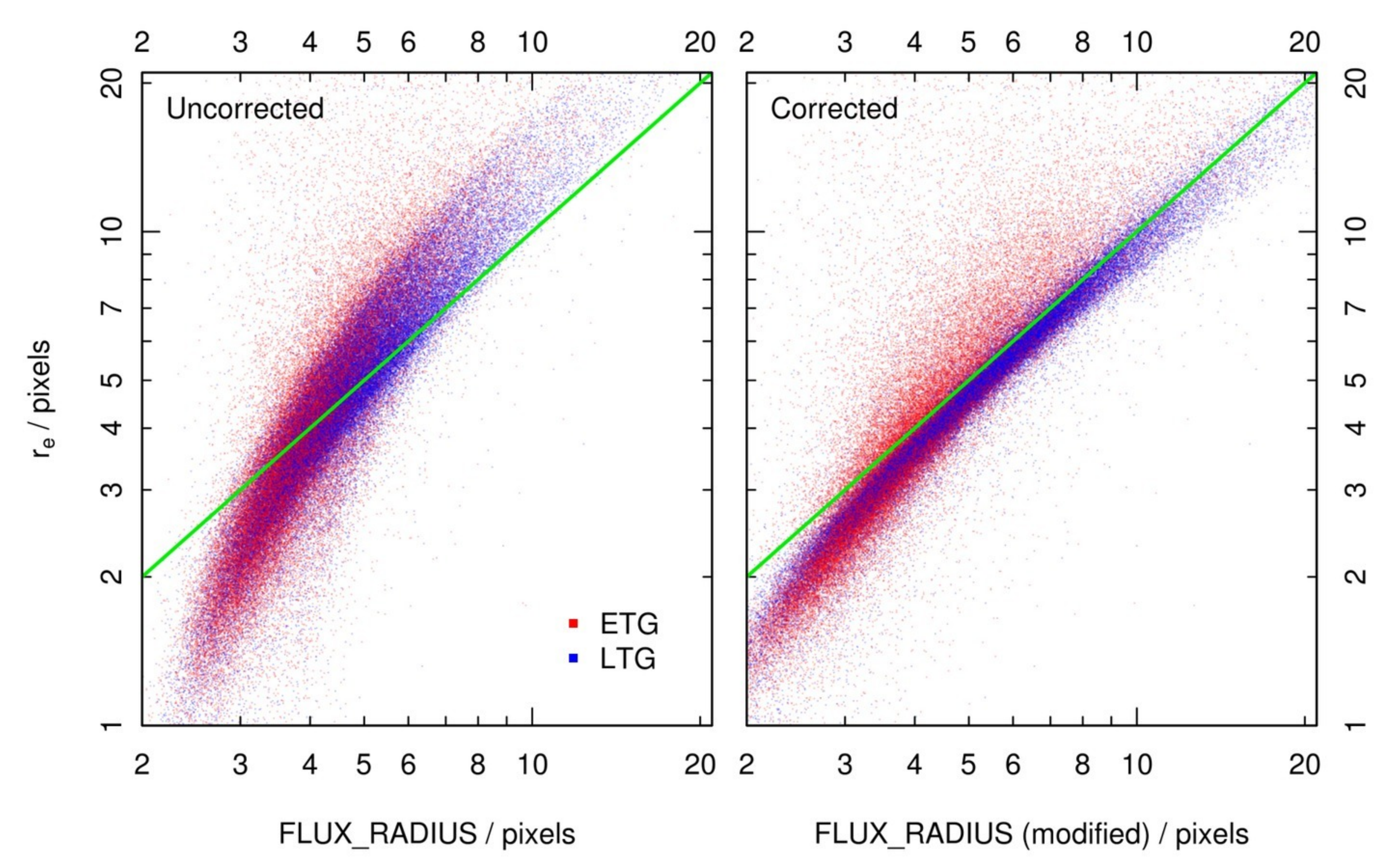}

\caption{\label{fig:radiicomp}A comparison between Source Extractor half-light
radii (FLUX\_RADIUS) and modelled GALFIT half-light radii ($r{}_{\text{e}}$)
in the $r$-band, with data points coloured according to their predicted
morphological type as described in Section \ref{sub:bimodal}. (Left)
Uncorrected half-light radii from Source Extractor are a poor initial
condition for modelling an object in GALFIT, as Source Extractor radii
are circularised and make no attempt to correct for the effect of
seeing. (Right) Using Equation \ref{eq:effradfix} to correct for
ellipticity and the PSF produces a much better starting value for
GALFIT, hence reducing the chance of finding local minima during the
minimisation phase. The green line represents a 1:1 ratio, for reference.}
\end{figure*}

Half-light radii from Source Extractor are modified before being fed
into GALFIT. This is to account for the difference in radii definitions
between the two programs. Source Extractor's FLUX\_RADIUS parameter
outputs a circularised radius which is based on PSF convolved imaging
data. The format of GALFIT's initial estimate of the half-light radius
is that along the semi-major axis which is intrinsic to the object
(i.e. - deconvolved from the PSF). Equation \ref{eq:effradfix} converts
Source Extractor circularised radii into semi-major intrinsic radii
appropriate for GALFIT. Figure \ref{fig:radiicomp} displays the before
(uncorrected) and after (corrected) Source Extractor half-light radii
against their output modelled half-light radii from the SIGMA full
sample, coloured according to their predicted morphological type as
detailed in Section \ref{sub:bimodal}. Unmodified Source Extractor
half-light radii are a poor predictor of final modelled GALFIT half-light
radii, as expected. Large galaxies ($r_{e}>4$ pixels) tend to have
their sizes underestimated by Source Extractor by as much as $50\%$.
Following a turn off at $r_{e}\sim4$ pixels, small galaxies tend
to have their sizes overestimated by Source Extractor. Once these
data have been corrected, we find a marked increase in the agreement
between the two measures, notably so for late-type galaxies (blue
data points). Data above $r_{e}>4$ pixels has been reduced to minimal
scatter about a $1$:$1$ correlation. The effect of the turn-off
has been significantly mitigated, yet not entirely diminished. This
indicates the difficulty in accurate size estimation of galaxies that
only occupy a matter of a few pixels. Correcting radii in this manner
significantly reduces the chance of GALFIT finding a local-minima
during the minimisation phase, and consequently reduces the risk of
convergence on a non-physical solution.

\begin{figure*}
\includegraphics[width=1\textwidth]{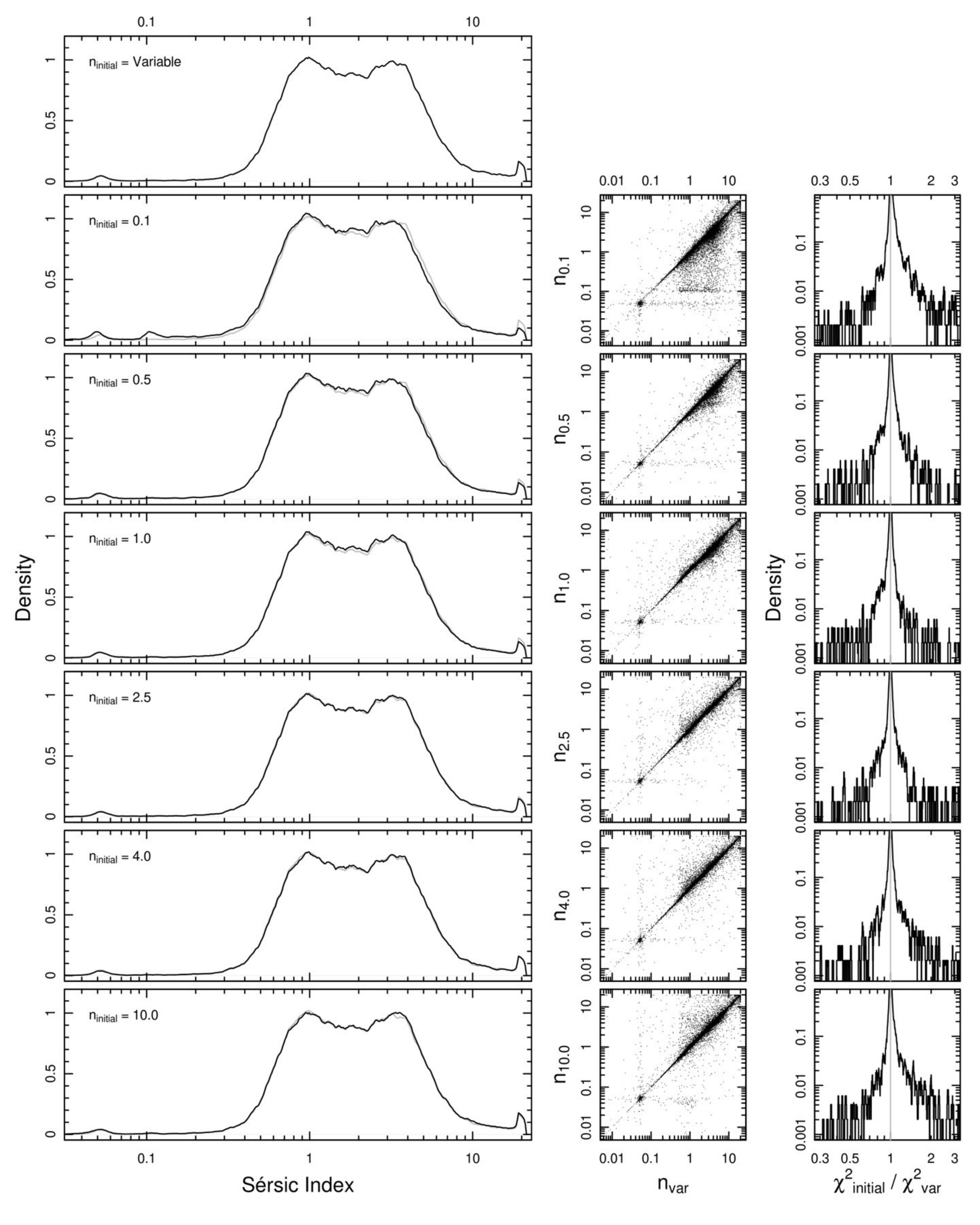}

\caption{\label{fig:initialindex}A plot comparing final modelled Sérsic indices
for a sample of $49,395$ galaxies in the $r$ band given different
initial Sérsic indices, as shown. From top to bottom, the initial
Sérsic indices fed into GALFIT are; variable (see Equation \ref{eq:varindex});
$0.1$; $0.5$; $1.0$; $2.5$; $4.0$; $10.0$. Underlying the fixed
initial Sérsic index distributions is the distribution for the variable
Sérsic index coloured in grey, for reference.}
\end{figure*}

There is no obvious proxy for Sérsic index in the default Source Extractor
parameters file. An approximation was created based on the trend between
the output Sérsic index and the ratio between the corrected half-light
radius to the Kron radius for a small test sample of trusted galaxies.
From this we were able to derive a relation for a predicted variable
Sérsic index:
\begin{equation}
n_{\mathrm{var}}=10^{-8.6\left(\frac{r_{e}}{r_{\mathrm{Kron}}}\right)+2.8}\label{eq:varindex}
\end{equation}
where $r_{e}$ is the corrected Source Extractor half-light radius
(Equation \ref{eq:effradfix}) and $r_{\mathrm{Kron}}$ is the Source
Extractor Kron radius. Figure \ref{fig:initialindex} shows the density
distributions between the variable Sérsic index, $n_{\mathrm{var}}$,
and several other initial conditions for a sample of $49,395$ galaxies
in the $r$ band. The other input parameters (size, position angle,
ellipticity, magnitude, position) are not modified. These results
show that the final recovered Sérsic index is largely independent
of its initial condition, with the notable exception of a bump in
the distribution at $n=0.1$ for $n_{initial}=0.1$ and a variable
height spike of failed objects at $n\sim20$. The $n=0.1$ bump represents
galaxies whose initial Sérsic index guess is placed too far away from
its true value, and so fails to successfully migrate away from the
initial parameter space using the Levenberg-Marquart method employed
by GALFIT. It appears that the minor fluctuations found in the main
body of the distributions directly correspond to the varying height
of the $n\sim20$ spike. Despite these features, it is clear that
the initial Sérsic index is afforded a great deal of variability in
order to achieve a successful and consistent fit. The majority of
distributions presented in Figure \ref{fig:initialindex} show little
variation, with similar levels of success and failure. It was therefore
felt that a simple $n_{\mathrm{initial}}=2.5$ would be an appropriate
initial condition as it lies in the middle of the expected probability
space, yet not at the value of either of the bimodal peaks.

\label{lastpage}
\end{document}